\documentclass[copyright]{eptcs}
\providecommand{\event}{TermGraph 2011}

\makeatletter
\@ifundefined{lhs2tex.lhs2tex.sty.read}%
  {\@namedef{lhs2tex.lhs2tex.sty.read}{}%
   \newcommand\SkipToFmtEnd{}%
   \newcommand\EndFmtInput{}%
   \long\def\SkipToFmtEnd#1\EndFmtInput{}%
  }\SkipToFmtEnd

\newcommand\ReadOnlyOnce[1]{\@ifundefined{#1}{\@namedef{#1}{}}\SkipToFmtEnd}
\usepackage{amstext}
\usepackage{amssymb}
\usepackage{stmaryrd}
\DeclareFontFamily{OT1}{cmtex}{}
\DeclareFontShape{OT1}{cmtex}{m}{n}
  {<5><6><7><8>cmtex8
   <9>cmtex9
   <10><10.95><12><14.4><17.28><20.74><24.88>cmtex10}{}
\DeclareFontShape{OT1}{cmtex}{m}{it}
  {<-> ssub * cmtt/m/it}{}

\DeclareFontShape{OT1}{cmtt}{bx}{n}
  {<5><6><7><8>cmtt8
   <9>cmbtt9
   <10><10.95><12><14.4><17.28><20.74><24.88>cmbtt10}{}
\DeclareFontShape{OT1}{cmtex}{bx}{n}
  {<-> ssub * cmtt/bx/n}{}

\newcommand{\Conid}[1]{\mathit{#1}}
\newcommand{\Varid}[1]{\mathit{#1}}
\newcommand{\anonymous}{\kern0.06em \vbox{\hrule\@width.5em}}
\newcommand{\plus}{\mathbin{+\!\!\!+}}

\usepackage{polytable}

\@ifundefined{mathindent}%
  {\newdimen\mathindent\mathindent\leftmargini}%
  {}%

\def\resethooks{%
  \global\let\SaveRestoreHook\empty
  \global\let\ColumnHook\empty}
\newcommand*{\savecolumns}[1][default]%
  {\g@addto@macro\SaveRestoreHook{\savecolumns[#1]}}
\newcommand*{\restorecolumns}[1][default]%
  {\g@addto@macro\SaveRestoreHook{\restorecolumns[#1]}}
\newcommand*{\aligncolumn}[2]%
  {\g@addto@macro\ColumnHook{\column{#1}{#2}}}

\resethooks

\newcommand{\onelinecommentchars}{\quad-{}- }
\newcommand{\commentbeginchars}{\enskip\{-}
\newcommand{\commentendchars}{-\}\enskip}

\newcommand{\visiblecomments}{%
  \let\onelinecomment=\onelinecommentchars
  \let\commentbegin=\commentbeginchars
  \let\commentend=\commentendchars}

\newcommand{\invisiblecomments}{%
  \let\onelinecomment=\empty
  \let\commentbegin=\empty
  \let\commentend=\empty}

\visiblecomments

\newlength{\blanklineskip}
\newcommand{\hsindent}[1]{\quad}
\let\hspre\empty
\let\hspost\empty

\EndFmtInput
\makeatother
\ReadOnlyOnce{polycode.fmt}%
\makeatletter

\newcommand{\hsnewpar}[1]%
  {{\parskip=0pt\parindent=0pt\par\vskip #1\noindent}}

\newcommand{\hscodestyle}{}

\newcommand{\sethscode}[1]%
  {\expandafter\let\expandafter\hscode\csname #1\endcsname
   \expandafter\let\expandafter\endhscode\csname end#1\endcsname}

  {\par\noindent
   \advance\leftskip\mathindent
   \hscodestyle
   \let\\=\@normalcr
   \let\hspre\(\let\hspost\)%
   \pboxed}%
  {\endpboxed\)%
   \par\noindent
   \ignorespacesafterend}

  {\hsnewpar\abovedisplayskip
   \advance\leftskip\mathindent
   \hscodestyle
   \let\hspre\(\let\hspost\)%
   \pboxed}%
  {\endpboxed%
   \hsnewpar\belowdisplayskip
   \ignorespacesafterend}

  {\hsnewpar\abovedisplayskip
   \advance\leftskip\mathindent
   \hscodestyle
   \let\\=\@normalcr
   \(\pboxed}%
  {\endpboxed\)%
   \hsnewpar\belowdisplayskip
   \ignorespacesafterend}

\newcommand{\plainhs}{\sethscode{plainhscode}}

\plainhs

  {\hsnewpar\abovedisplayskip
   \advance\leftskip\mathindent
   \hscodestyle
   \let\\=\@normalcr
   \(\parray}%
  {\endparray\)%
   \hsnewpar\belowdisplayskip
   \ignorespacesafterend}

  {\parray}{\endparray}

  {\(\parray}{\endparray\)}

\def\codeframewidth{\arrayrulewidth}
\RequirePackage{calc}

  {\parskip=\abovedisplayskip\par\noindent
   \hscodestyle
   \arrayrulewidth=\codeframewidth
   \tabular{@{}|p{\linewidth-2\arraycolsep-2\arrayrulewidth-2pt}|@{}}%
   \hline\framedhslinecorrect\\{-1.5ex}%
   \let\endoflinesave=\\
   \let\\=\@normalcr
   \(\pboxed}%
  {\endpboxed\)%
   \framedhslinecorrect\endoflinesave{.5ex}\hline
   \endtabular
   \parskip=\belowdisplayskip\par\noindent
   \ignorespacesafterend}

\newcommand{\framedhslinecorrect}[2]%
  {#1[#2]}

  {\(\def\column##1##2{}%
   \let\>\undefined\let\<\undefined\let\\\undefined
   \newcommand\>[1][]{}\newcommand\<[1][]{}\newcommand\\[1][]{}%
   \def\fromto##1##2##3{##3}%
   }{\) }%

  {\let\orighscode=\hscode
   \let\origendhscode=\endhscode
   \def\endhscode{\def\hscode{\endgroup\def\@currenvir{hscode}\\}\begingroup}
   \orighscode\def\hscode{\endgroup\def\@currenvir{hscode}}}%
  {\origendhscode
   \global\let\hscode=\orighscode
   \global\let\endhscode=\origendhscode}%

\makeatother
\EndFmtInput
\ReadOnlyOnce{agda.fmt}%

\RequirePackage[T1]{fontenc}
\RequirePackage[utf8x]{inputenc}
\RequirePackage{ucs}
\RequirePackage{amsfonts}

\DeclareUnicodeCharacter{737}{\textsuperscript{l}}
\DeclareUnicodeCharacter{8718}{\ensuremath{\blacksquare}}
\DeclareUnicodeCharacter{8759}{::}
\DeclareUnicodeCharacter{9669}{\ensuremath{\triangleleft}}
\DeclareUnicodeCharacter{8799}{\ensuremath{\stackrel{\scriptscriptstyle ?}{=}}}
\DeclareUnicodeCharacter{10214}{\ensuremath{\llbracket}}
\DeclareUnicodeCharacter{10215}{\ensuremath{\rrbracket}}

\renewcommand\Varid[1]{\mathord{\textsf{#1}}}
\let\Conid\Varid
\newcommand\Keyword[1]{\textsf{\textbf{#1}}}
\EndFmtInput

\DeclareUnicodeCharacter{10230}{\ensuremath{\rightarrow}} 
\DeclareUnicodeCharacter{8271}{\ensuremath{\fcmp}} 

\DeclareUnicodeCharacter{"228D}{\ensuremath{\kern-0.2em\scriptscriptstyle{\sqcup}\kern-0.2em}} 
\DeclareUnicodeCharacter{"2192}{\ensuremath{\rightarrow}} 
\DeclareUnicodeCharacter{"21C9}{\rightrightarrows} 
\DeclareUnicodeCharacter{"22A5}{\ensuremath{\RELbot}} 
\DeclareUnicodeCharacter{"03A3}{\ensuremath{\Sigma}} 
\DeclareUnicodeCharacter{"03BB}{\ensuremath{\lambda}} 

\DeclareUnicodeCharacter{"211B}{\ensuremath{\mathcal{R}\kern-0.2em}} 
\DeclareUnicodeCharacter{"212F}{{\it{e}}} 
\DeclareUnicodeCharacter{"2113}{{\it{l}}} 

\DeclareUnicodeCharacter{9585}{\ensuremath{\RELlresR}} 
\DeclareUnicodeCharacter{9586}{\ensuremath{\RELrresR}} 

\DeclareUnicodeCharacter{10231}{\ensuremath{\longleftrightarrow}} 

\DeclareUnicodeCharacter{8614}{\ensuremath{\mapsto}} 

\DeclareUnicodeCharacter{8636}{\ensuremath{\leftharpoonup}} 

\DeclareUnicodeCharacter{9001}{\ensuremath{\langle}} 

\DeclareUnicodeCharacter{9002}{\ensuremath{\rangle}} 

\DeclareUnicodeCharacter{9587}{\ensuremath{\RELsyqInfixOP}} 
\DeclareUnicodeCharacter{9659}{\ensuremath{\rhd}} 

\DeclareUnicodeCharacter{9661}{\ensuremath{\triangledown}} 
\DeclareUnicodeCharacter{9665}{\ensuremath{\triangleleft}} 

\DeclareUnicodeCharacter{9733}{\ensuremath{\star}} 

\DeclareUnicodeCharacter{739}{\ensuremath{{}^{*}}} 

\DeclareUnicodeCharacter{9312}{\ensuremath{\triv}} 

\DeclareUnicodeCharacter{8600}{\ensuremath{\searrow}} 

\DeclareUnicodeCharacter{8601}{\ensuremath{\swarrow}} 

\DeclareUnicodeCharacter{915}{\ensuremath{\Gamma}} 
\DeclareUnicodeCharacter{945}{\ensuremath{\alpha}} 
\DeclareUnicodeCharacter{946}{\ensuremath{\beta}} 
\DeclareUnicodeCharacter{947}{\ensuremath{\gamma}} 
\DeclareUnicodeCharacter{948}{\ensuremath{\delta}} 
\DeclareUnicodeCharacter{966}{\ensuremath{\phi}} 

\renewcommand{\hscodestyle}{\small}

\makeatletter
\def\mkcommand#1{\expandafter\gdef\csname #1\endcsname}
\makeatother

\title{Dependently-Typed Formalisation of Typed Term Graphs}
\def\titlerunning{Dependently-Typed Formalisation of Typed Term Graphs}

\def\authorrunning{Wolfram Kahl}
\author{Wolfram Kahl
\institute{McMaster University,
Hamilton, Ontario, Canada,
\email{kahl@cas.mcmaster.ca}
}
}

       \usepackage{latexsym,amssymb}
       \usepackage{graphicx}

\def\CGpicRaw#1#2{\includegraphics#2{pictures/#1.ps}}
\def\CGpic#1{\CGpicRaw{#1}{[scale=0.3]}}
\def\CGpicO#1#2{\CGpicRaw{#1}{[scale=0.3,#2]}}

\usepackage{breakurl} 

\usepackage{comment}

\usepackage{rmthm}
\renewcommand{\theoremclosing}{}
\makeatletter
\def\@thmcountersep{.}
\makeatother

\newtheorem{Def}{Definition}[section]

\usepackage{theref}
\def\TheRefWithPageRef#1{}

\newref{Def}{Def.\null{}}

\newdef{The}{Theorem}
\newdef{Lem}{Lemma}
\newdef{Prop}{Proposition}
\newref{Prop}{Prop.\null{}}
\newdef{Cor}{Corollary}
\newdef{Conv}{Convention}
\newdef{Fact}{Fact}

\def\sectionname{Sect.\null{}}
\usepackage{sectref}

\usepackage{slimfuzz}
\def\fcmp{{\scriptstyle\comp}}

\usepackage{RelNotation}
\RELsquareSymbols

\usepackage{formac}
\usepackage{qed}
\let\QED=\qed

\usepackage{objA}
\usepackage{CalABC}

\renewcommand{\arraystretch}{1.1}

\newcommand{\tlowername}[2]%
{$\stackrel{\makebox[1pt]{#1}}%
{\begin{picture}(0,0)%
\put(0,0){\makebox(0,6)[t]{\makebox[1pt]{$#2$}}}%
\end{picture}}$}%

\newcommand{\AR}[1]%
{\begin{picture}(#1,0)%
\put(0,0){\vector(1,0){#1}}%
\end{picture}}%

\newcommand{\DOTAR}[1]%
{\NUMBEROFDOTS=#1%
\divide\NUMBEROFDOTS by 3%
\begin{picture}(#1,0)%
\multiput(0,0)(3,0){\NUMBEROFDOTS}{\circle*{1}}%
\put(#1,0){\vector(1,0){0}}%
\end{picture}}%

\newcommand{\MONO}[1]%
{\begin{picture}(#1,0)%
\put(0,0){\vector(1,0){#1}}%
\put(2,-2){\line(0,1){4}}%
\end{picture}}%

\newcommand{\EPI}[1]%
{\begin{picture}(#1,0)(-#1,0)%
\put(-#1,0){\vector(1,0){#1}}%
\put(-6,-2){\line(0,1){4}}%
\end{picture}}%

\newcommand{\BIMO}[1]%
{\begin{picture}(#1,0)(-#1,0)%
\put(-#1,0){\vector(1,0){#1}}%
\put(-6,-2){\line(0,1){4}}%
\put(-#1,-2){\hspace{2pt}\line(0,1){4}}%
\end{picture}}%

\newcommand{\BIAR}[1]%
{\begin{picture}(#1,4)%
\put(0,0){\vector(1,0){#1}}%
\put(0,4){\vector(1,0){#1}}%
\end{picture}}%

\newcommand{\EQL}[1]%
{\begin{picture}(#1,0)%
\put(0,1){\line(1,0){#1}}%
\put(0,-1){\line(1,0){#1}}%
\end{picture}}%

\newcommand{\ADJAR}[1]%
{\begin{picture}(#1,4)%
\put(0,0){\vector(1,0){#1}}%
\put(#1,4){\vector(-1,0){#1}}%
\end{picture}}%

\newcommand{\BKAR}[1]%
{\begin{picture}(#1,0)%
\put(#1,0){\vector(-1,0){#1}}%
\end{picture}}%

\newcommand{\BKDOTAR}[1]%
{\NUMBEROFDOTS=#1%
\divide\NUMBEROFDOTS by 3%
\begin{picture}(#1,0)%
\multiput(#1,0)(-3,0){\NUMBEROFDOTS}{\circle*{1}}%
\put(0,0){\vector(-1,0){0}}%
\end{picture}}%

\newcommand{\BKMONO}[1]%
{\begin{picture}(#1,0)(-#1,0)%
\put(0,0){\vector(-1,0){#1}}%
\put(-2,-2){\line(0,1){4}}%
\end{picture}}%

\newcommand{\BKEPI}[1]%
{\begin{picture}(#1,0)%
\put(#1,0){\vector(-1,0){#1}}%
\put(6,-2){\line(0,1){4}}%
\end{picture}}%

\newcommand{\BKBIMO}[1]%
{\begin{picture}(#1,0)%
\put(#1,0){\vector(-1,0){#1}}%
\put(6,-2){\line(0,1){4}}%
\put(#1,-2){\hspace{-2pt}\line(0,1){4}}%
\end{picture}}%

\newcommand{\BKBIAR}[1]%
{\begin{picture}(#1,4)%
\put(#1,0){\vector(-1,0){#1}}%
\put(#1,4){\vector(-1,0){#1}}%
\end{picture}}%

\newcommand{\BKADJAR}[1]%
{\begin{picture}(#1,4)%
\put(0,4){\vector(1,0){#1}}%
\put(#1,0){\vector(-1,0){#1}}%
\end{picture}}%

\newcommand{\lowername}[2]%
{$\stackrel{\makebox[1pt]{#1}}%
{\begin{picture}(0,0)%
\truex{600}%
\put(0,-600){\makebox(0,\value{x})[t]{\makebox[1pt]{$#2$}}}%
\end{picture}}$}%

\newcommand{\hcase}[2]%
{\makebox[0pt]%
{\raisebox{-1pt}[0pt][0pt]{#1{#2}}}}%

\newcommand{\Hcase}[3]%
{\makebox[0pt]
{\raisebox{-1pt}[0pt][0pt]%
{$\stackrel{\makebox[0pt]{$\textstyle{#2}$}}{#1{#3}}$}}}%

\newcommand{\hcasE}[3]%
{\makebox[0pt]%
{\raisebox{-2pt}[0pt][0pt]%
{\lowername{#1{#3}}{#2}}}}%

\newcommand{\hbicase}[2]%
{\makebox[0pt]%
{\raisebox{-2.5pt}[0pt][0pt]{#1{#2}}}}%

\newcommand{\Hbicase}[4]%
{\makebox[0pt]
{\raisebox{-10.5pt}[0pt][0pt]%
{$\stackrel{\makebox[0pt]{$\textstyle{#2}$}}%
{\mbox{\lowername{#1{#4}}{#3}}}$}}}%

\newcommand{\EAR}[1]%
{\begin{picture}(#1,0)%
\put(0,0){\vector(1,0){#1}}%
\end{picture}}%

\newcommand{\EDOTAR}[1]%
{\truex{100}\truey{300}%
\NUMBEROFDOTS=#1%
\divide\NUMBEROFDOTS by \value{y}%
\begin{picture}(#1,0)%
\multiput(0,0)(\value{y},0){\NUMBEROFDOTS}%
{\circle*{\value{x}}}%
\put(#1,0){\vector(1,0){0}}%
\end{picture}}%

\newcommand{\EMONO}[1]%
{\begin{picture}(#1,0)%
\put(0,0){\vector(1,0){#1}}%
\truex{300}\truey{600}%
\put(\value{x},-\value{x}){\line(0,1){\value{y}}}%
\end{picture}}%

\newcommand{\EEPI}[1]%
{\begin{picture}(#1,0)(-#1,0)%
\put(-#1,0){\vector(1,0){#1}}%
\truex{300}\truey{600}\truez{800}%
\put(-\value{z},-\value{x}){\line(0,1){\value{y}}}%
\end{picture}}%

\newcommand{\EBIMO}[1]%
{\begin{picture}(#1,0)(-#1,0)%
\put(-#1,0){\vector(1,0){#1}}%
\truex{300}\truey{600}\truez{800}%
\put(-\value{z},-\value{x}){\line(0,1){\value{y}}}%
\put(-#1,-\value{x}){\hspace{3pt}\line(0,1){\value{y}}}%
\end{picture}}%

\newcommand{\EBIAR}[1]%
{\truex{400}%
\begin{picture}(#1,\value{x})%
\put(0,0){\vector(1,0){#1}}%
\put(0,\value{x}){\vector(1,0){#1}}%
\end{picture}}%

\newcommand{\EEQL}[1]%
{\begin{picture}(#1,0)%
\truex{200}%
\put(0,\value{x}){\line(1,0){#1}}%
\put(0,0){\line(1,0){#1}}%
\end{picture}}%

\newcommand{\EADJAR}[1]%
{\truex{400}%
\begin{picture}(#1,\value{x})%
\put(0,0){\vector(1,0){#1}}%
\put(#1,\value{x}){\vector(-1,0){#1}}%
\end{picture}}%

\newcommand{\ear}%
{\hspace{\SOURCE\unitlength}%
\hcase{\EAR}{\ARROWLENGTH}}%

\newcommand{\Ear}[1]%
{\hspace{\SOURCE\unitlength}%
\Hcase{\EAR}{#1}{\ARROWLENGTH}}%

\newcommand{\eaR}[1]%
{\hspace{\SOURCE\unitlength}%
\hcasE{\EAR}{#1}{\ARROWLENGTH}}%

\newcommand{\edotar}%
{\hspace{\SOURCE\unitlength}%
\hcase{\EDOTAR}{\ARROWLENGTH}}%

\newcommand{\Edotar}[1]%
{\hspace{\SOURCE\unitlength}%
\Hcase{\EDOTAR}{#1}{\ARROWLENGTH}}%

\newcommand{\edotaR}[1]%
{\hspace{\SOURCE\unitlength}%
\hcasE{\EDOTAR}{#1}{\ARROWLENGTH}}%

\newcommand{\emono}%
{\hspace{\SOURCE\unitlength}%
\hcase{\EMONO}{\ARROWLENGTH}}%

\newcommand{\Emono}[1]%
{\hspace{\SOURCE\unitlength}%
\Hcase{\EMONO}{#1}{\ARROWLENGTH}}%

\newcommand{\emonO}[1]%
{\hspace{\SOURCE\unitlength}%
\hcasE{\EMONO}{#1}{\ARROWLENGTH}}%

\newcommand{\eepi}%
{\hspace{\SOURCE\unitlength}%
\hcase{\EEPI}{\ARROWLENGTH}}%

\newcommand{\Eepi}[1]%
{\hspace{\SOURCE\unitlength}%
\Hcase{\EEPI}{#1}{\ARROWLENGTH}}%

\newcommand{\eepI}[1]%
{\hspace{\SOURCE\unitlength}%
\hcasE{\EEPI}{#1}{\ARROWLENGTH}}%

\newcommand{\ebimo}%
{\hspace{\SOURCE\unitlength}%
\hcase{\EBIMO}{\ARROWLENGTH}}%

\newcommand{\Ebimo}[1]%
{\hspace{\SOURCE\unitlength}%
\Hcase{\EBIMO}{#1}{\ARROWLENGTH}}%

\newcommand{\ebimO}[1]%
{\hspace{\SOURCE\unitlength}%
\hcasE{\EBIMO}{#1}{\ARROWLENGTH}}%

\newcommand{\eiso}%
{\hspace{\SOURCE\unitlength}%
\Hcase{\EAR}{\cong}{\ARROWLENGTH}}%

\newcommand{\Eiso}[1]%
{\hspace{\SOURCE\unitlength}%
\Hcase{\EAR}{\cong#1}{\ARROWLENGTH}}%

\newcommand{\eisO}[1]%
{\hspace{\SOURCE\unitlength}%
\hcasE{\EAR}{\cong#1}{\ARROWLENGTH}}%

\newcommand{\ebiar}%
{\hspace{\SOURCE\unitlength}%
\hbicase{\EBIAR}{\ARROWLENGTH}}%

\newcommand{\Ebiar}[2]%
{\hspace{\SOURCE\unitlength}%
\Hbicase{\EBIAR}{#1}{#2}{\ARROWLENGTH}}%

\newcommand{\eeql}%
{\hspace{\SOURCE\unitlength}%
\hbicase{\EEQL}{\ARROWLENGTH}}%

\newcommand{\eadjar}%
{\hspace{\SOURCE\unitlength}%
\hbicase{\EADJAR}{\ARROWLENGTH}}%

\newcommand{\Eadjar}[2]%
{\hspace{\SOURCE\unitlength}%
\Hbicase{\EADJAR}{#1}{#2}{\ARROWLENGTH}}%

\newcommand{\WAR}[1]%
{\begin{picture}(#1,0)%
\put(#1,0){\vector(-1,0){#1}}%
\end{picture}}%

\newcommand{\WDOTAR}[1]%
{\truex{100}\truey{300}%
\NUMBEROFDOTS=#1%
\divide\NUMBEROFDOTS by \value{y}%
\begin{picture}(#1,0)%
\multiput(#1,0)(-\value{y},0){\NUMBEROFDOTS}%
{\circle*{\value{x}}}%
\put(0,0){\vector(-1,0){0}}%
\end{picture}}%

\newcommand{\WMONO}[1]%
{\begin{picture}(#1,0)(-#1,0)%
\put(0,0){\vector(-1,0){#1}}%
\truex{300}\truey{600}%
\put(-\value{x},-\value{x}){\line(0,1){\value{y}}}%
\end{picture}}%

\newcommand{\WEPI}[1]%
{\begin{picture}(#1,0)%
\put(#1,0){\vector(-1,0){#1}}%
\truex{300}\truey{600}\truez{800}%
\put(\value{z},-\value{x}){\line(0,1){\value{y}}}%
\end{picture}}%

\newcommand{\WBIMO}[1]%
{\begin{picture}(#1,0)%
\put(#1,0){\vector(-1,0){#1}}%
\truex{300}\truey{600}\truez{800}%
\put(\value{z},-\value{x}){\line(0,1){\value{y}}}%
\put(#1,-\value{x}){\hspace{-3pt}\line(0,1){\value{y}}}%
\end{picture}}%

\newcommand{\WBIAR}[1]%
{\truex{400}%
\begin{picture}(#1,\value{x})%
\put(#1,0){\vector(-1,0){#1}}%
\put(#1,\value{x}){\vector(-1,0){#1}}%
\end{picture}}%

\newcommand{\WADJAR}[1]%
{\truex{400}%
\begin{picture}(#1,\value{x})%
\put(0,\value{x}){\vector(1,0){#1}}%
\put(#1,0){\vector(-1,0){#1}}%
\end{picture}}%

\newcommand{\war}%
{\hspace{\SOURCE\unitlength}%
\hcase{\WAR}{\ARROWLENGTH}}%

\newcommand{\War}[1]%
{\hspace{\SOURCE\unitlength}%
\Hcase{\WAR}{#1}{\ARROWLENGTH}}%

\newcommand{\waR}[1]%
{\hspace{\SOURCE\unitlength}%
\hcasE{\WAR}{#1}{\ARROWLENGTH}}%

\newcommand{\wdotar}%
{\hspace{\SOURCE\unitlength}%
\hcase{\WDOTAR}{\ARROWLENGTH}}%

\newcommand{\Wdotar}[1]%
{\hspace{\SOURCE\unitlength}%
\Hcase{\WDOTAR}{#1}{\ARROWLENGTH}}%

\newcommand{\wdotaR}[1]%
{\hspace{\SOURCE\unitlength}%
\hcasE{\WDOTAR}{#1}{\ARROWLENGTH}}%

\newcommand{\wmono}%
{\hspace{\SOURCE\unitlength}%
\hcase{\WMONO}{\ARROWLENGTH}}%

\newcommand{\Wmono}[1]%
{\hspace{\SOURCE\unitlength}%
\Hcase{\WMONO}{#1}{\ARROWLENGTH}}%

\newcommand{\wmonO}[1]%
{\hspace{\SOURCE\unitlength}%
\hcasE{\WMONO}{#1}{\ARROWLENGTH}}%

\newcommand{\wepi}%
{\hspace{\SOURCE\unitlength}%
\hcase{\WEPI}{\ARROWLENGTH}}%

\newcommand{\Wepi}[1]%
{\hspace{\SOURCE\unitlength}%
\Hcase{\WEPI}{#1}{\ARROWLENGTH}}%

\newcommand{\wepI}[1]%
{\hspace{\SOURCE\unitlength}%
\hcasE{\WEPI}{#1}{\ARROWLENGTH}}%

\newcommand{\wbimo}%
{\hspace{\SOURCE\unitlength}%
\hcase{\WBIMO}{\ARROWLENGTH}}%

\newcommand{\Wbimo}[1]%
{\hspace{\SOURCE\unitlength}%
\Hcase{\WBIMO}{#1}{\ARROWLENGTH}}%

\newcommand{\wbimO}[1]%
{\hspace{\SOURCE\unitlength}%
\hcasE{\WBIMO}{#1}{\ARROWLENGTH}}%

\newcommand{\wiso}%
{\hspace{\SOURCE\unitlength}%
\Hcase{\WAR}{\cong}{\ARROWLENGTH}}%

\newcommand{\Wiso}[1]%
{\hspace{\SOURCE\unitlength}%
\Hcase{\WAR}{#1}{\ARROWLENGTH}}%

\newcommand{\wisO}[1]%
{\hspace{\SOURCE\unitlength}%
\hcasE{\WAR}{#1}{\ARROWLENGTH}}%

\newcommand{\wbiar}%
{\hspace{\SOURCE\unitlength}%
\hbicase{\WBIAR}{\ARROWLENGTH}}%

\newcommand{\Wbiar}[2]%
{\hspace{\SOURCE\unitlength}%
\Hbicase{\WBIAR}{#1}{#2}{\ARROWLENGTH}}%

\newcommand{\weql}%
{\hspace{\SOURCE\unitlength}%
\hbicase{\EEQL}{\ARROWLENGTH}}%

\newcommand{\wadjar}%
{\hspace{\SOURCE\unitlength}%
\hbicase{\WADJAR}{\ARROWLENGTH}}%

\newcommand{\Wadjar}[2]%
{\hspace{\SOURCE\unitlength}%
\Hbicase{\WADJAR}{#1}{#2}{\ARROWLENGTH}}%

\newcommand{\Vcase}[3]{\makebox[0pt]%
{\makebox[0pt][r]{\raisebox{0pt}[0pt][0pt]{${#2}\hspace{2pt}$}}}#1{#3}}%

\newcommand{\vcasE}[3]{\makebox[0pt]%
{#1{#3}\makebox[0pt][l]{\raisebox{0pt}[0pt][0pt]{\hspace{2pt}$#2$}}}}%

\newcommand{\Vbicase}[4]{\makebox[0pt]%
{\makebox[0pt][r]{\raisebox{0pt}[0pt][0pt]{$#2$\hspace{4pt}}}#1{#4}%
\makebox[0pt][l]{\raisebox{0pt}[0pt][0pt]{\hspace{5pt}$#3$}}}}%

\newcommand{\SAR}[1]%
{\begin{picture}(0,0)%
\put(0,0){\makebox(0,0)%
{\begin{picture}(0,#1)%
\put(0,#1){\vector(0,-1){#1}}%
\end{picture}}}\end{picture}}%

\newcommand{\SDOTAR}[1]%
{\truex{100}\truey{300}%
\NUMBEROFDOTS=#1%
\divide\NUMBEROFDOTS by \value{y}%
\begin{picture}(0,0)%
\put(0,0){\makebox(0,0)%
{\begin{picture}(0,#1)%
\multiput(0,#1)(0,-\value{y}){\NUMBEROFDOTS}%
{\circle*{\value{x}}}%
\put(0,0){\vector(0,-1){0}}%
\end{picture}}}\end{picture}}%

\newcommand{\SMONO}[1]%
{\begin{picture}(0,0)%
\put(0,0){\makebox(0,0)%
{\begin{picture}(0,#1)%
\put(0,#1){\vector(0,-1){#1}}%
\truex{300}\truey{600}%
\put(0,#1){\begin{picture}(0,0)%
\put(-\value{x},-\value{x}){\line(1,0){\value{y}}}\end{picture}}%
\end{picture}}}\end{picture}}%

\newcommand{\SEPI}[1]%
{\begin{picture}(0,0)%
\put(0,0){\makebox(0,0)%
{\begin{picture}(0,#1)%
\put(0,#1){\vector(0,-1){#1}}%
\truex{300}\truey{600}\truez{800}%
\put(-\value{x},\value{z}){\line(1,0){\value{y}}}%
\end{picture}}}\end{picture}}%

\newcommand{\SBIMO}[1]%
{\begin{picture}(0,0)%
\put(0,0){\makebox(0,0)%
{\begin{picture}(0,#1)%
\put(0,#1){\vector(0,-1){#1}}%
\truex{300}\truey{600}\truez{800}%
\put(0,#1){\begin{picture}(0,0)%
\put(-\value{x},-\value{x}){\line(1,0){\value{y}}}\end{picture}}%
\put(-\value{x},\value{z}){\line(1,0){\value{y}}}%
\end{picture}}}\end{picture}}%

\newcommand{\SBIAR}[1]%
{\begin{picture}(0,0)%
\truex{200}%
\put(0,0){\makebox(0,0)%
{\begin{picture}(0,#1)\put(-\value{x},#1){\vector(0,-1){#1}}%
\put(\value{x},#1){\vector(0,-1){#1}}%
\end{picture}}}\end{picture}}%

\newcommand{\SEQL}[1]%
{\begin{picture}(0,0)%
\truex{100}%
\put(0,0){\makebox(0,0)%
{\begin{picture}(0,#1)\put(-\value{x},#1){\line(0,-1){#1}}%
\put(\value{x},#1){\line(0,-1){#1}}%
\end{picture}}}\end{picture}}%

\newcommand{\Sarv}[2]{\Vcase{\SAR}{#1}{#200}}%

\newcommand{\Sar}[1]{\Sarv{#1}{50}}%

\newcommand{\saRv}[2]{\vcasE{\SAR}{#1}{#200}}%

\newcommand{\saR}[1]{\saRv{#1}{50}}%

\newcommand{\Sisov}[2]%
{\Vbicase{\SAR}{#1\hspace{-2pt}}{\hspace{-2pt}\cong}{#200}}%

\newcommand{\NAR}[1]%
{\begin{picture}(0,0)%
\put(0,0){\makebox(0,0)%
{\begin{picture}(0,#1)\put(0,0){\vector(0,1){#1}}%
\end{picture}}}\end{picture}}%

\newcommand{\NDOTAR}[1]%
{\truex{100}\truey{300}%
\NUMBEROFDOTS=#1%
\divide\NUMBEROFDOTS by \value{y}%
\begin{picture}(0,0)%
\put(0,0){\makebox(0,0)%
{\begin{picture}(0,#1)%
\multiput(0,0)(0,\value{y}){\NUMBEROFDOTS}%
{\circle*{\value{x}}}%
\put(0,#1){\vector(0,1){0}}%
\end{picture}}}\end{picture}}%

\newcommand{\NMONO}[1]%
{\begin{picture}(0,0)%
\put(0,0){\makebox(0,0)%
{\begin{picture}(0,#1)%
\put(0,0){\vector(0,1){#1}}%
\truex{300}\truey{600}%
\put(-\value{x},\value{x}){\line(1,0){\value{y}}}%
\end{picture}}}%
\end{picture}}%

\newcommand{\NEPI}[1]%
{\begin{picture}(0,0)%
\put(0,0){\makebox(0,0)%
{\begin{picture}(0,#1)%
\put(0,0){\vector(0,1){#1}}%
\truex{300}\truey{600}\truez{800}%
\put(0,#1){\begin{picture}(0,0)%
\put(-\value{x},-\value{z}){\line(1,0){\value{y}}}\end{picture}}%
\end{picture}}}\end{picture}}%

\newcommand{\NBIMO}[1]%
{\begin{picture}(0,0)%
\put(0,0){\makebox(0,0)%
{\begin{picture}(0,#1)%
\put(0,0){\vector(0,1){#1}}%
\truex{300}\truey{600}\truez{800}%
\put(-\value{x},\value{x}){\line(1,0){\value{y}}}%
\put(0,#1){\begin{picture}(0,0)%
\put(-\value{x},-\value{z}){\line(1,0){\value{y}}}\end{picture}}%
\end{picture}}}\end{picture}}%

\newcommand{\NBIAR}[1]%
{\begin{picture}(0,0)%
\truex{200}%
\put(0,0){\makebox(0,0)%
{\begin{picture}(0,#1)\put(-\value{x},0){\vector(0,1){#1}}%
\put(\value{x},0){\vector(0,1){#1}}%
\end{picture}}}\end{picture}}%

\newcommand{\Narv}[2]{\Vcase{\NAR}{#1}{#200}}%

\newcommand{\Nar}[1]{\Narv{#1}{50}}%

\newcommand{\Nisov}[2]%
{\Vbicase{\NAR}{#1\hspace{-2pt}}{\hspace{-2pt}\cong}{#200}}%

\newcommand{\NEDOTAR}%
{\truex{100}\truey{212}%
\NUMBEROFDOTS=5800%
\divide\NUMBEROFDOTS by \value{y}%
\begin{picture}(0,0)%
\multiput(-2900,-2900)(\value{y},\value{y}){\NUMBEROFDOTS}%
{\circle*{\value{x}}}%
\put(2900,2900){\vector(1,1){0}}%
\end{picture}}%

\newcommand{\SWDOTAR}%
{\truex{100}\truey{212}%
\NUMBEROFDOTS=5800%
\divide\NUMBEROFDOTS by \value{y}%
\begin{picture}(0,0)%
\multiput(2900,2900)(-\value{y},-\value{y}){\NUMBEROFDOTS}%
{\circle*{\value{x}}}%
\put(-2900,-2900){\vector(-1,-1){0}}%
\end{picture}}%

\newcommand{\sdcase}[3]{\begin{picture}(0,0)%
\put(0,-150){#1}%
\truex{100}\truez{600}%
\put(\value{x},\value{x}){\makebox(0,\value{z})[l]{${#2}$}}%
\truex{300}\truey{800}%
\put(-\value{x},-\value{y}){\makebox(0,\value{z})[r]{${#3}$}}%
\end{picture}}%

\newcommand{\SEDOTAR}%
{\truex{100}\truey{212}%
\NUMBEROFDOTS=5800%
\divide\NUMBEROFDOTS by \value{y}%
\begin{picture}(0,0)%
\multiput(-2900,2900)(\value{y},-\value{y}){\NUMBEROFDOTS}%
{\circle*{\value{x}}}%
\put(2900,-2900){\vector(1,-1){0}}%
\end{picture}}%

\newcommand{\Sedotar}[1]{\sdcase{\SEDOTAR}{#1}{}}%

\newcommand{\NWDOTAR}%
{\truex{100}\truey{212}%
\NUMBEROFDOTS=5800%
\divide\NUMBEROFDOTS by \value{y}%
\begin{picture}(0,0)%
\multiput(2900,-2900)(-\value{y},\value{y}){\NUMBEROFDOTS}%
{\circle*{\value{x}}}%
\put(-2900,2900){\vector(-1,1){0}}%
\end{picture}}%

\newcommand{\Nwdotar}[1]{\sdcase{\NWDOTAR}{#1}{}}%

\newcommand{\ENEAR}[2]%
{\makebox[0pt]{\begin{picture}(0,0)%
\put(0,-150){\makebox(0,0){\begin{picture}(0,0)%
\put(-6600,-3300){\vector(2,1){13200}}%
\truex{200}\truey{800}\truez{600}%
\put(-\value{x},\value{x}){\makebox(0,\value{z})[r]{${#1}$}}%
\put(\value{x},-\value{y}){\makebox(0,\value{z})[l]{${#2}$}}%
\end{picture}}}\end{picture}}}%

\newcommand{\ESEAR}[2]%
{\makebox[0pt]{\begin{picture}(0,0)%
\put(0,-150){\makebox(0,0){\begin{picture}(0,0)%
\put(-6600,3300){\vector(2,-1){13200}}%
\truex{200}\truey{800}\truez{600}%
\put(\value{x},\value{x}){\makebox(0,\value{z})[l]{${#1}$}}%
\put(-\value{x},-\value{y}){\makebox(0,\value{z})[r]{${#2}$}}%
\end{picture}}}\end{picture}}}%

\newcommand{\Esear}[1]{\ESEAR{#1}{}}%

\newcommand{\eseaR}[1]{\ESEAR{}{#1}}%

\newcommand{\WNWAR}[2]%
{\makebox[0pt]{\begin{picture}(0,0)%
\put(0,-150){\makebox(0,0){\begin{picture}(0,0)%
\put(6600,-3300){\vector(-2,1){13200}}%
\truex{200}\truey{800}\truez{600}%
\put(\value{x},\value{x}){\makebox(0,\value{z})[l]{${#1}$}}%
\put(-\value{x},-\value{y}){\makebox(0,\value{z})[r]{${#2}$}}%
\end{picture}}}\end{picture}}}%

\newcommand{\wnwaR}[1]{\WNWAR{}{#1}}%

\newcommand{\WSWAR}[2]%
{\makebox[0pt]{\begin{picture}(0,0)%
\put(0,-150){\makebox(0,0){\begin{picture}(0,0)%
\put(6600,3300){\vector(-2,-1){13200}}%
\truex{200}\truey{800}\truez{600}%
\put(-\value{x},\value{x}){\makebox(0,\value{z})[r]{${#1}$}}%
\put(\value{x},-\value{y}){\makebox(0,\value{z})[l]{${#2}$}}%
\end{picture}}}\end{picture}}}%

\newcommand{\NNEAR}[2]%
{\raisebox{-1pt}[0pt][0pt]{\begin{picture}(0,0)%
\put(0,0){\makebox(0,0){\begin{picture}(0,0)%
\put(-3300,-6600){\vector(1,2){6600}}%
\truex{100}\truez{600}%
\put(-\value{x},\value{x}){\makebox(0,\value{z})[r]{${#1}$}}%
\put(\value{x},-\value{z}){\makebox(0,\value{z})[l]{${#2}$}}%
\end{picture}}}\end{picture}}}%

\newcommand{\SSWAR}[2]%
{\raisebox{-1pt}[0pt][0pt]{\begin{picture}(0,0)%
\put(0,0){\makebox(0,0){\begin{picture}(0,0)%
\put(3300,6600){\vector(-1,-2){6600}}%
\truex{100}\truez{600}%
\put(-\value{x},\value{x}){\makebox(0,\value{z})[r]{${#1}$}}%
\put(\value{x},-\value{z}){\makebox(0,\value{z})[l]{${#2}$}}%
\end{picture}}}\end{picture}}}%

\newcommand{\SSEAR}[2]%
{\raisebox{-1pt}[0pt][0pt]{\begin{picture}(0,0)%
\put(0,0){\makebox(0,0){\begin{picture}(0,0)%
\put(-3300,6600){\vector(1,-2){6600}}%
\truex{200}\truez{600}%
\put(\value{x},\value{x}){\makebox(0,\value{z})[l]{${#1}$}}%
\put(-\value{x},-\value{z}){\makebox(0,\value{z})[r]{${#2}$}}%
\end{picture}}}\end{picture}}}%

\newcommand{\Ssear}[1]{\SSEAR{#1}{}}%

\newcommand{\sseaR}[1]{\SSEAR{}{#1}}%

\newcommand{\NNWAR}[2]%
{\raisebox{-1pt}[0pt][0pt]{\begin{picture}(0,0)%
\put(0,0){\makebox(0,0){\begin{picture}(0,0)%
\put(3300,-6600){\vector(-1,2){6600}}%
\truex{200}\truez{600}%
\put(\value{x},\value{x}){\makebox(0,\value{z})[l]{${#1}$}}%
\put(-\value{x},-\value{z}){\makebox(0,\value{z})[r]{${#2}$}}%
\end{picture}}}\end{picture}}}%

\newcommand{\Nnwar}[1]{\NNWAR{#1}{}}%

\newcommand{\Necurve}[2]%
{\begin{picture}(0,0)%
\truex{1300}\truey{2000}\truez{200}%
\put(0,\value{x}){\oval(#200,\value{y})[t]}%
\put(0,\value{x}){\makebox(0,0){\begin{picture}(#200,0)%
\put(#200,0){\vector(0,-1){\value{z}}}%
\put(0,0){\line(0,-1){\value{z}}}\end{picture}}}%
\truex{2500}%
\put(0,\value{x}){\makebox(0,0)[b]{${#1}$}}%
\end{picture}}%

\newcommand{\Nwcurve}[2]%
{\begin{picture}(0,0)%
\truex{1300}\truey{2000}\truez{200}%
\put(0,\value{x}){\oval(#200,\value{y})[t]}%
\put(0,\value{x}){\makebox(0,0){\begin{picture}(#200,0)%
\put(#200,0){\line(0,-1){\value{z}}}%
\put(0,0){\vector(0,-1){\value{z}}}\end{picture}}}%
\truex{2500}%
\put(0,\value{x}){\makebox(0,0)[b]{${#1}$}}%
\end{picture}}%

\newcommand{\Securve}[2]%
{\begin{picture}(0,0)%
\truex{1300}\truey{2000}\truez{200}%
\put(0,-\value{x}){\oval(#200,\value{y})[b]}%
\put(0,-\value{x}){\makebox(0,0){\begin{picture}(#200,0)%
\put(#200,0){\vector(0,1){\value{z}}}%
\put(0,0){\line(0,1){\value{z}}}\end{picture}}}%
\truex{2500}%
\put(0,-\value{x}){\makebox(0,0)[t]{${#1}$}}%
\end{picture}}%

\newcommand{\Swcurve}[2]%
{\begin{picture}(0,0)%
\truex{1300}\truey{2000}\truez{200}%
\put(0,-\value{x}){\oval(#200,\value{y})[b]}%
\put(0,-\value{x}){\makebox(0,0){\begin{picture}(#200,0)%
\put(#200,0){\line(0,1){\value{z}}}%
\put(0,0){\vector(0,1){\value{z}}}\end{picture}}}%
\truex{2500}%
\put(0,-\value{x}){\makebox(0,0)[t]{${#1}$}}%
\end{picture}}%

\newcommand{\Escurve}[2]%
{\begin{picture}(0,0)%
\truex{1400}\truey{2000}\truez{200}%
\put(\value{x},0){\oval(\value{y},#200)[r]}%
\put(\value{x},0){\makebox(0,0){\begin{picture}(0,#200)%
\put(0,0){\vector(-1,0){\value{z}}}%
\put(0,#200){\line(-1,0){\value{z}}}\end{picture}}}%
\truex{2500}%
\put(\value{x},0){\makebox(0,0)[l]{${#1}$}}%
\end{picture}}%

\newcommand{\Encurve}[2]%
{\begin{picture}(0,0)%
\truex{1400}\truey{2000}\truez{200}%
\put(\value{x},0){\oval(\value{y},#200)[r]}%
\put(\value{x},0){\makebox(0,0){\begin{picture}(0,#200)%
\put(0,0){\line(-1,0){\value{z}}}%
\put(0,#200){\vector(-1,0){\value{z}}}\end{picture}}}%
\truex{2500}%
\put(\value{x},0){\makebox(0,0)[l]{${#1}$}}%
\end{picture}}%

\newcommand{\Wscurve}[2]%
{\begin{picture}(0,0)%
\truex{1300}\truey{2000}\truez{200}%
\put(-\value{x},0){\oval(\value{y},#200)[l]}%
\put(-\value{x},0){\makebox(0,0){\begin{picture}(0,#200)%
\put(0,0){\vector(1,0){\value{z}}}%
\put(0,#200){\line(1,0){\value{z}}}\end{picture}}}%
\truex{2400}%
\put(-\value{x},0){\makebox(0,0)[r]{${#1}$}}%
\end{picture}}%

\newcommand{\Wncurve}[2]%
{\begin{picture}(0,0)%
\truex{1300}\truey{2000}\truez{200}%
\put(-\value{x},0){\oval(\value{y},#200)[l]}%
\put(-\value{x},0){\makebox(0,0){\begin{picture}(0,#200)%
\put(0,0){\line(1,0){\value{z}}}%
\put(0,#200){\vector(1,0){\value{z}}}\end{picture}}}%
\truex{2400}%
\put(-\value{x},0){\makebox(0,0)[r]{${#1}$}}%
\end{picture}}%

\newcount\SCALE%

\newcount\NUMBER%

\newcount\LINE%

\newcount\COLUMN%

\newcount\WIDTH%

\newcount\SOURCE%

\newcount\ARROW%

\newcount\TARGET%

\newcount\ARROWLENGTH%

\newcount\NUMBEROFDOTS%

\newcounter{x}%

\newcounter{y}%

\newcounter{z}%

\newcounter{horizontal}%

\newcounter{vertical}%

\newskip\itemlength%

\newskip\firstitem%

\newskip\seconditem%

\newcommand{\printarrow}{}%

\newcommand{\truex}[1]{%
\NUMBER=#1%
\multiply\NUMBER by 100%
\divide\NUMBER by \SCALE%
\setcounter{x}{\NUMBER}}%

\newcommand{\truey}[1]{%
\NUMBER=#1%
\multiply\NUMBER by 100%
\divide\NUMBER by \SCALE%
\setcounter{y}{\NUMBER}}%

\newcommand{\truez}[1]{%
\NUMBER=#1%
\multiply\NUMBER by 100%
\divide\NUMBER by \SCALE%
\setcounter{z}{\NUMBER}}%

\newcommand{\changecounters}[1]{%
\SOURCE=\ARROW%
\ARROW=\TARGET%
\settowidth{\itemlength}{#1}%
\ifdim \itemlength > 2800\unitlength%
\addtolength{\itemlength}{-2800\unitlength}%
\TARGET=\itemlength%
\divide\TARGET by 1310%
\multiply\TARGET by 100%
\divide\TARGET by \SCALE%
\else%
\TARGET=0%
\fi%
\ARROWLENGTH=5000%
\advance\ARROWLENGTH by -\SOURCE%
\advance\ARROWLENGTH by -\TARGET%
\advance\SOURCE by -\TARGET}%

\newcommand{\initialize}[1]{%
\LINE=0%
\COLUMN=0%
\WIDTH=0%
\ARROW=0%
\TARGET=0%
\changecounters{#1}%
\renewcommand{\printarrow}{#1}%
\begin{center}%
\vspace{10pt}%
\begin{picture}(0,0)}%

\newcommand{\DIAGV}[2]{%
\SCALE=#1%
\setlength{\unitlength}{655sp}%
\multiply\unitlength by \SCALE%
\divide\unitlength by 100%
\initialize{\mbox{$#2$}}}%

\newcommand{\n}[1]{%
\changecounters{\mbox{$#1$}}%
\put(\COLUMN,\LINE){\makebox(0,0){\printarrow}}%
\thinlines%
\renewcommand{\printarrow}{\mbox{$#1$}}%
\advance\COLUMN by 4000}%

\newcommand{\nn}[1]{%
\put(\COLUMN,\LINE){\makebox(0,0){\printarrow}}%
\thinlines%
\ifnum \WIDTH < \COLUMN%
\WIDTH=\COLUMN%
\else%
\fi%
\advance\LINE by -4000%
\COLUMN=0%
\ARROW=0%
\TARGET=0%
\changecounters{\mbox{$#1$}}%
\renewcommand{\printarrow}{\mbox{$#1$}}}%

\newcommand{\conclude}{%
\put(\COLUMN,\LINE){\makebox(0,0){\printarrow}}%
\thinlines%
\ifnum \WIDTH < \COLUMN%
\WIDTH=\COLUMN%
\else%
\fi%
\setcounter{horizontal}{\WIDTH}%
\setcounter{vertical}{-\LINE}%
\end{picture}}%

\newcommand{\diag}{%
\conclude%
\raisebox{0pt}[0pt][\value{vertical}\unitlength]{}%
\hspace*{\value{horizontal}\unitlength}%
\vspace{10pt}%
\end{center}%
\setlength{\unitlength}{1pt}}%

\newcommand{\diagv}[3]{%
\conclude%
\NUMBER=#1%
\rule{0pt}{\NUMBER pt}%
\hspace*{-#2pt}%
\raisebox{0pt}[0pt][\value{vertical}\unitlength]{}%
\hspace*{\value{horizontal}\unitlength}
\NUMBER=#3%
\advance\NUMBER by 10%
\vspace*{\NUMBER pt}%
\end{center}%
\setlength{\unitlength}{1pt}}%

\newcommand{\N}[1]%
{\raisebox{0pt}[7pt][0pt]{$#1$}}%

\newcommand{\crosslength}[2]{%
\settowidth{\firstitem}{#1}%
\settowidth{\seconditem}{#2}%
\ifdim\firstitem < \seconditem%
\itemlength=\seconditem%
\else%
\itemlength=\firstitem%
\fi%
\divide\itemlength by 2%
\hspace{\itemlength}}%

\usepackage{CAT}
\usepackage{ALG}

\usepackage{squareDiags}
\def\Chi{X}

\def\exch{\mathbb{X}}
\def\exchU#1{\mathbb{X}_{#1}}
\def\munit{\triv}
\Tdef{Nabla}{\nabla}
\TdefA{NablaU}{#1}{\nabla_{\kern-.2ex{#1}}}
\Tdef{bang}{!}
\TdefA{bangU}{#1}{\bang_{#1}}

\usepackage{dsfont}
\def\triv{\mathds{1}}

\usepackage{edcomms}

\makeatletter
\def\@listI{\leftmargin\leftmargini
            \labelsep 0.3em
            \labelwidth\leftmargini
            \advance\labelwidth-\labelsep
            \parsep 0\p@ \@plus1\p@ \@minus\p@
            \topsep 2\p@ \@plus1\p@ \@minus1\p@
            \itemsep1\p@}
\let\@listi\@listI
\@listi
\makeatother

\def\Cleft{\big\{\;}\def\Cright{\;\big\}}

\def\BMAkern{\kern-5.7pt}

\def\bbbone{{\mathchoice {\rm 1\mskip-4mu l} {\rm 1\mskip-4mu l}
{\rm 1\mskip-4.5mu l} {\rm 1\mskip-5mu l}}}

\def\triv{\bbbone}

\makeindex

\begin{document}

\maketitle
\pagestyle{headings}

\begin{abstract}
We employ the dependently-typed programming language Agda2
to explore formalisation of untyped and typed term graphs
directly as set-based graph structures,
via the gs-monoidal categories of Corradini and Gadducci,
and as nested \ensuremath{\Keyword{let}}-expressions using Pouillard and Pottier's
\ensuremath{\Conid{NotSoFresh}} library of variable-binding abstractions.
\end{abstract}

\ignore{%
}

\section{Introduction}

The Coconut project \cite{Anand-Kahl-2009b,Anand-Kahl-2009a}
uses ``code graphs'' \cite{Kahl-Anand-Carette-2005},
a variant of term graphs in the spirit of
``jungles'' \cite{Hoffmann-Plump-1991,Corradini-Rossi-1991},
as intermediate presentation for the generation of
highly optimised assembly code.
This is currently implemented in Haskell,
and we use the Haskell type system in an embedded domain-specific
language (EDSL) for creating such code graphs
via what appears to be standard Haskell function definitions,
with \ensuremath{\Keyword{let}}-definitions introducing sharing,
and with functions representing assembly-level operations
constructing hyperedges \cite{Anand-Kahl-2009b}.
However, since Haskell does not support full dependent typing,
the intermediate term graph datatype interface,
supporting graph navigation, traversal, and manipulation operations,
cannot preserve the connection with the Haskell-level typing of the assembly
operations.
Therefore, although EDSL-created code graphs are
\emph{well-typed by construction, as certified by the type checker},
this does not hold anymore for code graphs that are the result
of internal operations.
Those internal operations either require separate proof
that they preserve well-typedness,
or they need to perform run-time checks, at considerable run-time cost.

In addition, our code-graph-creation EDSL has a second
``simulator'' implementation, which turns the EDSL expressions
into Haskell functions that implement a ``machine simulation''.
Since the code graph representation has lost its connection
with the Haskell-level typing,
it is ``unintuitively hard'' to use the simulation machinery
for code graphs that result from code graph manipulation operations.

Mainly for these reasons, we are now exploring implementation
of code graphs in a dependently typed programming language,
where there is no need to ``loose'' the type information
when moving to a graph representation,
and where even stronger assertions about operations on code graphs
than just type preservation can be proven \emph{inside} the
implementing system.

We start, in \sectref{AgdaIntro}, with a quick introduction to
the dependently typed programming language
(and proof checker) Agda \cite{Norell-2007}.
This is followed by formalisations of set-based mathematical
definitions
of untyped (\sectref{Jungle}) and typed (\sectref{CodeGraph}) term
graphs,
and then a summary of the gs-monoidal category view on these term
graphs in \sectref{GSMonCat}.
Finally, we present two formalisations of acyclic term graphs
as (differently structured) nested \ensuremath{\Keyword{let}}-expressions
(Sections \sectrawref{Let} and \sectrawref{SeqLet}).


\section{Introduction to Agda: Types, Sets, Equality}\sectlabel{AgdaIntro}

The Agda home page\footnote{\url{http://wiki.portal.chalmers.se/agda/}} states:

\begin{quotation}
\textbf{Agda is a dependently typed functional programming language.} It has inductive families, i.e., data types which depend on values, such as the type of vectors of a given length. It also has parametrised modules, mixfix operators, Unicode characters, and an interactive Emacs interface which can assist the programmer in writing the program.

\textbf{Agda is a proof assistant.} It is an interactive system for writing and checking proofs. Agda is based on intuitionistic type theory, a foundational system for constructive mathematics developed by the Swedish logician Per Martin-Löf. It has many similarities with other proof assistants based on dependent types, such as Coq, Epigram, Matita and NuPRL. 
\end{quotation}

\noindent
Syntactically and ``culturally'', Agda is quite close to Haskell.
However, since Agda is strongly normalising and has no \ensuremath{\bot }
values, the underlying semantics is quite different.
Also, since Agda is dependently typed,
it does not have the distinction that Haskell has between
terms, types, and kinds (the ``types of the types'').
The Agda constant \ensuremath{\Conid{Set}} corresponds to the Haskell kind \ensuremath{\Varid{*}};
it is the type of all ``normal'' datatypes.
For example, the Agda standard library defines the type \ensuremath{\Conid{Bool}} as follows:

\begin{hscode}\SaveRestoreHook
\column{B}{@{}>{\hspre}l<{\hspost}@{}}%
\column{20}{@{}>{\hspre}l<{\hspost}@{}}%
\column{29}{@{}>{\hspre}l<{\hspost}@{}}%
\column{35}{@{}>{\hspre}l<{\hspost}@{}}%
\column{E}{@{}>{\hspre}l<{\hspost}@{}}%
\>[B]{}\Keyword{data}\;\Conid{Bool}\;\mathbin{:}\;\Conid{Set}\;{}\<[20]%
\>[20]{}\Keyword{where}\;{}\<[29]%
\>[29]{}\Varid{true}\;{}\<[35]%
\>[35]{}\mathbin{:}\;\Conid{Bool}{}\<[E]%
\\
\>[29]{}\Varid{false}\;\mathbin{:}\;\Conid{Bool}{}\<[E]%
\ColumnHook
\end{hscode}\resethooks

\noindent
Since \ensuremath{\Conid{Set}} needs again a type, there is \ensuremath{\Conid{Set₁}},
with \ensuremath{\Conid{Set}\;\mathbin{:}\;\Conid{Set₁}}, etc., resulting in a hierarchy of ``universes''.
Since version 2.2.8, Agda supports \emph{universe polymorphism},
with universes \ensuremath{\Conid{Set}\;\Varid{i}} where \ensuremath{\Varid{i}} is an element of the following
special-purpose variant of the natural numbers:

\begin{hscode}\SaveRestoreHook
\column{B}{@{}>{\hspre}l<{\hspost}@{}}%
\column{20}{@{}>{\hspre}l<{\hspost}@{}}%
\column{28}{@{}>{\hspre}l<{\hspost}@{}}%
\column{33}{@{}>{\hspre}l<{\hspost}@{}}%
\column{E}{@{}>{\hspre}l<{\hspost}@{}}%
\>[B]{}\Keyword{data}\;\Conid{Level}\;\mathbin{:}\;\Conid{Set}\;{}\<[20]%
\>[20]{}\Keyword{where}\;{}\<[28]%
\>[28]{}\Varid{zero}\;\mathbin{:}\;\Conid{Level}{}\<[E]%
\\
\>[28]{}\Varid{suc}\;{}\<[33]%
\>[33]{}\mathbin{:}\;(\Varid{i}\;\mathbin{:}\;\Conid{Level})\;\Varid{→}\;\Conid{Level}{}\<[E]%
\ColumnHook
\end{hscode}\resethooks

\noindent
With this, the conventional usage turns into syntactic sugar,
so that \ensuremath{\Conid{Set}} is now \ensuremath{\Conid{Set}\;\Varid{zero}}, and \ensuremath{\Conid{Set₁}\;\mathrel{=}\;\Conid{Set}\;(\Varid{suc}\;\Varid{zero})}.
%
\ignore{
With universe polymorphism enabled, we may quantify over
\ensuremath{\Conid{Level}}-typed variables that occur as \ensuremath{\Conid{Level}} arguments of \ensuremath{\Conid{Set}}.
Universe polymorphism is essential for being able to talk about
both ``small'' and ``large'' categories or relation algebras,
or, for another example, also
for being able to treat diagrams of graphs and graph homomorphisms
as graphs again.
We therefore use universe polymorphism throughout this paper.
}
For example, the standard library includes the following
universe-polymorphic definition for the parameterised \ensuremath{\Conid{Maybe}} type: 

\begin{hscode}\SaveRestoreHook
\column{B}{@{}>{\hspre}l<{\hspost}@{}}%
\column{45}{@{}>{\hspre}l<{\hspost}@{}}%
\column{53}{@{}>{\hspre}l<{\hspost}@{}}%
\column{61}{@{}>{\hspre}l<{\hspost}@{}}%
\column{E}{@{}>{\hspre}l<{\hspost}@{}}%
\>[B]{}\Keyword{data}\;\Conid{Maybe}\;\{\mskip0.5mu \Varid{a}\;\mathbin{:}\;\Conid{Level}\mskip0.5mu\}\;(\Conid{A}\;\mathbin{:}\;\Conid{Set}\;\Varid{a})\;\mathbin{:}\;\Conid{Set}\;\Varid{a}\;{}\<[45]%
\>[45]{}\Keyword{where}\;{}\<[53]%
\>[53]{}\Varid{just}\;{}\<[61]%
\>[61]{}\mathbin{:}\;(\Varid{x}\;\mathbin{:}\;\Conid{A})\;\Varid{→}\;\Conid{Maybe}\;\Conid{A}{}\<[E]%
\\
\>[53]{}\Varid{nothing}\;\mathbin{:}\;\Conid{Maybe}\;\Conid{A}{}\<[E]%
\ColumnHook
\end{hscode}\resethooks

\noindent
\ensuremath{\Conid{Maybe}} has two parameters, \ensuremath{\Varid{a}} and \ensuremath{\Conid{A}},
where dependent typing is used since the type of the second parameter
depends on the first parameter.
The use of \ensuremath{\{\mskip0.5mu \Varid{...}\mskip0.5mu\}} flags \ensuremath{\Varid{a}} as an \emph{implicit parameter}
that can be elided where its type is implied by the call site of \ensuremath{\Conid{Maybe}}.
This happens in the occurrences of \ensuremath{\Conid{Maybe}\;\Conid{A}} in the types of the data
constructors \ensuremath{\Varid{just}} and \ensuremath{\Varid{nothing}}:
In \ensuremath{\Conid{Maybe}\;\Conid{A}}, the value of the first, implicit parameter of \ensuremath{\Conid{Maybe}}
can only be \ensuremath{\Varid{a}}, the level of the set \ensuremath{\Conid{A}}.

The same applies to implicit function arguments,
and in most cases, implicit arguments or parameters
are determined by later arguments respectively parameters.
Frequently, implicit arguments correspond quite precisely
to that part of the context of mathematical statements
that is frequently left implicit by mathematicians,
so that the reader may be advised to skip implicit arguments at first
reading of a type,
and return to them for clarification where necessary for understanding
the types of the explicit parameters.

While the Hindley-Milner typing of Haskell and ML
allows function definitions without declaration of the function type,
and type signatures without declaration of the universally quantified
type variables,
in Agda, almost all types and variables need to be declared,
but implicit parameters and the type checking machinery used to
resolve them alleviate that burden significantly.
For example, the original definition writes only
\ensuremath{\Conid{Maybe}\;\{\mskip0.5mu \Varid{a}\mskip0.5mu\}\;(\Conid{A}\;\mathbin{:}\;\Conid{Set}\;\Varid{a})\;\mathbin{:}\;\Conid{Set}\;\Varid{a}},
since the type of \ensuremath{\Varid{a}} will be inferred from \ensuremath{\Varid{a}}'s use
as argument to \ensuremath{\Conid{Set}}.

\medskip
The ``programming types'' like \ensuremath{\Conid{Maybe}}
can be freely mixed with ``formula types'',
inspired by the Curry-Howard-correspondence of ``formulae as types,
proofs as terms''.
The formula types of true formulae contain their proofs,
while the formula types of false formulae are empty.

The standard library type of propositional equality
has (besides two implicit parameters) one explicit parameter
and one explicit argument; the definition therefore gives rise to
types like the type ``\ensuremath{\Varid{2}\;\Varid{≡}\;\Varid{1}\;\Varid{+}\;\Varid{1}}'', which can be shown to be inhabited
using the definition of natural numbers \ensuremath{\Varid{1}} and \ensuremath{\Varid{2}} and natural number
addition \ensuremath{\Varid{+}},
and the type ``\ensuremath{\Varid{2}\;\Varid{≡}\;\Varid{3}}'', which is an empty type,
since it has no proof.

\begin{hscode}\SaveRestoreHook
\column{B}{@{}>{\hspre}l<{\hspost}@{}}%
\column{3}{@{}>{\hspre}l<{\hspost}@{}}%
\column{58}{@{}>{\hspre}l<{\hspost}@{}}%
\column{68}{@{}>{\hspre}l<{\hspost}@{}}%
\column{E}{@{}>{\hspre}l<{\hspost}@{}}%
\>[3]{}\Keyword{data}\;\Varid{\char95 ≡\char95 }\;\{\mskip0.5mu \Varid{a}\;\mathbin{:}\;\Conid{Level}\mskip0.5mu\}\;\{\mskip0.5mu \Conid{A}\;\mathbin{:}\;\Conid{Set}\;\Varid{a}\mskip0.5mu\}\;(\Varid{x}\;\mathbin{:}\;\Conid{A})\;\mathbin{:}\;\Conid{A}\;\Varid{→}\;\Conid{Set}\;\Varid{a}\;{}\<[58]%
\>[58]{}\Keyword{where}\;{}\<[68]%
\>[68]{}\Varid{refl}\;\mathbin{:}\;\Varid{x}\;\Varid{≡}\;\Varid{x}{}\<[E]%
\ColumnHook
\end{hscode}\resethooks

\noindent
The underscore characters occurring in the name \ensuremath{\Varid{\char95 ≡\char95 }}
declare mixfix syntax with argument positions for
explicit parameters and arguments;
this mixfix syntax is already used in the type of the single
constructor.
The definition introduces types \ensuremath{\Varid{x}\;\Varid{≡}\;\Varid{y}} for any \ensuremath{\Varid{x}} and \ensuremath{\Varid{y}} of type \ensuremath{\Conid{A}},
but only the types \ensuremath{\Varid{x}\;\Varid{≡}\;\Varid{x}} are inhabited,
and they contain the single element \ensuremath{\Varid{refl}\;\{\mskip0.5mu \Varid{a}\mskip0.5mu\}\;\{\mskip0.5mu \Conid{A}\mskip0.5mu\}\;\{\mskip0.5mu \Varid{x}\mskip0.5mu\}}.

\medskip
In Agda, as in other type theories without quotient types,
sets with equality are typically modelled as \emph{setoids},
that is, carrier types equipped with an equivalence.
This closely corresponds to the non-primitive nature of the
``equality'' test \ensuremath{(\mathrel{=}}{}\ensuremath{\mathrel{=})\;\mathbin{:}\;\Conid{Eq}\;\Varid{a}\;\Rightarrow \;\Varid{a}\;\to \;\Varid{a}\;\to \;\Conid{Bool}} in Haskell.
A setoid is a dependent record consisting of
a \ensuremath{\Conid{Carrier}} set, a relation \ensuremath{\Varid{\char95 ≈\char95 }} on that carrier,
and a proof that the relation \ensuremath{\Varid{\char95 ≈\char95 }} is an equivalence relation:
\begin{hscode}\SaveRestoreHook
\column{B}{@{}>{\hspre}l<{\hspost}@{}}%
\column{3}{@{}>{\hspre}l<{\hspost}@{}}%
\column{11}{@{}>{\hspre}l<{\hspost}@{}}%
\column{25}{@{}>{\hspre}l<{\hspost}@{}}%
\column{E}{@{}>{\hspre}l<{\hspost}@{}}%
\>[B]{}\Keyword{record}\;\Conid{Setoid}\;\Varid{c}\;\Varid{ℓ}\;\mathbin{:}\;\Conid{Set}\;(\Varid{suc}\;(\Varid{c}\;\Varid{⊍}\;\Varid{ℓ}))\;\Keyword{where}{}\<[E]%
\\
\>[B]{}\hsindent{3}{}\<[3]%
\>[3]{}\Keyword{field}\;{}\<[11]%
\>[11]{}\Conid{Carrier}\;{}\<[25]%
\>[25]{}\mathbin{:}\;\Conid{Set}\;\Varid{c}{}\<[E]%
\\
\>[11]{}\Varid{\char95 ≈\char95 }\;{}\<[25]%
\>[25]{}\mathbin{:}\;\Conid{Rel}\;\Conid{Carrier}\;\Varid{ℓ}{}\<[E]%
\\
\>[11]{}\Varid{isEquivalence}\;\mathbin{:}\;\Conid{IsEquivalence}\;\Varid{\char95 ≈\char95 }{}\<[E]%
\\
\>[B]{}\hsindent{3}{}\<[3]%
\>[3]{}\Keyword{open}\;\Conid{IsEquivalence}\;\Varid{isEquivalence}\;\Keyword{public}{}\<[E]%
\ColumnHook
\end{hscode}\resethooks

\noindent
An Agda record is also a module that may contain other material
besides its \ensuremath{\Keyword{field}}s;
the ``\ensuremath{\Keyword{open}}'' clause makes the fields of the equivalence proof
available as if they were fields of \ensuremath{\Conid{Setoid}}.
This language feature enables incremental extension of smaller theories
to larger theories at very low notational cost.

Whenever we allow arbitrary node or edge sets,
and we want to prove, for example, isomorphism of certain graphs,
we actually need setoids and not just sets.
For such contexts, we introduce the following abbreviation
for extracting the carrier set from a setoid:
\savecolumns
\begin{hscode}\SaveRestoreHook
\column{B}{@{}>{\hspre}l<{\hspost}@{}}%
\column{E}{@{}>{\hspre}l<{\hspost}@{}}%
\>[B]{}\Varid{⌊\char95 ⌋}\;\mathbin{:}\;\{\mskip0.5mu \Varid{c}\;\Varid{ℓ}\;\mathbin{:}\;\Conid{Level}\mskip0.5mu\}\;\Varid{→}\;\Conid{Setoid}\;\Varid{c}\;\Varid{ℓ}\;\Varid{→}\;\Conid{Set}\;\Varid{c}{}\<[E]%
\\
\>[B]{}\Varid{⌊}\;\Varid{s}\;\Varid{⌋}\;\mathrel{=}\;\Conid{Setoid.Carrier}\;\Varid{s}{}\<[E]%
\ColumnHook
\end{hscode}\resethooks

\ignore{
\begin{hscode}\SaveRestoreHook
\column{B}{@{}>{\hspre}l<{\hspost}@{}}%
\column{3}{@{}>{\hspre}l<{\hspost}@{}}%
\column{5}{@{}>{\hspre}l<{\hspost}@{}}%
\column{E}{@{}>{\hspre}l<{\hspost}@{}}%
\>[B]{}\Keyword{record}\;\Conid{Σ}\;\{\mskip0.5mu \Varid{a}\;\Varid{b}\;\mathbin{:}\;\Conid{Level}\mskip0.5mu\}\;(\Conid{A}\;\mathbin{:}\;\Conid{Set}\;\Varid{a})\;(\Conid{B}\;\mathbin{:}\;\Conid{A}\;\Varid{→}\;\Conid{Set}\;\Varid{b})\;\mathbin{:}\;\Conid{Set}\;(\Varid{a}\;\Varid{⊔}\;\Varid{b})\;\Keyword{where}{}\<[E]%
\\
\>[B]{}\hsindent{3}{}\<[3]%
\>[3]{}\Varid{constructor}\;\anonymous ,\anonymous {}\<[E]%
\\
\>[B]{}\hsindent{3}{}\<[3]%
\>[3]{}\Keyword{field}{}\<[E]%
\\
\>[3]{}\hsindent{2}{}\<[5]%
\>[5]{}\Varid{proj₁}\;\mathbin{:}\;\Conid{A}{}\<[E]%
\\
\>[3]{}\hsindent{2}{}\<[5]%
\>[5]{}\Varid{proj₂}\;\mathbin{:}\;\Conid{B}\;\Varid{proj₁}{}\<[E]%
\ColumnHook
\end{hscode}\resethooks

\begin{hscode}\SaveRestoreHook
\column{B}{@{}>{\hspre}l<{\hspost}@{}}%
\column{E}{@{}>{\hspre}l<{\hspost}@{}}%
\>[B]{}\Varid{∃}\;\mathbin{:}\;\Varid{∀}\;\{\mskip0.5mu \Varid{a}\;\Varid{b}\mskip0.5mu\}\;\{\mskip0.5mu \Conid{A}\;\mathbin{:}\;\Conid{Set}\;\Varid{a}\mskip0.5mu\}\;\Varid{→}\;(\Conid{A}\;\Varid{→}\;\Conid{Set}\;\Varid{b})\;\Varid{→}\;\Conid{Set}\;(\Varid{a}\;\Varid{⊔}\;\Varid{b}){}\<[E]%
\\
\>[B]{}\Varid{∃}\;\mathrel{=}\;\Conid{Σ}\;\anonymous {}\<[E]%
\ColumnHook
\end{hscode}\resethooks
}


\section{Set-Based Term Graphs}\sectlabel{Jungle}

We now present a simple definition of term graphs
that is intentionally kept close to conventional mathematical
formulations.
To reduce complexity and improve readability of this initial
formalisation, we present untyped term graphs here;
a typed variant will be shown in \sectref{CodeGraph}.

In the context of an arity-indexed label type \ensuremath{\Conid{Label}\;\mathbin{:}\;\Conid{ℕ}\;\Varid{→}\;\Conid{Set}},
we first define a type \ensuremath{\Conid{DHG₁}} of directed hypergraphs with one putput
per edge, indexed by input and output arities of the whole graph,
with the following components (since Agda records are also modules,
they can contain additional material besides their \ensuremath{\Keyword{field}}s):
\begin{itemize}
\item A setoid \ensuremath{\Conid{Inner}} of non-input nodes. (For simplicity, we do not
  emply universe polymorphism here, and all our setoids are of type
  \ensuremath{\Conid{Setoid}\;\Varid{zero}\;\Varid{zero}}.)

  For technical reasons, we find it more convenient
  to have the non-input nodes separate from the input nodes.
  Otherwise we would have had to include an explicit injection
  from the input positions to the complete node set.

\item The setoid \ensuremath{\Conid{Node}} of all nodes is then derived as the disjoint
  union of \ensuremath{\Conid{Inner}} with the setoid of input positions,
  which is obtained from \ensuremath{\Conid{Fin}\;\Varid{m}}, the set of natural numbers smaller
  than \ensuremath{\Varid{m}}.

\item The second \ensuremath{\Keyword{field}} is the \ensuremath{\Varid{n}}-element vector of \ensuremath{\Varid{output}} nodes,
  which can be either input nodes or inner nodes.

\item For symmetry, we also provide the \ensuremath{\Varid{m}}-element vector of \ensuremath{\Varid{input}}
  nodes, constructed using \ensuremath{\Varid{allFin}\;\Varid{m}} which is the vector (i.e.,
  array) containing all \ensuremath{\Varid{m}} elements of the set \ensuremath{\Conid{Fin}\;\Varid{m}} in sequence,
  i.e., \ensuremath{\Varid{0}}, \ensuremath{\Varid{1}}, \ldots, \ensuremath{\Varid{m}\;\Varid{-}\;\Varid{1}}.

\item \ensuremath{\Conid{Edge}} is the setoid of hyperedges.

\item \ensuremath{\Varid{eInfo}} maps each edge to a dependent tuple
  consisting of an arity \ensuremath{\Varid{k}}, a \ensuremath{\Varid{k}}-ary label, and a \ensuremath{\Varid{k}}-element
  vector of edge input nodes.

\item \ensuremath{\Varid{eOut}} maps each edge to its output node,
  which cannot be an input node of the \ensuremath{\Conid{Jungle}},
  and therefore has to be an \ensuremath{\Conid{Inner}} node.
  (The function arrow between setoids is optically not distinguishable
  from the general function type arrow, but is technically a different
  symbol. Since setoids cannot be used as types, no confusion can arise.)

\item We derive the function \ensuremath{\Varid{eLabel}} that maps each edge \ensuremath{\Varid{e}}
  to its edge label. Since the arity of that label is not known in
  advance,
  the function \ensuremath{\Varid{eLabel}} returns a dependent pair consisting
  of the label arity \ensuremath{\Varid{k}} and a \ensuremath{\Varid{k}}-ary label.

\item We also derive the function \ensuremath{\Varid{eIn}} that maps each edge \ensuremath{\Varid{e}}
  to the vector of input nodes of \ensuremath{\Varid{e}};
  the type of this vector depends on the arity of \ensuremath{\Varid{e}},
  which is the first component (\ensuremath{\Varid{proj₁}}) of the dependent tuple
  \ensuremath{\Varid{eLabel}\;\Varid{e}}.

\end{itemize}

\kern-2.9ex
\noindent
\begin{minipage}[t]{0.49\textwidth}
\savecolumns
\begin{hscode}\SaveRestoreHook
\column{B}{@{}>{\hspre}l<{\hspost}@{}}%
\column{3}{@{}>{\hspre}l<{\hspost}@{}}%
\column{5}{@{}>{\hspre}l<{\hspost}@{}}%
\column{12}{@{}>{\hspre}l<{\hspost}@{}}%
\column{18}{@{}>{\hspre}l<{\hspost}@{}}%
\column{E}{@{}>{\hspre}l<{\hspost}@{}}%
\>[3]{}\Keyword{record}\;\Conid{DHG₁}\;(\Varid{m}\;\Varid{n}\;\mathbin{:}\;\Conid{ℕ})\;\mathbin{:}\;\Conid{Set₁}\;\Keyword{where}{}\<[E]%
\\
\>[3]{}\hsindent{2}{}\<[5]%
\>[5]{}\Keyword{field}\;{}\<[12]%
\>[12]{}\Conid{Inner}\;\mathbin{:}\;\Conid{Setoid}\;\Varid{zero}\;\Varid{zero}{}\<[E]%
\\
\>[3]{}\hsindent{2}{}\<[5]%
\>[5]{}\Conid{Node}\;\mathrel{=}\;\Conid{Fin.setoid}\;\Varid{m}\;\Varid{⊎⊎}\;\Conid{Inner}{}\<[E]%
\\
\>[3]{}\hsindent{2}{}\<[5]%
\>[5]{}\Keyword{field}\;{}\<[12]%
\>[12]{}\Varid{output}\;\mathbin{:}\;\Conid{Vec}\;\Varid{⌊}\;\Conid{Node}\;\Varid{⌋}\;\Varid{n}{}\<[E]%
\\[\blanklineskip]%
\>[3]{}\hsindent{2}{}\<[5]%
\>[5]{}\Varid{input}\;\mathbin{:}\;\Conid{Vec}\;\Varid{⌊}\;\Conid{Node}\;\Varid{⌋}\;\Varid{m}{}\<[E]%
\\
\>[3]{}\hsindent{2}{}\<[5]%
\>[5]{}\Varid{input}\;\mathrel{=}\;\Conid{Vec.map}\;\Varid{inj₁}\;(\Varid{allFin}\;\Varid{m}){}\<[E]%
\\[\blanklineskip]%
\>[3]{}\hsindent{2}{}\<[5]%
\>[5]{}\Keyword{field}\;{}\<[12]%
\>[12]{}\Conid{Edge}\;\mathbin{:}\;\Conid{Setoid}\;\Varid{zero}\;\Varid{zero}{}\<[E]%
\\
\>[12]{}\Varid{eInfo}\;\mathbin{:}\;\Varid{⌊}\;\Conid{Edge}\;\Varid{⌋}\;{}\<[E]%
\\
\>[12]{}\hsindent{6}{}\<[18]%
\>[18]{}\Varid{→}\;\Conid{Σ}\;[\mskip1.5mu \Varid{k}\;\Varid{∶}\;\Conid{ℕ}\mskip1.5mu]\;(\Conid{Label}\;\Varid{k}\;\Varid{×}\;\Conid{Vec}\;\Varid{⌊}\;\Conid{Node}\;\Varid{⌋}\;\Varid{k}){}\<[E]%
\ColumnHook
\end{hscode}\resethooks

\kern-1.8ex
\restorecolumns
\begin{hscode}\SaveRestoreHook
\column{B}{@{}>{\hspre}l<{\hspost}@{}}%
\column{12}{@{}>{\hspre}l<{\hspost}@{}}%
\column{18}{@{}>{\hspre}l<{\hspost}@{}}%
\column{E}{@{}>{\hspre}l<{\hspost}@{}}%
\>[12]{}\Varid{eOut}\;{}\<[18]%
\>[18]{}\mathbin{:}\;\Conid{Edge}\;\Varid{⟶}\;\Conid{Inner}{}\<[E]%
\ColumnHook
\end{hscode}\resethooks

\kern3.8ex
\restorecolumns
\begin{hscode}\SaveRestoreHook
\column{B}{@{}>{\hspre}l<{\hspost}@{}}%
\column{5}{@{}>{\hspre}l<{\hspost}@{}}%
\column{E}{@{}>{\hspre}l<{\hspost}@{}}%
\>[5]{}\Varid{eLabel}\;\mathbin{:}\;\Varid{⌊}\;\Conid{Edge}\;\Varid{⌋}\;\Varid{→}\;\Conid{Σ}\;[\mskip1.5mu \Varid{k}\;\Varid{∶}\;\Conid{ℕ}\mskip1.5mu]\;\Conid{Label}\;\Varid{k}{}\<[E]%
\\
\>[5]{}\Varid{eLabel}\;\Varid{e}\;\mathrel{=}\;\Conid{Product.map}\;\Varid{id}\;\Varid{proj₁}\;(\Varid{eInfo}\;\Varid{e}){}\<[E]%
\\[\blanklineskip]%
\>[5]{}\Varid{eIn}\;\mathbin{:}\;(\Varid{e}\;\mathbin{:}\;\Varid{⌊}\;\Conid{Edge}\;\Varid{⌋})\;\Varid{→}\;\Conid{Vec}\;\Varid{⌊}\;\Conid{Node}\;\Varid{⌋}\;(\Varid{proj₁}\;(\Varid{eLabel}\;\Varid{e})){}\<[E]%
\\
\>[5]{}\Varid{eIn}\;\mathrel{=}\;\Varid{proj₂}\;\Varid{∘}\;\Varid{proj₂}\;\Varid{∘}\;\Varid{eInfo}{}\<[E]%
\ColumnHook
\end{hscode}\resethooks
\end{minipage}
\hfill
\begin{minipage}[t]{0.49\textwidth}
\savecolumns
\begin{hscode}\SaveRestoreHook
\column{B}{@{}>{\hspre}l<{\hspost}@{}}%
\column{3}{@{}>{\hspre}l<{\hspost}@{}}%
\column{5}{@{}>{\hspre}l<{\hspost}@{}}%
\column{12}{@{}>{\hspre}l<{\hspost}@{}}%
\column{18}{@{}>{\hspre}l<{\hspost}@{}}%
\column{E}{@{}>{\hspre}l<{\hspost}@{}}%
\>[3]{}\Keyword{record}\;\Conid{Jungle}\;(\Varid{m}\;\Varid{n}\;\mathbin{:}\;\Conid{ℕ})\;\mathbin{:}\;\Conid{Set₁}\;\Keyword{where}{}\<[E]%
\\
\>[3]{}\hsindent{2}{}\<[5]%
\>[5]{}\Keyword{field}\;{}\<[12]%
\>[12]{}\Conid{Inner}\;\mathbin{:}\;\Conid{Setoid}\;\Varid{zero}\;\Varid{zero}{}\<[E]%
\\
\>[3]{}\hsindent{2}{}\<[5]%
\>[5]{}\Conid{Node}\;\mathrel{=}\;\Conid{Fin.setoid}\;\Varid{m}\;\Varid{⊎⊎}\;\Conid{Inner}{}\<[E]%
\\
\>[3]{}\hsindent{2}{}\<[5]%
\>[5]{}\Keyword{field}\;{}\<[12]%
\>[12]{}\Varid{output}\;\mathbin{:}\;\Conid{Vec}\;\Varid{⌊}\;\Conid{Node}\;\Varid{⌋}\;\Varid{n}{}\<[E]%
\\[\blanklineskip]%
\>[3]{}\hsindent{2}{}\<[5]%
\>[5]{}\Varid{input}\;\mathbin{:}\;\Conid{Vec}\;\Varid{⌊}\;\Conid{Node}\;\Varid{⌋}\;\Varid{m}{}\<[E]%
\\
\>[3]{}\hsindent{2}{}\<[5]%
\>[5]{}\Varid{input}\;\mathrel{=}\;\Conid{Vec.map}\;\Varid{inj₁}\;(\Varid{allFin}\;\Varid{m}){}\<[E]%
\\[\blanklineskip]%
\>[3]{}\hsindent{2}{}\<[5]%
\>[5]{}\Keyword{field}\;{}\<[12]%
\>[12]{}\Conid{Edge}\;\mathbin{:}\;\Conid{Setoid}\;\Varid{zero}\;\Varid{zero}{}\<[E]%
\\
\>[12]{}\Varid{eInfo}\;\mathbin{:}\;\Varid{⌊}\;\Conid{Edge}\;\Varid{⌋}\;{}\<[E]%
\\
\>[12]{}\hsindent{6}{}\<[18]%
\>[18]{}\Varid{→}\;\Conid{Σ}\;[\mskip1.5mu \Varid{k}\;\Varid{∶}\;\Conid{ℕ}\mskip1.5mu]\;(\Conid{Label}\;\Varid{k}\;\Varid{×}\;\Conid{Vec}\;\Varid{⌊}\;\Conid{Node}\;\Varid{⌋}\;\Varid{k}){}\<[E]%
\\
\>[12]{}\Conid{EOut}\;{}\<[18]%
\>[18]{}\mathbin{:}\;\Conid{Inverse}\;\Conid{Edge}\;\Conid{Inner}{}\<[E]%
\\
\>[3]{}\hsindent{2}{}\<[5]%
\>[5]{}\Varid{eOut}\;\mathbin{:}\;\Conid{Edge}\;\Varid{⟶}\;\Conid{Inner}{}\<[E]%
\\
\>[3]{}\hsindent{2}{}\<[5]%
\>[5]{}\Varid{eOut}\;\mathrel{=}\;\Conid{Inverse.to}\;\Conid{EOut}{}\<[E]%
\\
\>[3]{}\hsindent{2}{}\<[5]%
\>[5]{}\Varid{producer}\;\mathbin{:}\;\Conid{Inner}\;\Varid{⟶}\;\Conid{Edge}{}\<[E]%
\\
\>[3]{}\hsindent{2}{}\<[5]%
\>[5]{}\Varid{producer}\;\mathrel{=}\;\Conid{Inverse.from}\;\Conid{EOut}{}\<[E]%
\\[\blanklineskip]%
\>[3]{}\hsindent{2}{}\<[5]%
\>[5]{}\Varid{eLabel}\;\mathbin{:}\;\Varid{⌊}\;\Conid{Edge}\;\Varid{⌋}\;\Varid{→}\;\Conid{Σ}\;[\mskip1.5mu \Varid{k}\;\Varid{∶}\;\Conid{ℕ}\mskip1.5mu]\;\Conid{Label}\;\Varid{k}{}\<[E]%
\\
\>[3]{}\hsindent{2}{}\<[5]%
\>[5]{}\Varid{eLabel}\;\Varid{e}\;\mathrel{=}\;\Conid{Product.map}\;\Varid{id}\;\Varid{proj₁}\;(\Varid{eInfo}\;\Varid{e}){}\<[E]%
\\[\blanklineskip]%
\>[3]{}\hsindent{2}{}\<[5]%
\>[5]{}\Varid{eIn}\;\mathbin{:}\;(\Varid{e}\;\mathbin{:}\;\Varid{⌊}\;\Conid{Edge}\;\Varid{⌋})\;\Varid{→}\;\Conid{Vec}\;\Varid{⌊}\;\Conid{Node}\;\Varid{⌋}\;(\Varid{proj₁}\;(\Varid{eLabel}\;\Varid{e})){}\<[E]%
\\
\>[3]{}\hsindent{2}{}\<[5]%
\>[5]{}\Varid{eIn}\;\mathrel{=}\;\Varid{proj₂}\;\Varid{∘}\;\Varid{proj₂}\;\Varid{∘}\;\Varid{eInfo}{}\<[E]%
\ColumnHook
\end{hscode}\resethooks
\end{minipage}

\noindent
In this \ensuremath{\Conid{DHG₁}} definition,
\ensuremath{\Varid{eOut}} does not have to be surjective,
which means that there may be ``undefined nodes'',
and
\ensuremath{\Varid{eOut}} also does not have to be injective,
which means that there may be ``join nodes''
in the sense of \cite{Kahl-Anand-Carette-2005}.
If bijectivity of \ensuremath{\Varid{eOut}} is desired,
we can replace the setoid mapping with an inverse pair of mappings,
and extract \ensuremath{\Varid{eOut}} and the \ensuremath{\Varid{producer}} mapping for inner nodes from
that, as shown above to the right.

These jungles are isomorphic to conventional termgraphs,
where inputs (as arguments) and labels are attached directly to inner nodes:

\savecolumns
\begin{hscode}\SaveRestoreHook
\column{B}{@{}>{\hspre}l<{\hspost}@{}}%
\column{3}{@{}>{\hspre}l<{\hspost}@{}}%
\column{5}{@{}>{\hspre}l<{\hspost}@{}}%
\column{12}{@{}>{\hspre}l<{\hspost}@{}}%
\column{18}{@{}>{\hspre}l<{\hspost}@{}}%
\column{E}{@{}>{\hspre}l<{\hspost}@{}}%
\>[3]{}\Keyword{record}\;\Conid{TermGraph}\;(\Varid{m}\;\Varid{n}\;\mathbin{:}\;\Conid{ℕ})\;\mathbin{:}\;\Conid{Set₁}\;\Keyword{where}{}\<[E]%
\\
\>[3]{}\hsindent{2}{}\<[5]%
\>[5]{}\Keyword{field}\;{}\<[12]%
\>[12]{}\Conid{Inner}\;\mathbin{:}\;\Conid{Setoid}\;\Varid{zero}\;\Varid{zero}{}\<[E]%
\\
\>[3]{}\hsindent{2}{}\<[5]%
\>[5]{}\Conid{Node}\;\mathrel{=}\;\Conid{Fin.setoid}\;\Varid{m}\;\Varid{⊎⊎}\;\Conid{Inner}{}\<[E]%
\\
\>[3]{}\hsindent{2}{}\<[5]%
\>[5]{}\Keyword{field}\;{}\<[12]%
\>[12]{}\Varid{output}\;\mathbin{:}\;\Conid{Vec}\;\Varid{⌊}\;\Conid{Node}\;\Varid{⌋}\;\Varid{n}{}\<[E]%
\\
\>[3]{}\hsindent{2}{}\<[5]%
\>[5]{}\Varid{input}\;\mathbin{:}\;\Conid{Vec}\;\Varid{⌊}\;\Conid{Node}\;\Varid{⌋}\;\Varid{m}{}\<[E]%
\\
\>[3]{}\hsindent{2}{}\<[5]%
\>[5]{}\Varid{input}\;\mathrel{=}\;\Conid{Vec.map}\;\Varid{inj₁}\;(\Varid{allFin}\;\Varid{m}){}\<[E]%
\\
\>[3]{}\hsindent{2}{}\<[5]%
\>[5]{}\Keyword{field}\;{}\<[12]%
\>[12]{}\Varid{label}\;\mathbin{:}\;\Varid{⌊}\;\Conid{Inner}\;\Varid{⌋}\;\Varid{→}\;\Conid{Σ}\;[\mskip1.5mu \Varid{k}\;\Varid{∶}\;\Conid{ℕ}\mskip1.5mu]\;\Conid{Label}\;\Varid{k}{}\<[E]%
\\
\>[12]{}\Varid{args}\;{}\<[18]%
\>[18]{}\mathbin{:}\;(\Varid{n}\;\mathbin{:}\;\Varid{⌊}\;\Conid{Inner}\;\Varid{⌋})\;\Varid{→}\;\Conid{Vec}\;\Varid{⌊}\;\Conid{Node}\;\Varid{⌋}\;(\Varid{proj₁}\;(\Varid{label}\;\Varid{n})){}\<[E]%
\ColumnHook
\end{hscode}\resethooks

\noindent
The following basic constructor functions are highly similar for
\ensuremath{\Conid{DHG₁}}, \ensuremath{\Conid{Jungle}}, and \ensuremath{\Conid{TermGraph}}; we show them here for \ensuremath{\Conid{Jungle}}.

Using the one-element setoid \ensuremath{\Varid{⊤}} (with element \ensuremath{\Varid{tt}}),
we can define primitive jungles consisting of a single hyperedge:

\restorecolumns
\begin{hscode}\SaveRestoreHook
\column{B}{@{}>{\hspre}l<{\hspost}@{}}%
\column{3}{@{}>{\hspre}l<{\hspost}@{}}%
\column{5}{@{}>{\hspre}l<{\hspost}@{}}%
\column{15}{@{}>{\hspre}l<{\hspost}@{}}%
\column{E}{@{}>{\hspre}l<{\hspost}@{}}%
\>[3]{}\Varid{prim}\;\mathbin{:}\;\{\mskip0.5mu \Varid{k}\;\mathbin{:}\;\Conid{ℕ}\mskip0.5mu\}\;\Varid{→}\;\Conid{Label}\;\Varid{k}\;\Varid{→}\;\Conid{Jungle}\;\Varid{k}\;\Varid{1}{}\<[E]%
\\
\>[3]{}\Varid{prim}\;\{\mskip0.5mu \Varid{k}\mskip0.5mu\}\;\Varid{f}\;\mathrel{=}\;\Keyword{record}{}\<[E]%
\\
\>[3]{}\hsindent{2}{}\<[5]%
\>[5]{}\{\mskip0.5mu \Conid{Inner}\;{}\<[15]%
\>[15]{}\mathrel{=}\;\Varid{⊤}{}\<[E]%
\\
\>[3]{}\hsindent{2}{}\<[5]%
\>[5]{};\Varid{output}\;{}\<[15]%
\>[15]{}\mathrel{=}\;[\mskip1.5mu \Varid{inj₂}\;\Varid{tt}\mskip1.5mu]{}\<[E]%
\\
\>[3]{}\hsindent{2}{}\<[5]%
\>[5]{};\Conid{Edge}\;{}\<[15]%
\>[15]{}\mathrel{=}\;\Varid{⊤}{}\<[E]%
\\
\>[3]{}\hsindent{2}{}\<[5]%
\>[5]{};\Varid{eInfo}\;{}\<[15]%
\>[15]{}\mathrel{=}\;\Varid{λ}\;\anonymous \;\Varid{→}\;(\Varid{k},(\Varid{f},\Conid{Vec.map}\;\Varid{inj₁}\;(\Varid{allFin}\;\Varid{k}))){}\<[E]%
\\
\>[3]{}\hsindent{2}{}\<[5]%
\>[5]{};\Conid{EOut}\;{}\<[15]%
\>[15]{}\mathrel{=}\;\Conid{Inverse.id}{}\<[E]%
\\
\>[3]{}\hsindent{2}{}\<[5]%
\>[5]{}\mskip0.5mu\}{}\<[E]%
\ColumnHook
\end{hscode}\resethooks

\noindent
For wiring graphs, we need empty sets (\ensuremath{\Varid{⊥}}) of edges and inner nodes:
\restorecolumns
\begin{hscode}\SaveRestoreHook
\column{B}{@{}>{\hspre}l<{\hspost}@{}}%
\column{3}{@{}>{\hspre}l<{\hspost}@{}}%
\column{5}{@{}>{\hspre}l<{\hspost}@{}}%
\column{15}{@{}>{\hspre}l<{\hspost}@{}}%
\column{23}{@{}>{\hspre}l<{\hspost}@{}}%
\column{E}{@{}>{\hspre}l<{\hspost}@{}}%
\>[3]{}\Varid{wire}\;\mathbin{:}\;\{\mskip0.5mu \Varid{m}\;\Varid{n}\;\mathbin{:}\;\Conid{ℕ}\mskip0.5mu\}\;\Varid{→}\;{}\<[23]%
\>[23]{}\Conid{Vec}\;(\Conid{Fin}\;\Varid{m})\;\Varid{n}\;\Varid{→}\;\Conid{Jungle}\;\Varid{m}\;\Varid{n}{}\<[E]%
\\
\>[3]{}\Varid{wire}\;\{\mskip0.5mu \Varid{m}\mskip0.5mu\}\;\{\mskip0.5mu \Varid{n}\mskip0.5mu\}\;\Varid{v}\;\mathrel{=}\;\Keyword{record}{}\<[E]%
\\
\>[3]{}\hsindent{2}{}\<[5]%
\>[5]{}\{\mskip0.5mu \Conid{Inner}\;{}\<[15]%
\>[15]{}\mathrel{=}\;\Varid{⊥}{}\<[E]%
\\
\>[3]{}\hsindent{2}{}\<[5]%
\>[5]{};\Varid{output}\;{}\<[15]%
\>[15]{}\mathrel{=}\;\Conid{Vec.map}\;\Varid{inj₁}\;\Varid{v}{}\<[E]%
\\
\>[3]{}\hsindent{2}{}\<[5]%
\>[5]{};\Conid{Edge}\;{}\<[15]%
\>[15]{}\mathrel{=}\;\Varid{⊥}{}\<[E]%
\\
\>[3]{}\hsindent{2}{}\<[5]%
\>[5]{};\Varid{eInfo}\;{}\<[15]%
\>[15]{}\mathrel{=}\;\Conid{E.⊥-elim}{}\<[E]%
\\
\>[3]{}\hsindent{2}{}\<[5]%
\>[5]{};\Conid{EOut}\;{}\<[15]%
\>[15]{}\mathrel{=}\;\Conid{Inverse.id}{}\<[E]%
\\
\>[3]{}\hsindent{2}{}\<[5]%
\>[5]{}\mskip0.5mu\}{}\<[E]%
\ColumnHook
\end{hscode}\resethooks

\noindent
With this, we can easily construct the standard wiring graphs
required for defining a gs-monoidal category (see \sectref{GSMonCat})
of \ensuremath{\Conid{Jungle}}s:

\restorecolumns
\begin{hscode}\SaveRestoreHook
\column{B}{@{}>{\hspre}l<{\hspost}@{}}%
\column{3}{@{}>{\hspre}l<{\hspost}@{}}%
\column{E}{@{}>{\hspre}l<{\hspost}@{}}%
\>[3]{}\Varid{idJungle}\;\mathbin{:}\;\{\mskip0.5mu \Varid{m}\;\mathbin{:}\;\Conid{ℕ}\mskip0.5mu\}\;\Varid{→}\;\Conid{Jungle}\;\Varid{m}\;\Varid{m}{}\<[E]%
\\
\>[3]{}\Varid{idJungle}\;\mathrel{=}\;\Varid{wire}\;(\Varid{allFin}\;\anonymous ){}\<[E]%
\\[\blanklineskip]%
\>[3]{}\Varid{dupJungle}\;\mathbin{:}\;\{\mskip0.5mu \Varid{m}\;\mathbin{:}\;\Conid{ℕ}\mskip0.5mu\}\;\Varid{→}\;\Conid{Jungle}\;\Varid{m}\;(\Varid{m}\;\Varid{+}\;\Varid{m}){}\<[E]%
\\
\>[3]{}\Varid{dupJungle}\;\{\mskip0.5mu \Varid{m}\mskip0.5mu\}\;\mathrel{=}\;\Varid{wire}\;(\Varid{allFin}\;\Varid{m}\;\plus \;\Varid{allFin}\;\Varid{m}){}\<[E]%
\\[\blanklineskip]%
\>[3]{}\Varid{termJungle}\;\mathbin{:}\;\{\mskip0.5mu \Varid{m}\;\mathbin{:}\;\Conid{ℕ}\mskip0.5mu\}\;\Varid{→}\;\Conid{Jungle}\;\Varid{m}\;\Varid{0}{}\<[E]%
\\
\>[3]{}\Varid{termJungle}\;\mathrel{=}\;\Varid{wire}\;[\mskip1.5mu \mskip1.5mu]{}\<[E]%
\\[\blanklineskip]%
\>[3]{}\Varid{exchJungle}\;\mathbin{:}\;(\Varid{m}\;\Varid{n}\;\mathbin{:}\;\Conid{ℕ})\;\Varid{→}\;\Conid{Jungle}\;(\Varid{m}\;\Varid{+}\;\Varid{n})\;(\Varid{n}\;\Varid{+}\;\Varid{m}){}\<[E]%
\\
\>[3]{}\Varid{exchJungle}\;\Varid{m}\;\Varid{n}\;\mathrel{=}\;\Varid{wire}\;(\Conid{Vec.map}\;(\Varid{raise}\;\Varid{m})\;(\Varid{allFin}\;\Varid{n})\;\plus \;\Conid{Vec.map}\;(\Varid{inject+}\;\Varid{n})\;(\Varid{allFin}\;\Varid{m})){}\<[E]%
\ColumnHook
\end{hscode}\resethooks

\noindent
Separating the inner nodes from the inputs
in particular has the advantage
that for sequential composition,
we can just use the disjoint union of the two \ensuremath{\Conid{Inner}} node sets:

\restorecolumns
\begin{hscode}\SaveRestoreHook
\column{B}{@{}>{\hspre}l<{\hspost}@{}}%
\column{3}{@{}>{\hspre}l<{\hspost}@{}}%
\column{5}{@{}>{\hspre}l<{\hspost}@{}}%
\column{7}{@{}>{\hspre}l<{\hspost}@{}}%
\column{13}{@{}>{\hspre}l<{\hspost}@{}}%
\column{15}{@{}>{\hspre}l<{\hspost}@{}}%
\column{53}{@{}>{\hspre}l<{\hspost}@{}}%
\column{E}{@{}>{\hspre}l<{\hspost}@{}}%
\>[3]{}\Varid{seqJungle}\;\mathbin{:}\;\{\mskip0.5mu \Varid{k}\;\Varid{m}\;\Varid{n}\;\mathbin{:}\;\Conid{ℕ}\mskip0.5mu\}\;\Varid{→}\;\Conid{Jungle}\;\Varid{k}\;\Varid{m}\;\Varid{→}\;\Conid{Jungle}\;\Varid{m}\;\Varid{n}\;\Varid{→}\;\Conid{Jungle}\;\Varid{k}\;\Varid{n}{}\<[E]%
\\
\>[3]{}\Varid{seqJungle}\;\{\mskip0.5mu \Varid{k}\mskip0.5mu\}\;\{\mskip0.5mu \Varid{m}\mskip0.5mu\}\;\{\mskip0.5mu \Varid{n}\mskip0.5mu\}\;\Varid{g₁}\;\Varid{g₂}\;\mathrel{=}\;\Keyword{let}{}\<[E]%
\\
\>[3]{}\hsindent{4}{}\<[7]%
\>[7]{}\Keyword{open}\;\Conid{Jungle}{}\<[E]%
\\
\>[3]{}\hsindent{4}{}\<[7]%
\>[7]{}\Varid{h₁}\;\mathbin{:}\;{}\<[13]%
\>[13]{}\Varid{⌊}\;\Conid{Node}\;\Varid{g₁}\;\Varid{⌋}\;\Varid{→}\;\Conid{Fin}\;\Varid{k}\;\Varid{⊎}\;(\Varid{⌊}\;\Conid{Inner}\;\Varid{g₁}\;\Varid{⌋}\;\Varid{⊎}\;{}\<[53]%
\>[53]{}\Varid{⌊}\;\Conid{Inner}\;\Varid{g₂}\;\Varid{⌋}){}\<[E]%
\\
\>[3]{}\hsindent{4}{}\<[7]%
\>[7]{}\Varid{h₁}\;\mathrel{=}\;\Conid{Sum.map}\;\Varid{id}\;\Varid{inj₁}{}\<[E]%
\\
\>[3]{}\hsindent{4}{}\<[7]%
\>[7]{}\Varid{h₂}\;\mathbin{:}\;\Varid{⌊}\;{}\<[15]%
\>[15]{}\Conid{Node}\;\Varid{g₂}\;\Varid{⌋}\;\Varid{→}\;\Conid{Fin}\;\Varid{k}\;\Varid{⊎}\;(\Varid{⌊}\;\Conid{Inner}\;\Varid{g₁}\;\Varid{⌋}\;\Varid{⊎}\;{}\<[53]%
\>[53]{}\Varid{⌊}\;\Conid{Inner}\;\Varid{g₂}\;\Varid{⌋}){}\<[E]%
\\
\>[3]{}\hsindent{4}{}\<[7]%
\>[7]{}\Varid{h₂}\;\mathrel{=}\;[\mskip1.5mu (\Varid{λ}\;\Varid{i}\;\Varid{→}\;\Varid{h₁}\;(\Conid{Vec.lookup}\;\Varid{i}\;(\Varid{output}\;\Varid{g₁}))),\Varid{inj₂}\;\Varid{∘}\;\Varid{inj₂}\mskip1.5mu]\;\Varid{′}{}\<[E]%
\\
\>[3]{}\hsindent{2}{}\<[5]%
\>[5]{}\Keyword{in}\;\Keyword{record}{}\<[E]%
\\
\>[5]{}\hsindent{2}{}\<[7]%
\>[7]{}\{\mskip0.5mu \Conid{Inner}\;\mathrel{=}\;\Conid{Inner}\;\Varid{g₁}\;\Varid{⊎⊎}\;\Conid{Inner}\;\Varid{g₂}{}\<[E]%
\\
\>[5]{}\hsindent{2}{}\<[7]%
\>[7]{};\Varid{output}\;\mathrel{=}\;\Conid{Vec.map}\;\Varid{h₂}\;(\Varid{output}\;\Varid{g₂}){}\<[E]%
\\
\>[5]{}\hsindent{2}{}\<[7]%
\>[7]{};\Conid{Edge}\;\mathrel{=}\;\Conid{Edge}\;\Varid{g₁}\;\Varid{⊎⊎}\;\Conid{Edge}\;\Varid{g₂}{}\<[E]%
\\
\>[5]{}\hsindent{2}{}\<[7]%
\>[7]{};\Varid{eInfo}\;\mathrel{=}\;[\mskip1.5mu \Varid{productMap₂₂}\;(\Conid{Vec.map}\;\Varid{h₁})\;\Varid{∘}\;\Varid{eInfo}\;\Varid{g₁},\Varid{productMap₂₂}\;(\Conid{Vec.map}\;\Varid{h₂})\;\Varid{∘}\;\Varid{eInfo}\;\Varid{g₂}\mskip1.5mu]\;\Varid{′}{}\<[E]%
\\
\>[5]{}\hsindent{2}{}\<[7]%
\>[7]{};\Conid{EOut}\;\mathrel{=}\;\Conid{EOut}\;\Varid{g₁}\;\Varid{⊕⊕}\;\Conid{EOut}\;\Varid{g₂}{}\<[E]%
\\
\>[5]{}\hsindent{2}{}\<[7]%
\>[7]{}\mskip0.5mu\}{}\<[E]%
\ColumnHook
\end{hscode}\resethooks

\noindent
Parallel composition works similarly;
here the input positions need to be adapted.

\restorecolumns
\begin{hscode}\SaveRestoreHook
\column{B}{@{}>{\hspre}l<{\hspost}@{}}%
\column{3}{@{}>{\hspre}l<{\hspost}@{}}%
\column{5}{@{}>{\hspre}l<{\hspost}@{}}%
\column{7}{@{}>{\hspre}l<{\hspost}@{}}%
\column{E}{@{}>{\hspre}l<{\hspost}@{}}%
\>[3]{}\Varid{parJungle}\;\mathbin{:}\;\{\mskip0.5mu \Varid{m₁}\;\Varid{n₁}\;\Varid{m₂}\;\Varid{n₂}\;\mathbin{:}\;\Conid{ℕ}\mskip0.5mu\}\;\Varid{→}\;\Conid{Jungle}\;\Varid{m₁}\;\Varid{n₁}\;\Varid{→}\;\Conid{Jungle}\;\Varid{m₂}\;\Varid{n₂}\;\Varid{→}\;\Conid{Jungle}\;(\Varid{m₁}\;\Varid{+}\;\Varid{m₂})\;(\Varid{n₁}\;\Varid{+}\;\Varid{n₂}){}\<[E]%
\\
\>[3]{}\Varid{parJungle}\;\{\mskip0.5mu \Varid{m₁}\mskip0.5mu\}\;\{\mskip0.5mu \Varid{n₁}\mskip0.5mu\}\;\{\mskip0.5mu \Varid{m₂}\mskip0.5mu\}\;\{\mskip0.5mu \Varid{n₂}\mskip0.5mu\}\;\Varid{g₁}\;\Varid{g₂}\;\mathrel{=}\;\Keyword{let}{}\<[E]%
\\
\>[3]{}\hsindent{4}{}\<[7]%
\>[7]{}\Keyword{open}\;\Conid{Jungle}{}\<[E]%
\\
\>[3]{}\hsindent{4}{}\<[7]%
\>[7]{}\Varid{h₁}\;\mathbin{:}\;\Varid{⌊}\;\Conid{Node}\;\Varid{g₁}\;\Varid{⌋}\;\Varid{→}\;\Conid{Fin}\;(\Varid{m₁}\;\Varid{+}\;\Varid{m₂})\;\Varid{⊎}\;(\Varid{⌊}\;\Conid{Inner}\;\Varid{g₁}\;\Varid{⌋}\;\Varid{⊎}\;\Varid{⌊}\;\Conid{Inner}\;\Varid{g₂}\;\Varid{⌋}){}\<[E]%
\\
\>[3]{}\hsindent{4}{}\<[7]%
\>[7]{}\Varid{h₁}\;\mathrel{=}\;\Conid{Sum.map}\;(\Varid{inject+}\;\Varid{m₂})\;\Varid{inj₁}{}\<[E]%
\\
\>[3]{}\hsindent{4}{}\<[7]%
\>[7]{}\Varid{h₂}\;\mathbin{:}\;\Varid{⌊}\;\Conid{Node}\;\Varid{g₂}\;\Varid{⌋}\;\Varid{→}\;\Conid{Fin}\;(\Varid{m₁}\;\Varid{+}\;\Varid{m₂})\;\Varid{⊎}\;(\Varid{⌊}\;\Conid{Inner}\;\Varid{g₁}\;\Varid{⌋}\;\Varid{⊎}\;\Varid{⌊}\;\Conid{Inner}\;\Varid{g₂}\;\Varid{⌋}){}\<[E]%
\\
\>[3]{}\hsindent{4}{}\<[7]%
\>[7]{}\Varid{h₂}\;\mathrel{=}\;\Conid{Sum.map}\;(\Varid{raise}\;\Varid{m₁})\;\Varid{inj₂}{}\<[E]%
\\
\>[3]{}\hsindent{2}{}\<[5]%
\>[5]{}\Keyword{in}\;\Keyword{record}{}\<[E]%
\\
\>[5]{}\hsindent{2}{}\<[7]%
\>[7]{}\{\mskip0.5mu \Conid{Inner}\;\mathrel{=}\;\Conid{Inner}\;\Varid{g₁}\;\Varid{⊎⊎}\;\Conid{Inner}\;\Varid{g₂}{}\<[E]%
\\
\>[5]{}\hsindent{2}{}\<[7]%
\>[7]{};\Varid{output}\;\mathrel{=}\;\Conid{Vec.map}\;\Varid{h₁}\;(\Varid{output}\;\Varid{g₁})\;\plus \;\Conid{Vec.map}\;\Varid{h₂}\;(\Varid{output}\;\Varid{g₂}){}\<[E]%
\\
\>[5]{}\hsindent{2}{}\<[7]%
\>[7]{};\Conid{Edge}\;\mathrel{=}\;\Conid{Edge}\;\Varid{g₁}\;\Varid{⊎⊎}\;\Conid{Edge}\;\Varid{g₂}{}\<[E]%
\\
\>[5]{}\hsindent{2}{}\<[7]%
\>[7]{};\Varid{eInfo}\;\mathrel{=}\;[\mskip1.5mu \Varid{productMap₂₂}\;(\Conid{Vec.map}\;\Varid{h₁})\;\Varid{∘}\;\Varid{eInfo}\;\Varid{g₁},\Varid{productMap₂₂}\;(\Conid{Vec.map}\;\Varid{h₂})\;\Varid{∘}\;\Varid{eInfo}\;\Varid{g₂}\mskip1.5mu]\;\Varid{′}{}\<[E]%
\\
\>[5]{}\hsindent{2}{}\<[7]%
\>[7]{};\Conid{EOut}\;\mathrel{=}\;\Conid{EOut}\;\Varid{g₁}\;\Varid{⊕⊕}\;\Conid{EOut}\;\Varid{g₂}{}\<[E]%
\\
\>[5]{}\hsindent{2}{}\<[7]%
\>[7]{}\mskip0.5mu\}{}\<[E]%
\ColumnHook
\end{hscode}\resethooks

\section{Typed Code Graphs}\sectlabel{CodeGraph}

Coconut code graphs \cite{Kahl-Anand-Carette-2005} have types
associated with nodes,
and hyperedges may have not only multiple inputs,
but also multiple outputs, to be able to model operations
that yield multiple results;
the typing of the input and output nodes
needs to be compatible with the operations indicated by the edge labels.

For simplicity, we assume here a global set \ensuremath{\Conid{Type}\;\mathbin{:}\;\Conid{Set}} of node
types, and dispense with using setoids in this section.
An edge label is now indexed by vectors of input and output types,
so we assume \ensuremath{\Conid{Label}\;\mathbin{:}\;\{\mskip0.5mu \Varid{m}\;\Varid{n}\;\mathbin{:}\;\Conid{ℕ}\mskip0.5mu\}\;\Varid{→}\;\Conid{Vec}\;\Conid{Type}\;\Varid{m}\;\Varid{→}\;\Conid{Vec}\;\Conid{Type}\;\Varid{n}\;\Varid{→}\;\Conid{Set}},
and also define the dependent record type \ensuremath{\Conid{EdgeType}} for collecting these indices:

\restorecolumns
\begin{hscode}\SaveRestoreHook
\column{B}{@{}>{\hspre}l<{\hspost}@{}}%
\column{3}{@{}>{\hspre}l<{\hspost}@{}}%
\column{6}{@{}>{\hspre}l<{\hspost}@{}}%
\column{13}{@{}>{\hspre}l<{\hspost}@{}}%
\column{E}{@{}>{\hspre}l<{\hspost}@{}}%
\>[3]{}\Keyword{record}\;\Conid{EdgeType}\;\mathbin{:}\;\Conid{Set}\;\Keyword{where}{}\<[E]%
\\
\>[3]{}\hsindent{3}{}\<[6]%
\>[6]{}\Keyword{field}\;{}\<[13]%
\>[13]{}\Varid{inArity}\;\mathbin{:}\;\Conid{ℕ}{}\<[E]%
\\
\>[13]{}\Varid{outArity}\;\mathbin{:}\;\Conid{ℕ}{}\<[E]%
\\
\>[13]{}\Varid{inTypes}\;\mathbin{:}\;\Conid{Vec}\;\Conid{Type}\;\Varid{inArity}{}\<[E]%
\\
\>[13]{}\Varid{outTypes}\;\mathbin{:}\;\Conid{Vec}\;\Conid{Type}\;\Varid{outArity}{}\<[E]%
\ColumnHook
\end{hscode}\resethooks

\noindent
An edge label then is such an index collection together
with a label drawn from the corresponding label set;
the \ensuremath{\Keyword{open}} declaration makes the \ensuremath{\Conid{EdgeType}} fields available
for \ensuremath{\Conid{EdgeLabel}} elements as if this was a record extension:

\restorecolumns
\begin{hscode}\SaveRestoreHook
\column{B}{@{}>{\hspre}l<{\hspost}@{}}%
\column{3}{@{}>{\hspre}l<{\hspost}@{}}%
\column{5}{@{}>{\hspre}l<{\hspost}@{}}%
\column{12}{@{}>{\hspre}l<{\hspost}@{}}%
\column{E}{@{}>{\hspre}l<{\hspost}@{}}%
\>[3]{}\Keyword{record}\;\Conid{EdgeLabel}\;\mathbin{:}\;\Conid{Set}\;\Keyword{where}{}\<[E]%
\\
\>[3]{}\hsindent{2}{}\<[5]%
\>[5]{}\Keyword{field}\;{}\<[12]%
\>[12]{}\Varid{eType}\;\mathbin{:}\;\Conid{EdgeType}{}\<[E]%
\\
\>[12]{}\Varid{label}\;\mathbin{:}\;\Conid{Label}\;(\Conid{EdgeType.inTypes}\;\Varid{eType})\;(\Conid{EdgeType.outTypes}\;\Varid{eType}){}\<[E]%
\\
\>[3]{}\hsindent{2}{}\<[5]%
\>[5]{}\Keyword{open}\;{}\<[12]%
\>[12]{}\Conid{EdgeType}\;\Varid{eType}\;\Keyword{public}{}\<[E]%
\ColumnHook
\end{hscode}\resethooks

\noindent
For typed term graphs,
there are many different ways to deal with node typing,
and for any given way, different views are useful in different
contexts.
We will keep a node typing function as a \ensuremath{\Keyword{field}},
and derive from this an indexed view of typed nodes,
using the following general construct:
Given a set \ensuremath{\Conid{A}} and a typing function \ensuremath{\Varid{type}} for \ensuremath{\Conid{A}},
the \ensuremath{\Conid{Type}}-indexed set \ensuremath{\Conid{Typed}\;\Conid{A}\;\Varid{type}}
associates with every type \ensuremath{\Varid{ty}}
all elements of \ensuremath{\Conid{A}} that have type \ensuremath{\Varid{ty}};
formally, an element of \ensuremath{\Conid{Typed}\;\Conid{A}\;\Varid{type}\;\Varid{ty}}
is a dependent pair consisting of an element \ensuremath{\Varid{a}\;\mathbin{:}\;\Conid{A}}
together with a proof that \ensuremath{\Varid{type}\;\Varid{a}\;\Varid{≡}\;\Varid{ty}}:

\restorecolumns
\begin{hscode}\SaveRestoreHook
\column{B}{@{}>{\hspre}l<{\hspost}@{}}%
\column{3}{@{}>{\hspre}l<{\hspost}@{}}%
\column{E}{@{}>{\hspre}l<{\hspost}@{}}%
\>[3]{}\Conid{Typed}\;\mathbin{:}\;(\Conid{A}\;\mathbin{:}\;\Conid{Set})\;\Varid{→}\;(\Conid{A}\;\Varid{→}\;\Conid{Type})\;\Varid{→}\;\Conid{Type}\;\Varid{→}\;\Conid{Set}{}\<[E]%
\\
\>[3]{}\Conid{Typed}\;\Conid{A}\;\Varid{type}\;\Varid{ty}\;\mathrel{=}\;\Conid{Σ}\;[\mskip1.5mu \Varid{a}\;\Varid{∶}\;\Conid{A}\mskip1.5mu]\;(\Varid{type}\;\Varid{a}\;\Varid{≡}\;\Varid{ty}){}\<[E]%
\ColumnHook
\end{hscode}\resethooks

\noindent
Since the Agda standard library does not provide a variant of \ensuremath{\Conid{Vec}}
where the element types may depend on their positions,
we directly use dependently typed functions starting from these
positions instead,
producing ``typed vectors'' with elements type according to the
argument type vector \ensuremath{\Varid{v}}:

\restorecolumns
\begin{hscode}\SaveRestoreHook
\column{B}{@{}>{\hspre}l<{\hspost}@{}}%
\column{3}{@{}>{\hspre}l<{\hspost}@{}}%
\column{E}{@{}>{\hspre}l<{\hspost}@{}}%
\>[3]{}\Conid{TypedVec}\;\mathbin{:}\;(\Conid{A}\;\mathbin{:}\;\Conid{Set})\;\Varid{→}\;(\Conid{A}\;\Varid{→}\;\Conid{Type})\;\Varid{→}\;\{\mskip0.5mu \Varid{k}\;\mathbin{:}\;\Conid{ℕ}\mskip0.5mu\}\;\Varid{→}\;\Conid{Vec}\;\Conid{Type}\;\Varid{k}\;\Varid{→}\;\Conid{Set}{}\<[E]%
\\
\>[3]{}\Conid{TypedVec}\;\Conid{A}\;\Varid{type}\;\{\mskip0.5mu \Varid{k}\mskip0.5mu\}\;\Varid{v}\;\mathrel{=}\;(\Varid{i}\;\mathbin{:}\;\Conid{Fin}\;\Varid{k})\;\Varid{→}\;\Conid{Typed}\;\Conid{A}\;\Varid{type}\;(\Conid{Vec.lookup}\;\Varid{i}\;\Varid{v}){}\<[E]%
\ColumnHook
\end{hscode}\resethooks

\noindent
The \ensuremath{\Conid{EdgeInfo}} associated with each hyperedge
then contains, besides an \ensuremath{\Conid{EdgeLabel}},
two such ``typed node vectors'',
typed according to the label's typing information
(for modularity, this definition is kept outside the code graph
 definition and parameterised with the type \ensuremath{\Conid{Nodes}} for ``typed node
 vectors'' to be supplied there):

\restorecolumns
\begin{hscode}\SaveRestoreHook
\column{B}{@{}>{\hspre}l<{\hspost}@{}}%
\column{3}{@{}>{\hspre}l<{\hspost}@{}}%
\column{5}{@{}>{\hspre}l<{\hspost}@{}}%
\column{12}{@{}>{\hspre}l<{\hspost}@{}}%
\column{21}{@{}>{\hspre}l<{\hspost}@{}}%
\column{E}{@{}>{\hspre}l<{\hspost}@{}}%
\>[3]{}\Keyword{record}\;\Conid{EdgeInfo}\;(\Conid{Nodes}\;\mathbin{:}\;\{\mskip0.5mu \Varid{k}\;\mathbin{:}\;\Conid{ℕ}\mskip0.5mu\}\;\Varid{→}\;\Conid{Vec}\;\Conid{Type}\;\Varid{k}\;\Varid{→}\;\Conid{Set})\;\mathbin{:}\;\Conid{Set}\;\Keyword{where}{}\<[E]%
\\
\>[3]{}\hsindent{2}{}\<[5]%
\>[5]{}\Keyword{field}\;{}\<[12]%
\>[12]{}\Varid{eLab}\;{}\<[21]%
\>[21]{}\mathbin{:}\;\Conid{EdgeLabel}{}\<[E]%
\\
\>[12]{}\Varid{eInput}\;{}\<[21]%
\>[21]{}\mathbin{:}\;\Conid{Nodes}\;(\Conid{EdgeLabel.inTypes}\;\Varid{eLab}){}\<[E]%
\\
\>[12]{}\Varid{eOutput}\;{}\<[21]%
\>[21]{}\mathbin{:}\;\Conid{Nodes}\;(\Conid{EdgeLabel.outTypes}\;\Varid{eLab}){}\<[E]%
\\
\>[3]{}\hsindent{2}{}\<[5]%
\>[5]{}\Keyword{open}\;{}\<[12]%
\>[12]{}\Conid{EdgeLabel}\;\Varid{eLab}\;\Keyword{public}{}\<[E]%
\ColumnHook
\end{hscode}\resethooks

\noindent
A \ensuremath{\Conid{CodeGraph}} is now defined roughly analogous to a \ensuremath{\Conid{Jungle}},
with the following differences worth pointing out:
\begin{itemize}
\item Code graphs can be considered as ``generalised hyperedges'',
  and therefore have an \ensuremath{\Conid{EdgeType}} derived from the \ensuremath{\Conid{CodeGraph}} type
  parameters. Keeping the current parameters eases the implementation
  of the categorical view, in comparison with using the \ensuremath{\Conid{EdgeType}} as
  a parameter instead.

\item We only need to explicitly represent the typing of the inner
  nodes; from this we can derive the typing of all \ensuremath{\Conid{Node}}s
  by looking up the typing of the input positions in \ensuremath{\Varid{inTypes}}.

\item A \ensuremath{\Conid{TypedNode}\;\Varid{ty}} is a \ensuremath{\Conid{Node}} with type \ensuremath{\Varid{ty}};
  an element of \ensuremath{\Conid{TypedNodes}\;\Varid{v}} is a ``typed node vector''
  according to the type vector \ensuremath{\Varid{v}}.

\item The \ensuremath{\Conid{CodeGraph}} field \ensuremath{\Varid{output}} and each individual edge
  interface use \ensuremath{\Conid{TypedNode}} ``vectors''.

\item We can still provide lower-level interfaces to edges;
  we show functions that extract the edge label,
  edge input arity,
  and edge input \ensuremath{\Conid{Node}} vectors (discarding the type information),
  both dependently-typed and existentially-typed with respect to the
  vector length.
  (The corresponding functions \ensuremath{\Varid{eOut}} etc.\null are not shown.)
\end{itemize}

\restorecolumns
\begin{hscode}\SaveRestoreHook
\column{B}{@{}>{\hspre}l<{\hspost}@{}}%
\column{3}{@{}>{\hspre}l<{\hspost}@{}}%
\column{5}{@{}>{\hspre}l<{\hspost}@{}}%
\column{10}{@{}>{\hspre}l<{\hspost}@{}}%
\column{11}{@{}>{\hspre}l<{\hspost}@{}}%
\column{12}{@{}>{\hspre}l<{\hspost}@{}}%
\column{22}{@{}>{\hspre}l<{\hspost}@{}}%
\column{34}{@{}>{\hspre}l<{\hspost}@{}}%
\column{E}{@{}>{\hspre}l<{\hspost}@{}}%
\>[3]{}\Keyword{record}\;\Conid{CodeGraph}\;\{\mskip0.5mu \Varid{m}\;\Varid{n}\;\mathbin{:}\;\Conid{ℕ}\mskip0.5mu\}\;(\Varid{inTypes}\;\mathbin{:}\;\Conid{Vec}\;\Conid{Type}\;\Varid{m})\;(\Varid{outTypes}\;\mathbin{:}\;\Conid{Vec}\;\Conid{Type}\;\Varid{n})\;\mathbin{:}\;\Conid{Set₁}\;\Keyword{where}{}\<[E]%
\\
\>[3]{}\hsindent{2}{}\<[5]%
\>[5]{}\Varid{cgType}\;\mathbin{:}\;\Conid{EdgeType}{}\<[E]%
\\
\>[3]{}\hsindent{2}{}\<[5]%
\>[5]{}\Varid{cgType}\;\mathrel{=}\;\Keyword{record}\;{}\<[22]%
\>[22]{}\{\mskip0.5mu \Varid{inArity}\;{}\<[34]%
\>[34]{}\mathrel{=}\;\Varid{m}{}\<[E]%
\\
\>[22]{};\Varid{outArity}\;{}\<[34]%
\>[34]{}\mathrel{=}\;\Varid{n}{}\<[E]%
\\
\>[22]{};\Varid{inTypes}\;{}\<[34]%
\>[34]{}\mathrel{=}\;\Varid{inTypes}{}\<[E]%
\\
\>[22]{};\Varid{outTypes}\;{}\<[34]%
\>[34]{}\mathrel{=}\;\Varid{outTypes}\mskip0.5mu\}{}\<[E]%
\\[\blanklineskip]%
\>[3]{}\hsindent{2}{}\<[5]%
\>[5]{}\Keyword{field}\;{}\<[12]%
\>[12]{}\Conid{Inner}\;\mathbin{:}\;\Conid{Set}{}\<[E]%
\\
\>[12]{}\Varid{iType}\;\mathbin{:}\;\Conid{Inner}\;\Varid{→}\;\Conid{Type}{}\<[E]%
\\[\blanklineskip]%
\>[3]{}\hsindent{2}{}\<[5]%
\>[5]{}\Conid{Node}\;\mathrel{=}\;\Conid{Fin}\;\Varid{m}\;\Varid{⊎}\;\Conid{Inner}{}\<[E]%
\\[\blanklineskip]%
\>[3]{}\hsindent{2}{}\<[5]%
\>[5]{}\Varid{nType}\;\mathbin{:}\;\Conid{Node}\;\Varid{→}\;\Conid{Type}{}\<[E]%
\\
\>[3]{}\hsindent{2}{}\<[5]%
\>[5]{}\Varid{nType}\;\mathrel{=}\;[\mskip1.5mu (\Varid{λ}\;\Varid{i}\;\Varid{→}\;\Conid{Vec.lookup}\;\Varid{i}\;\Varid{inTypes}),\Varid{iType}\mskip1.5mu]\;\Varid{′}{}\<[E]%
\\[\blanklineskip]%
\>[3]{}\hsindent{2}{}\<[5]%
\>[5]{}\Conid{TypedNode}\;\mathbin{:}\;\Conid{Type}\;\Varid{→}\;\Conid{Set}{}\<[E]%
\\
\>[3]{}\hsindent{2}{}\<[5]%
\>[5]{}\Conid{TypedNode}\;\mathrel{=}\;\Conid{Typed}\;\Conid{Node}\;\Varid{nType}{}\<[E]%
\\[\blanklineskip]%
\>[3]{}\hsindent{2}{}\<[5]%
\>[5]{}\Conid{TypedNodes}\;\mathbin{:}\;\{\mskip0.5mu \Varid{k}\;\mathbin{:}\;\Conid{ℕ}\mskip0.5mu\}\;\Varid{→}\;\Conid{Vec}\;\Conid{Type}\;\Varid{k}\;\Varid{→}\;\Conid{Set}{}\<[E]%
\\
\>[3]{}\hsindent{2}{}\<[5]%
\>[5]{}\Conid{TypedNodes}\;\mathrel{=}\;\Conid{TypedVec}\;\Conid{Node}\;\Varid{nType}{}\<[E]%
\\[\blanklineskip]%
\>[3]{}\hsindent{2}{}\<[5]%
\>[5]{}\Keyword{field}\;{}\<[12]%
\>[12]{}\Varid{output}\;\mathbin{:}\;\Conid{TypedNodes}\;\Varid{outTypes}{}\<[E]%
\\[\blanklineskip]%
\>[3]{}\hsindent{2}{}\<[5]%
\>[5]{}\Varid{input}\;\mathbin{:}\;\Conid{TypedNodes}\;\Varid{inTypes}{}\<[E]%
\\
\>[3]{}\hsindent{2}{}\<[5]%
\>[5]{}\Varid{input}\;\mathrel{=}\;\Varid{λ}\;\Varid{i}\;\Varid{→}\;(\Varid{inj₁}\;\Varid{i},\Varid{refl}){}\<[E]%
\\[\blanklineskip]%
\>[3]{}\hsindent{2}{}\<[5]%
\>[5]{}\Keyword{field}\;{}\<[12]%
\>[12]{}\Conid{Edge}\;\mathbin{:}\;\Conid{Set}{}\<[E]%
\\
\>[12]{}\Varid{eInfo}\;\mathbin{:}\;\Conid{Edge}\;\Varid{→}\;\Conid{EdgeInfo}\;\Conid{TypedNodes}{}\<[E]%
\\[\blanklineskip]%
\>[3]{}\hsindent{2}{}\<[5]%
\>[5]{}\Varid{eLabel}\;\mathbin{:}\;\Conid{Edge}\;\Varid{→}\;\Conid{EdgeLabel}{}\<[E]%
\\
\>[3]{}\hsindent{2}{}\<[5]%
\>[5]{}\Varid{eLabel}\;\mathrel{=}\;\Conid{EdgeInfo.eLab}\;\Varid{∘}\;\Varid{eInfo}{}\<[E]%
\\[\blanklineskip]%
\>[3]{}\hsindent{2}{}\<[5]%
\>[5]{}\Varid{eInArity}\;\mathbin{:}\;\Conid{Edge}\;\Varid{→}\;\Conid{ℕ}{}\<[E]%
\\
\>[3]{}\hsindent{2}{}\<[5]%
\>[5]{}\Varid{eInArity}\;\mathrel{=}\;\Conid{EdgeInfo.inArity}\;\Varid{∘}\;\Varid{eInfo}{}\<[E]%
\\[\blanklineskip]%
\>[3]{}\hsindent{2}{}\<[5]%
\>[5]{}\Varid{eIn}\;{}\<[10]%
\>[10]{}\mathbin{:}\;(\Varid{e}\;\mathbin{:}\;\Conid{Edge})\;\Varid{→}\;\Conid{Vec}\;\Conid{Node}\;(\Varid{eInArity}\;\Varid{e}){}\<[E]%
\\
\>[3]{}\hsindent{2}{}\<[5]%
\>[5]{}\Varid{eIn}\;\Varid{e}\;\mathrel{=}\;\Varid{mkVec}\;(\Varid{proj₁}\;\Varid{∘}\;\Conid{EdgeInfo.eInput}\;(\Varid{eInfo}\;\Varid{e})){}\<[E]%
\\[\blanklineskip]%
\>[3]{}\hsindent{2}{}\<[5]%
\>[5]{}\Varid{eIn′}\;{}\<[11]%
\>[11]{}\mathbin{:}\;\Conid{Edge}\;\Varid{→}\;\Conid{Σ}\;[\mskip1.5mu \Varid{k}\;\Varid{∶}\;\Conid{ℕ}\mskip1.5mu]\;(\Conid{Vec}\;\Conid{Node}\;\Varid{k}){}\<[E]%
\\
\>[3]{}\hsindent{2}{}\<[5]%
\>[5]{}\Varid{eIn′}\;\Varid{e}\;\mathrel{=}\;\Varid{eInArity}\;\Varid{e},\Varid{eIn}\;\Varid{e}{}\<[E]%
\ColumnHook
\end{hscode}\resethooks

\noindent
Again, \ensuremath{\Varid{eOut}} is not guaranteed to reach all nodes,
and, due to the possibility of multi-output operations,
this cannot be amended by joining the \ensuremath{\Conid{Inner}} and \ensuremath{\Conid{Edge}} sets as in jungles.
This and other degrees of generality contained in this definition
can be useful for certain purposes,
but also can be forbidden for other purposes
by adding appropriate constraints.

We show the function for producing primitive one-edge code graphs:

\restorecolumns
\begin{hscode}\SaveRestoreHook
\column{B}{@{}>{\hspre}l<{\hspost}@{}}%
\column{3}{@{}>{\hspre}l<{\hspost}@{}}%
\column{5}{@{}>{\hspre}l<{\hspost}@{}}%
\column{15}{@{}>{\hspre}l<{\hspost}@{}}%
\column{31}{@{}>{\hspre}l<{\hspost}@{}}%
\column{42}{@{}>{\hspre}l<{\hspost}@{}}%
\column{67}{@{}>{\hspre}c<{\hspost}@{}}%
\column{67E}{@{}l@{}}%
\column{71}{@{}>{\hspre}c<{\hspost}@{}}%
\column{71E}{@{}l@{}}%
\column{E}{@{}>{\hspre}l<{\hspost}@{}}%
\>[3]{}\Varid{prim}\;\mathbin{:}\;(\Varid{ℓ}\;\mathbin{:}\;\Conid{EdgeLabel})\;\Varid{→}\;\Conid{CodeGraph}\;(\Conid{EdgeLabel.inTypes}\;\Varid{ℓ})\;(\Conid{EdgeLabel.outTypes}\;\Varid{ℓ}){}\<[E]%
\\
\>[3]{}\Varid{prim}\;\Varid{ℓ}\;\mathrel{=}\;\Keyword{record}{}\<[E]%
\\
\>[3]{}\hsindent{2}{}\<[5]%
\>[5]{}\{\mskip0.5mu \Conid{Inner}\;{}\<[15]%
\>[15]{}\mathrel{=}\;\Conid{Fin}\;(\Conid{EdgeLabel.outArity}\;\Varid{ℓ}){}\<[E]%
\\
\>[3]{}\hsindent{2}{}\<[5]%
\>[5]{};\Varid{output}\;{}\<[15]%
\>[15]{}\mathrel{=}\;\Varid{λ}\;\Varid{i}\;\Varid{→}\;(\Varid{inj₂}\;\Varid{i},\Varid{refl}){}\<[E]%
\\
\>[3]{}\hsindent{2}{}\<[5]%
\>[5]{};\Conid{Edge}\;{}\<[15]%
\>[15]{}\mathrel{=}\;\Varid{⊤}{}\<[E]%
\\
\>[3]{}\hsindent{2}{}\<[5]%
\>[5]{};\Varid{eInfo}\;{}\<[15]%
\>[15]{}\mathrel{=}\;\Varid{λ}\;\anonymous \;\Varid{→}\;\Keyword{record}\;{}\<[31]%
\>[31]{}\{\mskip0.5mu \Varid{eLab}\;{}\<[42]%
\>[42]{}\mathrel{=}\;\Varid{ℓ}{}\<[E]%
\\
\>[31]{};\Varid{eInput}\;{}\<[42]%
\>[42]{}\mathrel{=}\;\Varid{λ}\;\Varid{i}\;\Varid{→}\;(\Varid{inj₁}\;\Varid{i},\Varid{refl}){}\<[E]%
\\
\>[31]{};\Varid{eOutput}\;{}\<[42]%
\>[42]{}\mathrel{=}\;\Varid{λ}\;\Varid{i}\;\Varid{→}\;(\Varid{inj₂}\;\Varid{i},\Varid{refl}){}\<[67]%
\>[67]{}\mskip0.5mu\}{}\<[67E]%
\>[71]{}\mskip0.5mu\}{}\<[71E]%
\ColumnHook
\end{hscode}\resethooks

\noindent
While type-checking the three propositional equality proofs \ensuremath{\Varid{refl}} in here,
Agda actually proves that the mentioned types are indeed equal:
An Agda program can only produce \ensuremath{\Conid{CodeGraph}} values that are
correctly typed,
both on the external interface, and internally at each port of each edge.

\section{GS-Monoidal Categories}\sectlabel{GSMonCat}

Corradini and Gadducci
proposed \emph{gs-monoidal categories}
for modelling acyclic term graphs \cite{Corradini-Gadducci-1999-APTG};
extended discussion of how code graphs fit into this framework
is contained in \cite{Kahl-Anand-Carette-2005}.
Here we only present a quick summary,
and tie this into the formalisation in \sectref{Jungle}.

In a category theory context,
we write ``$f : \objA \fun \objB$''
to declare that morphism $f$ goes from object $\objA$ to object
$\objB$,
and use ``$\RELcomp$'' as the associative binary \emph{composition} operator;
composition of two morphisms $f : \objA \fun \objB$ and
$g : \objB' \fun \objC$ is defined iff $\objB = \objB'$,
and then $(f \RELcomp g) : \objA \fun \objC$.
Furthermore, the identity morphism for object $\objA$
is written $\RELid_{\objA}$.

\ensuremath{\Conid{Jungle}} can be seen to define morphisms
of an untyped term graph category
where objects are natural numbers.
(For \ensuremath{\Conid{CodeGraph}}, the collection of Objects is \ensuremath{\Conid{Σ}\;[\mskip1.5mu \Varid{k}\;\Varid{∶}\;\Conid{ℕ}\mskip1.5mu]\;(\Conid{Vec}\;\Conid{Type}\;\Varid{k})}.)

In the \ensuremath{\Conid{Jungle}} category,
a morphism from $m$ to $n$ is an element of \ensuremath{\Conid{Jungle}\;\Varid{m}\;\Varid{n}},
that is, a term graph with $m$ input nodes and $n$ output nodes.
More precisely, such a morphism is an isomorphism class of jungles,
since node and edge identities do not matter;
we will define a \ensuremath{\Conid{Setoid}} where the \ensuremath{\Conid{Carrier}} is \ensuremath{\Conid{Jungle}\;\Varid{m}\;\Varid{n}}
and equivalence proofs are \ensuremath{\Conid{Jungle}} isomorphisms.

Composition $F \RELcomp G$ ``glues'' together
the output nodes of $F$ with the respective input nodes of $G$,
as we have implemented in \ensuremath{\Varid{seqJungle}}.
The identity on $n$ consists only of $n$ input nodes which are also,
in the same sequence, output nodes, and no edges,
and is therefore constructed as a wiring graph:

\restorecolumns
\begin{hscode}\SaveRestoreHook
\column{B}{@{}>{\hspre}l<{\hspost}@{}}%
\column{3}{@{}>{\hspre}l<{\hspost}@{}}%
\column{25}{@{}>{\hspre}l<{\hspost}@{}}%
\column{E}{@{}>{\hspre}l<{\hspost}@{}}%
\>[3]{}\Varid{idJungle}\;\mathbin{:}\;\{\mskip0.5mu \Varid{m}\;\mathbin{:}\;\Conid{ℕ}\mskip0.5mu\}\;\Varid{→}\;{}\<[25]%
\>[25]{}\Conid{Jungle}\;\Varid{m}\;\Varid{m}{}\<[E]%
\\
\>[3]{}\Varid{idJungle}\;\mathrel{=}\;\Varid{wire}\;(\Varid{allFin}\;\anonymous ){}\<[E]%
\ColumnHook
\end{hscode}\resethooks

\begin{Def}\Deflabel{ssmc}
A \emph{symmetric strict monoidal category} \cite{MacLane-1971}
consists of a category $\catCo$,
a strictly associative monoidal bifunctor $\otimes$
with $\munit$ as its strict unit,
and a transformation $\exch$
that associates with every two objects $\objA$ and $\objB$
an arrow $\exch_{\objA,\objB} : \objA \otimes \objB \fun \objB \otimes \objA$
with:
\BCM
\begin{array}[b]{rcl@{\hskip1em}rcl}
    (F \otimes G) \RELcomp \exch_{\objC,\objD} &=&
    \exch_{\objA,\objB} \RELcomp (G \otimes F)
\enskip,&
    \exch_{\objA,\objB} \RELcomp \exch_{\objB,\objA} &=&
    \RELid_{\objA} \otimes \RELid_{\objB}
\enskip,\\[.3ex]
    \exch_{\objA\otimes\objB,\objC} &=&
    (\RELid_{\objA} \otimes \exch_{\objB,\objC}) \RELcomp
    (\exch_{\objA,\objC} \otimes \RELid_{\objB})
\enskip,&
    \exch_{\munit,\munit} &=& \RELid_{\munit}
\enskip.
\ECMAQ
\unskip
\end{Def}

\noindent
For \ensuremath{\Conid{Jungle}}, the unit object $\munit$ is the natural number \ensuremath{\Varid{0}},
and $\otimes$ on objects is addition.
On morphisms, $\otimes$ forms the disjoint union of code graphs,
concatenating the input and output node sequences,
as implemented in \ensuremath{\Varid{parJungle}}.
$\exch_{m,n}$ differs from $\RELid_{m + n}$
only in the fact that the two parts of the output node sequence are swapped:

\restorecolumns
\begin{hscode}\SaveRestoreHook
\column{B}{@{}>{\hspre}l<{\hspost}@{}}%
\column{3}{@{}>{\hspre}l<{\hspost}@{}}%
\column{29}{@{}>{\hspre}l<{\hspost}@{}}%
\column{E}{@{}>{\hspre}l<{\hspost}@{}}%
\>[3]{}\Varid{exchJungle}\;\mathbin{:}\;(\Varid{m}\;\Varid{n}\;\mathbin{:}\;\Conid{ℕ})\;\Varid{→}\;{}\<[29]%
\>[29]{}\Conid{Jungle}\;(\Varid{m}\;\Varid{+}\;\Varid{n})\;(\Varid{n}\;\Varid{+}\;\Varid{m}){}\<[E]%
\\
\>[3]{}\Varid{exchJungle}\;\Varid{m}\;\Varid{n}\;\mathrel{=}\;\Varid{wire}\;(\Conid{Vec.map}\;(\Varid{raise}\;\Varid{m})\;(\Varid{allFin}\;\Varid{n})\;\plus \;\Conid{Vec.map}\;(\Varid{inject+}\;\Varid{n})\;(\Varid{allFin}\;\Varid{m})){}\<[E]%
\ColumnHook
\end{hscode}\resethooks

\begin{Def}\Deflabel{gs-monoidal}
A {\em strict gs-monoidal category}
is a symmetric strict monoidal category
where in addition
$!$ associates with every object $\objA$ of $\catCo$
an arrow $!_{\objA} : \objA \fun \munit$, and
$\Nabla$ associates with every object $\objA$ of $\catCo$
 an arrow $\NablaU{\objA} : \objA \fun \objA \otimes \objA$,
such that $\RELid_{\munit} = !_{\munit} = \NablaU{\munit}$,
and the following axioms hold:

\BCM
\def\arraystretch{1.3}
\begin{array}[b]{l}
    \NablaU{\objA} \RELcomp (\RELid_{\objA} \otimes \NablaU{\objA})
  \sepA{=}
    \NablaU{\objA} \RELcomp (\NablaU{\objA} \otimes \RELid_{\objA})
\qquad
    \NablaU{\objA} \RELcomp \exchU{\objA,\objA}
  \sepA{=}
    \NablaU{\objA}
\qquad
    \NablaU{\objA} \RELcomp (\RELid_{\objA} \otimes !_{\objA})
  \sepA{=}
    \RELid_{\objA}
\\
    \NablaU{\objA \otimes \objB} \RELcomp
    (\RELid_{\objA} \otimes \exchU{\objB,\objA} \otimes \RELid_{\objB})
  \sepA{=}
    \NablaU{\objA} \otimes \NablaU{\objB}
\qquad
\qquad
    !_{\objA \otimes \objB}
  \sepA{=}
    !_{\objA} \otimes !_{\objB}
\ECMAQ
\end{Def}

\noindent
In \ensuremath{\Conid{Jungle}},
the ``terminator'' $!_n$ differs from $\RELid_n$
only in the fact that the output node sequence is empty.

\restorecolumns
\begin{hscode}\SaveRestoreHook
\column{B}{@{}>{\hspre}l<{\hspost}@{}}%
\column{3}{@{}>{\hspre}l<{\hspost}@{}}%
\column{27}{@{}>{\hspre}l<{\hspost}@{}}%
\column{E}{@{}>{\hspre}l<{\hspost}@{}}%
\>[3]{}\Varid{termJungle}\;\mathbin{:}\;\{\mskip0.5mu \Varid{n}\;\mathbin{:}\;\Conid{ℕ}\mskip0.5mu\}\;\Varid{→}\;{}\<[27]%
\>[27]{}\Conid{Jungle}\;\Varid{n}\;\Varid{0}{}\<[E]%
\\
\>[3]{}\Varid{termJungle}\;\mathrel{=}\;\Varid{wire}\;[\mskip1.5mu \mskip1.5mu]{}\<[E]%
\ColumnHook
\end{hscode}\resethooks
The ``g'' of ``gs-monoidal'' stands for ``garbage'':
all edges of a term graph $G : m \fun n$
are garbage in the term graph $G \RELcomp !_n$.

The duplicator $\Nabla_{n}$ in \ensuremath{\Conid{Jungle}} differs from $\RELid_{n}$
only in the fact that the output node sequence is
the concatenation of the input node sequence with itself:

\restorecolumns
\begin{hscode}\SaveRestoreHook
\column{B}{@{}>{\hspre}l<{\hspost}@{}}%
\column{3}{@{}>{\hspre}l<{\hspost}@{}}%
\column{26}{@{}>{\hspre}l<{\hspost}@{}}%
\column{E}{@{}>{\hspre}l<{\hspost}@{}}%
\>[3]{}\Varid{dupJungle}\;\mathbin{:}\;\{\mskip0.5mu \Varid{n}\;\mathbin{:}\;\Conid{ℕ}\mskip0.5mu\}\;\Varid{→}\;{}\<[26]%
\>[26]{}\Conid{Jungle}\;\Varid{n}\;(\Varid{n}\;\Varid{+}\;\Varid{n}){}\<[E]%
\\
\>[3]{}\Varid{dupJungle}\;\{\mskip0.5mu \Varid{n}\mskip0.5mu\}\;\mathrel{=}\;\Varid{wire}\;(\Varid{allFin}\;\Varid{n}\;\plus \;\Varid{allFin}\;\Varid{n}){}\<[E]%
\ColumnHook
\end{hscode}\resethooks
The ``s'' of ``gs-monoidal'' stands for ``sharing'':
every input of $\Nabla_k \RELcomp (F \otimes G)$ is shared by $F : k \fun m$
and $G : k \fun n$.

Code graphs (and term graphs) over a fixed edge label set
form a gs-monoidal category, but not a \emph{Cartesian} category,
where in addition $!$ and $\Nabla$ are \emph{natural} transformations,
i.e., for all $F : \objA \fun \objB$
we have
$F \RELcomp !_{\objB} = !_{\objA}$
and
\hbox{$F \RELcomp \NablaU{\objB} = \NablaU{\objA} \RELcomp (F \otimes F)$.}
\ignore{
\begin{Def}\Deflabel{cartesian}
A {\em strict Cartesian category} $\catC$
is a strict gs-monoidal category
$( \catCo, \otimes, \munit, \exch, \Nabla, !  )$,
where
\begin{itemize}
\item $!$ is a natural transformation from the identity functor to the
  constant-$\munit$ functor,
  i.e., $F \RELcomp !_{\objB} = !_{\objA}$ for all $F : \objA \tfun \objB$,
  and
\item $\Nabla$ is a natural transformation from the identity functor to
  $\otimes$,
  i.e., $F \RELcomp \NablaU{\objB} = \NablaU{\objA} \RELcomp (F \otimes F)$
  for all $F : \objA \tfun \objB$
\qed
\end{itemize}
\end{Def}
}
%
To see how these naturality conditions are violated by term graphs,
the following five \ensuremath{\Conid{Jungle}}s
correspond to the expressions below them
(we draw jungles and code graphs from the inputs on top to the outputs
 at the bottom,
 with numbered triangles marking input and output positions,
 and rectangles enclosing edge labels).

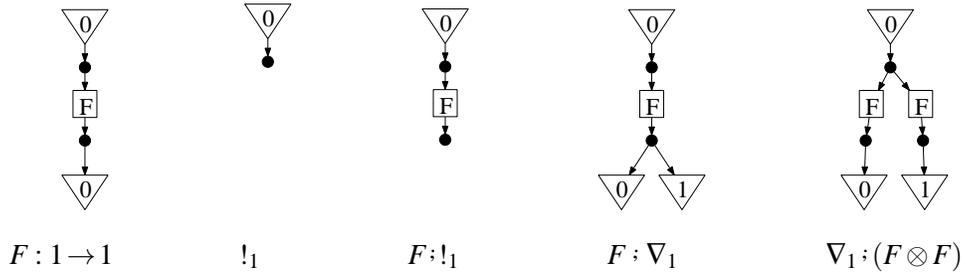
\phantom{.}\hfill
\CGpic{gF}\hfill 
\CGpicO{bang}{bb=35 -143 95 147}\hfill  
\CGpicO{gFbang}{bb=35 -45 95 245}\hfill 
\CGpic{gFdup}\hfill
\CGpic{dup_gF}\hfill
\phantom{.}

\kern1ex
\noindent
\phantom{.}\hfill
$F : 1 \fun 1$
\hfill
$!_1$\quad
\hfill
$F \,\RELcomp\, !_1$\quad
\hfill
$F \,\RELcomp\, \Nabla_1$\quad
\hfill
$\Nabla_1 \RELcomp (F \otimes F)$
\hfill
\strut

\bigskip
\noindent
Formalising (symmetric gs-) monoidal categories in Agda is
a straight-forward extension of the standard type-theoretic
formalisation of category theory deriving essentially from
Kanda's ``effective categories'' \cite{Kanda-1981};
this uses setoids of morphisms,
but not of objects.
This approach is also used by
Huet and Sa{\"{\i}}bi \cite{Huet-Saibi-1998,Huet-Saibi-2000}
for their formalisation of category theory in Coq,
and by Gonzal{\'{\i}}a  \cite{Gonzalia-2006}
for his formalisation of Freyd and Scedrov's allegory hierarchy \cite{Freyd-Scedrov-1990}
in Alf, a predecessor of Agda.

This approach also corresponds to the general practice
in category theory to consider objects only up to isomorphism,
not up to equality.
However, the definition of strict monoidal categories runs counter to
this approach, by assuming an object-level operation $(\otimes)$
satisfying non-trivial object-level equations.
Therefore we directly formalise what MacLane calls ``relaxed''
monoidal categories, with natural isomorphisms
$
\alpha : 
     \objA \otimes (\objB \otimes \objC)
   \fun
     (\objA \otimes \objB) \otimes \objC
$ and $\lambda : \triv \otimes \objA \fun \objA$
and $\rho : \objA \otimes \triv \fun \objA$.

This explicit approach also has advantages for moving
between different levels of data nesting without requiring additional
features;
this is important for example for reasoning about the effect of SIMD
operations together with SIMD vector manipulations on individual
scalar values, which is necessary for verifying numerous
high-performance ``tricks'', see e.g.\null \cite{Anand-Kahl-2007_AGTIVE}. 

\section{Term Graphs as \ensuremath{\Conid{Let}} Constructs}\sectlabel{Let}

The code graph representation of \sectref{CodeGraph}
essentially is a typed variant of the current internal representation
of Coconut code graphs, but, as mentioned in the introduction,
we essentially write Haskell definitions
to initially create code graphs.

In lazy functional programming implemented by graph reduction,
since at least KRC \cite{Turner-1982},
local definitions (via \ensuremath{\Keyword{let}} or \ensuremath{\Keyword{where}}) are 
understood to introduce \emph{sharing}.
In a mathematical context, 
\cite{Ariola-Klop-1994} represents cyclic term graphs
as systems of mutually recursive equations,
and \cite{Maraist-Oderski-Wadler-1998} presents sharing in the
call-by-need $\lambda$-calculus via \ensuremath{\Keyword{let}}-expressions.

In the following,
we present two formalisations of term graphs
defined by non-recursive nested \ensuremath{\Keyword{let}}-expressions.
For the sake of readability,
we restrict ourselves to untyped term graphs
and single-output primitives.

With \ensuremath{\Keyword{let}}-expressions,
we automatically have to deal with the complications
of bound variables,
involving scoping, renaming to avoid variable clashes, etc.
The Agda library \ensuremath{\Conid{NotSoFresh}} by Pouillard and Pottier
\cite{Pouillard-Pottier-2010} allows us to abstract
from these concerns to a large degree,
at the cost of following the discipline of their
\ensuremath{\Conid{World}}-based programming interface.
At the core of their approach,
there are \ensuremath{\Conid{World}}s in which different variables are in scope;
for a world \ensuremath{\Varid{α}}, the set of usable names is \ensuremath{\Conid{Name}\;\Varid{α}}.
Introducing a new name happens via a ``world extension link'';
an element of \ensuremath{\Varid{α}\;\Varid{↼}\;\Varid{β}} is a \emph{weak link} that provides a variable
in \ensuremath{\Varid{β}} that might be shadowing one of the variables in \ensuremath{\Varid{α}},
while an element of \ensuremath{\Varid{α}\;\Varid{↼→}\;\Varid{β}} is a \emph{strong link} that provides,
in \ensuremath{\Varid{β}}, a variable that is \emph{fresh} with respect to all variables in \ensuremath{\Varid{α}}.

For programming and in mathematics,
we are used to working in a context of weak links,
while symbol manipulation systems,
including theorem provers and compilers,
frequently disambiguate names so that they can work with strong links exclusively.
To enable both settings, we will parameterise over these
``world \ensuremath{\Conid{E}}xtension relation'' with a parameter \ensuremath{\Conid{E}\;\mathbin{:}\;\Conid{World}\;\Varid{→}\;\Conid{World}\;\Varid{→}\;\Conid{Set}}.

We first present the type \ensuremath{\Conid{TG}}
that formalises \ensuremath{\Keyword{let}}-expressions with arbitrary nesting;
this type is only a slight modification of the $\lambda$-term datatype
\ensuremath{\Conid{Tm}} from \cite{Pouillard-Pottier-2010}.

A value of type \ensuremath{\Conid{TG}\;\Conid{E}\;\Varid{α}\;\Varid{m}\;\Varid{n}}
is,
in the context of \ensuremath{\Varid{m}} input nodes
and of a world \ensuremath{\Varid{α}} providing already existing inner nodes,
a term graph ``suffix'' producing \ensuremath{\Varid{n}} output nodes:
\begin{itemize}
\item The input node at position \ensuremath{\Varid{i}} can be produced as an output node by
\ensuremath{\Conid{Input}\;\Varid{i}}.

\item An existing node \ensuremath{\Varid{x}\;\mathbin{:}\;\Conid{Name}\;\Varid{α}} is produced as an output node by
  \ensuremath{\Conid{V}\;\Varid{x}}.

\item The empty suffix is called \ensuremath{\Varid{ε}}.

\item Given two suffixes \ensuremath{\Varid{t}} and \ensuremath{\Varid{u}} of output lengths \ensuremath{\Varid{n₁}} and \ensuremath{\Varid{n₂}},
  their union, with concatenated output lists, is \ensuremath{\Varid{t}\;\Varid{▽}\;\Varid{u}}.
  The symbol \ensuremath{\Varid{▽}} reads ``fork'', as in the fork algebras of
  \cite{Haeberer-Frias-Baum-Veloso-1997}; it is related with the
  duplicator $\Nabla$ via the equation
  $\ensuremath{\Varid{t}\;\Varid{▽}\;\Varid{u}} = \NablaU{m} \RELcomp (\ensuremath{\Varid{t}} \otimes \ensuremath{\Varid{u}})$.

\item A primitive \ensuremath{\Varid{f}} can only be invoked while applying it to the outputs
of a term graph suffix \ensuremath{\Varid{t}} and while at the same time creating a new node
\ensuremath{\Varid{x}} in an expression of the shape \ensuremath{\Conid{Let}\;\Varid{x}\;\Varid{f}\;\Varid{t}\;\Varid{u}},
which, in more conventional notation, would read
``\textbf{let} \ensuremath{\Varid{x}\;\mathrel{=}\;\Varid{f}\;(\Varid{t})} \textbf{in} \ensuremath{\Varid{u}}''.

 If the primitive \ensuremath{\Varid{f}} expects \ensuremath{\Varid{k}} inputs,
 the argument term graph suffix \ensuremath{\Varid{t}},
 which may not use the new name \ensuremath{\Varid{x}} because it is in the ``old'' world \ensuremath{\Varid{α}},
 has to have \ensuremath{\Varid{k}} outputs.

 The term graph suffix \ensuremath{\Varid{u}} may use also the new name \ensuremath{\Varid{x}},
 and its outputs will be the outputs of the ``\ensuremath{\Conid{Let}\;\Varid{x}\;\Varid{f}\;\Varid{t}\;\Varid{u}}'' expression.
\end{itemize}

\begin{hscode}\SaveRestoreHook
\column{B}{@{}>{\hspre}l<{\hspost}@{}}%
\column{3}{@{}>{\hspre}l<{\hspost}@{}}%
\column{5}{@{}>{\hspre}l<{\hspost}@{}}%
\column{12}{@{}>{\hspre}l<{\hspost}@{}}%
\column{15}{@{}>{\hspre}l<{\hspost}@{}}%
\column{28}{@{}>{\hspre}l<{\hspost}@{}}%
\column{52}{@{}>{\hspre}l<{\hspost}@{}}%
\column{E}{@{}>{\hspre}l<{\hspost}@{}}%
\>[3]{}\Keyword{data}\;\Conid{TG}\;(\Conid{E}\;\mathbin{:}\;\Conid{World}\;\Varid{→}\;\Conid{World}\;\Varid{→}\;\Conid{Set})\;(\Varid{α}\;\mathbin{:}\;\Conid{World})\;(\Varid{m}\;\mathbin{:}\;\Conid{ℕ})\;\mathbin{:}\;\Conid{ℕ}\;\Varid{→}\;\Conid{Set}\;\Keyword{where}{}\<[E]%
\\
\>[3]{}\hsindent{2}{}\<[5]%
\>[5]{}\Conid{Input}\;{}\<[12]%
\>[12]{}\mathbin{:}\;(\Varid{i}\;\mathbin{:}\;\Conid{Fin}\;\Varid{m})\;{}\<[28]%
\>[28]{}\Varid{→}\;\Conid{TG}\;\Conid{E}\;\Varid{α}\;\Varid{m}\;\Varid{1}{}\<[E]%
\\
\>[3]{}\hsindent{2}{}\<[5]%
\>[5]{}\Conid{V}\;{}\<[12]%
\>[12]{}\mathbin{:}\;(\Varid{x}\;\mathbin{:}\;\Conid{Name}\;\Varid{α})\;{}\<[28]%
\>[28]{}\Varid{→}\;\Conid{TG}\;\Conid{E}\;\Varid{α}\;\Varid{m}\;\Varid{1}{}\<[E]%
\\
\>[3]{}\hsindent{2}{}\<[5]%
\>[5]{}\Varid{ε}\;{}\<[12]%
\>[12]{}\mathbin{:}\;\Conid{TG}\;\Conid{E}\;\Varid{α}\;\Varid{m}\;\Varid{0}{}\<[E]%
\\
\>[3]{}\hsindent{2}{}\<[5]%
\>[5]{}\Varid{\char95 ▽\char95 }\;{}\<[12]%
\>[12]{}\mathbin{:}\;\{\mskip0.5mu \Varid{n₁}\;\Varid{n₂}\;\mathbin{:}\;\Conid{ℕ}\mskip0.5mu\}\;\Varid{→}\;\Conid{TG}\;\Conid{E}\;\Varid{α}\;\Varid{m}\;\Varid{n₁}\;\Varid{→}\;\Conid{TG}\;\Conid{E}\;\Varid{α}\;\Varid{m}\;\Varid{n₂}\;\Varid{→}\;\Conid{TG}\;\Conid{E}\;\Varid{α}\;\Varid{m}\;(\Varid{n₁}\;\Varid{+}\;\Varid{n₂}){}\<[E]%
\\
\>[3]{}\hsindent{2}{}\<[5]%
\>[5]{}\Conid{Let}\;{}\<[12]%
\>[12]{}\mathbin{:}\;{}\<[15]%
\>[15]{}\{\mskip0.5mu \Varid{β}\;\mathbin{:}\;\Conid{World}\mskip0.5mu\}\;\{\mskip0.5mu \Varid{k}\;\Varid{n}\;\mathbin{:}\;\Conid{ℕ}\mskip0.5mu\}\;{}\<[E]%
\\
\>[12]{}\Varid{→}\;{}\<[15]%
\>[15]{}(\Varid{x}\;\mathbin{:}\;\Conid{E}\;\Varid{α}\;\Varid{β})\;{}\<[52]%
\>[52]{}\mbox{\onelinecomment  \textbf{let} \ensuremath{\Varid{x}}}{}\<[E]%
\\
\>[12]{}\Varid{→}\;{}\<[15]%
\>[15]{}(\Varid{f}\;\mathbin{:}\;\Conid{Label}\;\Varid{k})\;\Varid{→}\;(\Varid{t}\;\mathbin{:}\;\Conid{TG}\;\Conid{E}\;\Varid{α}\;\Varid{m}\;\Varid{k})\;{}\<[52]%
\>[52]{}\mbox{\onelinecomment  $\mathbf{=}$ \ensuremath{\Varid{f}\;(\Varid{t})} }{}\<[E]%
\\
\>[12]{}\Varid{→}\;{}\<[15]%
\>[15]{}(\Varid{u}\;\mathbin{:}\;\Conid{TG}\;\Conid{E}\;\Varid{β}\;\Varid{m}\;\Varid{n})\;{}\<[52]%
\>[52]{}\mbox{\onelinecomment  \textbf{in} \ensuremath{\Varid{u}}}{}\<[E]%
\\
\>[12]{}\Varid{→}\;{}\<[15]%
\>[15]{}\Conid{TG}\;\Conid{E}\;\Varid{α}\;\Varid{m}\;\Varid{n}{}\<[E]%
\ColumnHook
\end{hscode}\resethooks

\noindent
Without additional support,
defining term graphs using this interface is somewhat inconvenient ---
the following assumes a unary label \ensuremath{\Conid{F}}, a binary label \ensuremath{\Conid{G}},
and a ternary label \ensuremath{\Conid{H}}:

\noindent
\begin{minipage}[b]{0.65\textwidth}
\begin{hscode}\SaveRestoreHook
\column{B}{@{}>{\hspre}l<{\hspost}@{}}%
\column{3}{@{}>{\hspre}l<{\hspost}@{}}%
\column{5}{@{}>{\hspre}l<{\hspost}@{}}%
\column{7}{@{}>{\hspre}l<{\hspost}@{}}%
\column{9}{@{}>{\hspre}l<{\hspost}@{}}%
\column{10}{@{}>{\hspre}l<{\hspost}@{}}%
\column{20}{@{}>{\hspre}l<{\hspost}@{}}%
\column{47}{@{}>{\hspre}l<{\hspost}@{}}%
\column{E}{@{}>{\hspre}l<{\hspost}@{}}%
\>[3]{}\Conid{TG0}\;\mathbin{:}\;\Conid{Label}\;\Varid{1}\;\Varid{→}\;\Conid{Label}\;\Varid{2}\;\Varid{→}\;\Conid{Label}\;\Varid{3}\;\Varid{→}\;\Conid{TG}\;\Varid{\char95 ↼\char95 }\;\Varid{ø}\;\Varid{3}\;\Varid{1}{}\<[E]%
\\
\>[3]{}\Conid{TG0}\;\Conid{F}\;\Conid{G}\;\Conid{H}\;\mathrel{=}\;\Keyword{let}\;{}\<[20]%
\>[20]{}\Varid{f0}\;\mathrel{=}\;\Varid{freshø}{}\<[47]%
\>[47]{}\mbox{\onelinecomment  a strong link}{}\<[E]%
\\
\>[20]{}\Varid{x0}\;\mathrel{=}\;\Conid{FreshPack.weakOf}\;\Varid{f0}{}\<[47]%
\>[47]{}\mbox{\onelinecomment  weak view of \ensuremath{\Varid{f0}}}{}\<[E]%
\\
\>[20]{}\Varid{n0}\;\mathrel{=}\;\Conid{FreshPack.nameOf}\;\Varid{f0}{}\<[47]%
\>[47]{}\mbox{\onelinecomment  \ensuremath{\Conid{Name}} of \ensuremath{\Varid{f0}}}{}\<[E]%
\\
\>[3]{}\hsindent{2}{}\<[5]%
\>[5]{}\Keyword{in}\;\Conid{Let}\;\Varid{x0}\;\Conid{H}\;{}\<[E]%
\\
\>[5]{}\hsindent{4}{}\<[9]%
\>[9]{}(\Conid{Let}\;\Varid{x0}\;\Conid{F}\;(\Conid{Input}\;\Varid{zero})\;(\Conid{V}\;\Varid{n0}\;\Varid{▽}\;\Conid{V}\;\Varid{n0}){}\<[E]%
\\
\>[9]{}\hsindent{1}{}\<[10]%
\>[10]{}\Varid{▽}{}\<[E]%
\\
\>[9]{}\hsindent{1}{}\<[10]%
\>[10]{}\Conid{Let}\;\Varid{x0}\;\Conid{G}\;(\Conid{Input}\;(\Varid{suc}\;\Varid{zero})\;\Varid{▽}\;\Conid{Input}\;(\Varid{suc}\;(\Varid{suc}\;\Varid{zero})))\;(\Conid{V}\;\Varid{n0}){}\<[E]%
\\
\>[5]{}\hsindent{4}{}\<[9]%
\>[9]{}){}\<[E]%
\\
\>[5]{}\hsindent{2}{}\<[7]%
\>[7]{}(\Conid{V}\;\Varid{n0}){}\<[E]%
\ColumnHook
\end{hscode}\resethooks
\end{minipage}
\hfill
\raise5ex\hbox{\CGpic{tg0}}
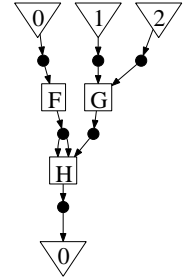
\hfill
\strut

\noindent
Using slightly more conventional notation,
this corresponds to the following, relatively readable version,
with ``\ensuremath{\Varid{i}}'' prefixing inputs and ``\ensuremath{\Varid{n}}'' prefixing node names:
\begin{hscode}\SaveRestoreHook
\column{B}{@{}>{\hspre}l<{\hspost}@{}}%
\column{13}{@{}>{\hspre}l<{\hspost}@{}}%
\column{16}{@{}>{\hspre}l<{\hspost}@{}}%
\column{E}{@{}>{\hspre}l<{\hspost}@{}}%
\>[B]{}\Keyword{let}\;\Varid{n0}\;\mathrel{=}\;\Conid{H}\;{}\<[13]%
\>[13]{}({}\<[16]%
\>[16]{}(\Keyword{let}\;\Varid{n0}\;\mathrel{=}\;\Conid{F}\;(\Varid{i0})\;\Keyword{in}\;(\Varid{n0}\;\Varid{▽}\;\Varid{n0})){}\<[E]%
\\
\>[16]{}\Varid{▽}{}\<[E]%
\\
\>[16]{}(\Keyword{let}\;\Varid{n0}\;\mathrel{=}\;\Conid{G}\;(\Varid{i1}\;\Varid{▽}\;\Varid{i2})\;\Keyword{in}\;\Varid{n0}){}\<[E]%
\\
\>[13]{})\;\Keyword{in}\;\Varid{n0}{}\<[E]%
\ColumnHook
\end{hscode}\resethooks

\ignore{
\begin{hscode}\SaveRestoreHook
\column{B}{@{}>{\hspre}l<{\hspost}@{}}%
\column{5}{@{}>{\hspre}l<{\hspost}@{}}%
\column{9}{@{}>{\hspre}l<{\hspost}@{}}%
\column{E}{@{}>{\hspre}l<{\hspost}@{}}%
\>[B]{}\Keyword{let}\;\Varid{n0}\;\mathrel{=}\;\Conid{H}\;\Varid{◅}{}\<[E]%
\\
\>[B]{}\hsindent{5}{}\<[5]%
\>[5]{}\Keyword{let}\;\Varid{n0}\;\mathrel{=}\;\Conid{F}\;\Varid{◅}{}\<[E]%
\\
\>[5]{}\hsindent{4}{}\<[9]%
\>[9]{}\Varid{i0}{}\<[E]%
\\
\>[B]{}\hsindent{5}{}\<[5]%
\>[5]{}\Keyword{in}{}\<[E]%
\\
\>[5]{}\hsindent{4}{}\<[9]%
\>[9]{}\Varid{n0}{}\<[E]%
\\
\>[5]{}\hsindent{4}{}\<[9]%
\>[9]{}\Varid{n0}{}\<[E]%
\\
\>[B]{}\hsindent{5}{}\<[5]%
\>[5]{}\Keyword{let}\;\Varid{n0}\;\mathrel{=}\;\Conid{G}\;\Varid{◅}{}\<[E]%
\\
\>[5]{}\hsindent{4}{}\<[9]%
\>[9]{}\Varid{i1}{}\<[E]%
\\
\>[5]{}\hsindent{4}{}\<[9]%
\>[9]{}\Varid{i2}{}\<[E]%
\\
\>[B]{}\hsindent{5}{}\<[5]%
\>[5]{}\Keyword{in}{}\<[E]%
\\
\>[5]{}\hsindent{4}{}\<[9]%
\>[9]{}\Varid{n0}{}\<[E]%
\\
\>[B]{}\Keyword{in}{}\<[E]%
\\
\>[B]{}\hsindent{5}{}\<[5]%
\>[5]{}\Varid{n0}{}\<[E]%
\ColumnHook
\end{hscode}\resethooks
}

\noindent
Either by adding more notational support,
or by defining a separate input language,
this can provide an interface that comes reasonably close to
Haskell-style programming.

The real point of the definition of \ensuremath{\Conid{TG}} however
is that it not only provides an input language,
but also a representation of term graphs
that can be manipulated and transformed by programs.
For example, we can turn a \ensuremath{\Conid{TG}} with name shadowing
(i.e., using weak links)
into one with strong links by replacing all node names
with fresh names relative to their respective worlds:

\begin{hscode}\SaveRestoreHook
\column{B}{@{}>{\hspre}l<{\hspost}@{}}%
\column{3}{@{}>{\hspre}l<{\hspost}@{}}%
\column{8}{@{}>{\hspre}l<{\hspost}@{}}%
\column{10}{@{}>{\hspre}l<{\hspost}@{}}%
\column{11}{@{}>{\hspre}l<{\hspost}@{}}%
\column{17}{@{}>{\hspre}l<{\hspost}@{}}%
\column{31}{@{}>{\hspre}l<{\hspost}@{}}%
\column{32}{@{}>{\hspre}l<{\hspost}@{}}%
\column{E}{@{}>{\hspre}l<{\hspost}@{}}%
\>[3]{}\Varid{strengthenTG}\;{}\<[17]%
\>[17]{}\mathbin{:}\;\{\mskip0.5mu \Varid{α}\;\Varid{α′}\;\mathbin{:}\;\Conid{World}\mskip0.5mu\}\;\Varid{→}\;\Conid{Fresh}\;\Varid{α′}\;\Varid{→}\;\Conid{CEnv}\;(\Conid{Name}\;\Varid{α′})\;\Varid{α}\;{}\<[E]%
\\
\>[17]{}\Varid{→}\;\{\mskip0.5mu \Varid{m}\;\Varid{n}\;\mathbin{:}\;\Conid{ℕ}\mskip0.5mu\}\;\Varid{→}\;\Conid{TG}\;\Varid{\char95 ↼\char95 }\;\Varid{α}\;\Varid{m}\;\Varid{n}\;\Varid{→}\;\Conid{TG}\;\Varid{\char95 ↼→\char95 }\;\Varid{α′}\;\Varid{m}\;\Varid{n}{}\<[E]%
\\
\>[3]{}\Varid{strengthenTG}\;\anonymous \;\anonymous \;\Varid{ε}\;{}\<[31]%
\>[31]{}\mathrel{=}\;\Varid{ε}{}\<[E]%
\\
\>[3]{}\Varid{strengthenTG}\;\Varid{fr}\;\Conid{Γ}\;(\Varid{t}\;\Varid{▽}\;\Varid{u})\;{}\<[32]%
\>[32]{}\mathrel{=}\;(\Varid{strengthenTG}\;\Varid{fr}\;\Conid{Γ}\;\Varid{t})\;\Varid{▽}\;(\Varid{strengthenTG}\;\Varid{fr}\;\Conid{Γ}\;\Varid{u}){}\<[E]%
\\
\>[3]{}\Varid{strengthenTG}\;\anonymous \;\anonymous \;(\Conid{Input}\;\Varid{i})\;{}\<[31]%
\>[31]{}\mathrel{=}\;\Conid{Input}\;\Varid{i}{}\<[E]%
\\
\>[3]{}\Varid{strengthenTG}\;\Varid{fr}\;\Conid{Γ}\;(\Conid{V}\;\Varid{x})\;{}\<[32]%
\>[32]{}\mathrel{=}\;\Conid{V}\;(\Varid{lookupCEnv}\;\Conid{Γ}\;\Varid{x}){}\<[E]%
\\
\>[3]{}\Varid{strengthenTG}\;\Varid{fr}\;\Conid{Γ}\;(\Conid{Let}\;\Varid{x}\;\Varid{f}\;\Varid{t}\;\Varid{u})\;{}\<[E]%
\\
\>[3]{}\hsindent{5}{}\<[8]%
\>[8]{}\mathrel{=}\;{}\<[11]%
\>[11]{}\Keyword{let}\;\Conid{Γ′}\;\mathrel{=}\;\Varid{mapCEnv}\;\Varid{importWith}\;\Conid{Γ},\Varid{x}\;\Varid{↦}\;\Varid{nameOf}{}\<[E]%
\\
\>[11]{}\Keyword{in}\;\Conid{Let}\;\Varid{strongOf}\;\Varid{f}\;(\Varid{strengthenTG}\;\Varid{fr}\;\Conid{Γ}\;\Varid{t})\;(\Varid{strengthenTG}\;\Varid{nextOf}\;\Conid{Γ′}\;\Varid{u}){}\<[E]%
\\
\>[8]{}\hsindent{2}{}\<[10]%
\>[10]{}\Keyword{where}\;\Keyword{open}\;\Conid{FreshPack}\;\Varid{fr}{}\<[E]%
\ColumnHook
\end{hscode}\resethooks

\noindent
Parallel composition is also easy to program,
using fork after embedding, respectively shifting, the inputs:

\begin{hscode}\SaveRestoreHook
\column{B}{@{}>{\hspre}l<{\hspost}@{}}%
\column{3}{@{}>{\hspre}l<{\hspost}@{}}%
\column{10}{@{}>{\hspre}l<{\hspost}@{}}%
\column{E}{@{}>{\hspre}l<{\hspost}@{}}%
\>[3]{}\Varid{parTG}\;{}\<[10]%
\>[10]{}\mathbin{:}\;\{\mskip0.5mu \Conid{E}\;\mathbin{:}\;\anonymous \mskip0.5mu\}\;\{\mskip0.5mu \Varid{α}\;\mathbin{:}\;\anonymous \mskip0.5mu\}\;\{\mskip0.5mu \Varid{m₁}\;\Varid{n₁}\;\Varid{m₂}\;\Varid{n₂}\;\mathbin{:}\;\Conid{ℕ}\mskip0.5mu\}\;{}\<[E]%
\\
\>[10]{}\Varid{→}\;\Conid{TG}\;\Conid{E}\;\Varid{α}\;\Varid{m₁}\;\Varid{n₁}\;\Varid{→}\;\Conid{TG}\;\Conid{E}\;\Varid{α}\;\Varid{m₂}\;\Varid{n₂}\;\Varid{→}\;\Conid{TG}\;\Conid{E}\;\Varid{α}\;(\Varid{m₁}\;\Varid{+}\;\Varid{m₂})\;(\Varid{n₁}\;\Varid{+}\;\Varid{n₂}){}\<[E]%
\\
\>[3]{}\Varid{parTG}\;\{\mskip0.5mu \Conid{E}\mskip0.5mu\}\;\{\mskip0.5mu \Varid{α}\mskip0.5mu\}\;\{\mskip0.5mu \Varid{m₁}\mskip0.5mu\}\;\{\mskip0.5mu \Varid{n₁}\mskip0.5mu\}\;\Varid{g₁}\;\{\mskip0.5mu \Varid{m₂}\mskip0.5mu\}\;\{\mskip0.5mu \Varid{n₂}\mskip0.5mu\}\;\Varid{g₂}\;\mathrel{=}\;\Varid{extendTG}\;\Varid{m₂}\;\Varid{g₁}\;\Varid{▽}\;\Varid{shiftTG}\;\Varid{m₁}\;\Varid{g₂}{}\<[E]%
\ColumnHook
\end{hscode}\resethooks

\noindent
Sequential composition is much harder to implement directly,
since the output nodes of the first argument
may have been defined in separate worlds and combined with fork,
and now need to be brought into a common world,
which in general requires renaming and restructuring.
A convenient ``canonical form'' for such \ensuremath{\Keyword{let}}-expressions
has no \ensuremath{\Conid{Let}} at argument positions,
and no \ensuremath{\Conid{Let}} below fork,
and therefore degenerates into a sequence of \ensuremath{\Conid{Let}} declarations
each binding a new node to the application of some primitive
to existing nodes.
When dealing with any kind of canonical forms,
especially in a dependently-typed setting,
it is frequently worth while declaring this as a separate datatype
so that it becomes easier to exploit its properties.
For this canonical form of \ensuremath{\Conid{TG}},
we introduce a separate datatype with additional restructuring below.

\section{Term Graphs with Sequential Node Declaration}\sectlabel{SeqLet}

According to our explanation of \ensuremath{\Conid{TG}} term graphs, \ensuremath{\Varid{▽}} with \ensuremath{\Varid{ε}}
obviously forms a monoid,
but the monoid laws do not come for free in \ensuremath{\Conid{TG}}.
Moving to the \ensuremath{\Conid{Vec}} container type instead
provides us with the monoid laws in the standard library,
and makes for a more canonical representation.
With this change, and with strictly linearised node declaration,
the term graph \ensuremath{\Conid{TG0}} shown above could be written in a somewhat
conventional notation as follows (without fully specifying the
number of inputs):

\begin{hscode}\SaveRestoreHook
\column{B}{@{}>{\hspre}l<{\hspost}@{}}%
\column{7}{@{}>{\hspre}l<{\hspost}@{}}%
\column{E}{@{}>{\hspre}l<{\hspost}@{}}%
\>[B]{}\Keyword{let}\;\Varid{n0}\;\mathrel{=}\;\Conid{F}\;\Varid{i0}{}\<[E]%
\\
\>[B]{}\Keyword{let}\;\Varid{n1}\;\mathrel{=}\;\Conid{G}\;\Varid{i1}\;\Varid{i2}{}\<[E]%
\\
\>[B]{}\Keyword{let}\;\Varid{n2}\;\mathrel{=}\;\Conid{H}\;\Varid{n0}\;\Varid{n0}\;\Varid{n1}{}\<[E]%
\\
\>[B]{}\Keyword{in}\;[\mskip1.5mu {}\<[7]%
\>[7]{}\Varid{n2}\mskip1.5mu]{}\<[E]%
\ColumnHook
\end{hscode}\resethooks

\noindent
We introduce the type \ensuremath{\Conid{Arg}} for individual nodes,
either existing inner nodes, or input positions,
and a type synonym \ensuremath{\Conid{Args}} for their vectors:

\savecolumns
\begin{hscode}\SaveRestoreHook
\column{B}{@{}>{\hspre}l<{\hspost}@{}}%
\column{3}{@{}>{\hspre}l<{\hspost}@{}}%
\column{5}{@{}>{\hspre}l<{\hspost}@{}}%
\column{12}{@{}>{\hspre}l<{\hspost}@{}}%
\column{E}{@{}>{\hspre}l<{\hspost}@{}}%
\>[3]{}\Keyword{data}\;\Conid{Arg}\;\Varid{α}\;(\Varid{m}\;\mathbin{:}\;\Conid{ℕ})\;\mathbin{:}\;\Conid{Set}\;\Keyword{where}{}\<[E]%
\\
\>[3]{}\hsindent{2}{}\<[5]%
\>[5]{}\Conid{Input}\;{}\<[12]%
\>[12]{}\mathbin{:}\;(\Varid{i}\;\mathbin{:}\;\Conid{Fin}\;\Varid{m})\;\Varid{→}\;\Conid{Arg}\;\Varid{α}\;\Varid{m}{}\<[E]%
\\
\>[3]{}\hsindent{2}{}\<[5]%
\>[5]{}\Conid{V}\;{}\<[12]%
\>[12]{}\mathbin{:}\;(\Varid{x}\;\mathbin{:}\;\Conid{Name}\;\Varid{α})\;\Varid{→}\;\Conid{Arg}\;\Varid{α}\;\Varid{m}{}\<[E]%
\\[\blanklineskip]%
\>[3]{}\Conid{Args}\;\Varid{α}\;\Varid{m}\;\Varid{n}\;\mathrel{=}\;\Conid{Vec}\;(\Conid{Arg}\;\Varid{α}\;\Varid{m})\;\Varid{n}{}\<[E]%
\ColumnHook
\end{hscode}\resethooks

\noindent
The datatype \ensuremath{\Conid{TG′}} has the same reading as \ensuremath{\Conid{TG}},
but a simpler structure:
\begin{itemize}
\item If all nodes have been declared, \ensuremath{\Conid{Output}\;\Varid{as}} assembles the
  vector of output nodes.

\item \ensuremath{\Conid{Let}\;\Varid{x}\;\Varid{f}\;\Varid{v}\;\Varid{u}},
  which, in more conventional notation, would read
  ``\textbf{let} \ensuremath{\Varid{x}\;\mathrel{=}\;\Varid{f}\;(\Varid{v})} \textbf{in} \ensuremath{\Varid{u}}'',
  binds a new node \ensuremath{\Varid{x}} to an edge labelled \ensuremath{\Varid{f}} with input nodes \ensuremath{\Varid{v}},
  and makes \ensuremath{\Varid{x}} visible in the remaining term graph suffix \ensuremath{\Varid{u}}.
\end{itemize}

\restorecolumns
\begin{hscode}\SaveRestoreHook
\column{B}{@{}>{\hspre}l<{\hspost}@{}}%
\column{3}{@{}>{\hspre}l<{\hspost}@{}}%
\column{5}{@{}>{\hspre}l<{\hspost}@{}}%
\column{12}{@{}>{\hspre}l<{\hspost}@{}}%
\column{15}{@{}>{\hspre}l<{\hspost}@{}}%
\column{49}{@{}>{\hspre}l<{\hspost}@{}}%
\column{E}{@{}>{\hspre}l<{\hspost}@{}}%
\>[3]{}\Keyword{data}\;\Conid{TG′}\;\Conid{E}\;\Varid{α}\;(\Varid{m}\;\mathbin{:}\;\Conid{ℕ})\;\mathbin{:}\;\Conid{ℕ}\;\Varid{→}\;\Conid{Set}\;\Keyword{where}{}\<[E]%
\\
\>[3]{}\hsindent{2}{}\<[5]%
\>[5]{}\Conid{Output}\;\mathbin{:}\;{}\<[15]%
\>[15]{}\{\mskip0.5mu \Varid{n}\;\mathbin{:}\;\Conid{ℕ}\mskip0.5mu\}\;\Varid{→}\;\Conid{Args}\;\Varid{α}\;\Varid{m}\;\Varid{n}\;\Varid{→}\;\Conid{TG′}\;\Conid{E}\;\Varid{α}\;\Varid{m}\;\Varid{n}{}\<[E]%
\\
\>[3]{}\hsindent{2}{}\<[5]%
\>[5]{}\Conid{Let}\;{}\<[12]%
\>[12]{}\mathbin{:}\;{}\<[15]%
\>[15]{}\{\mskip0.5mu \Varid{β}\;\mathbin{:}\;\Conid{World}\mskip0.5mu\}\;\{\mskip0.5mu \Varid{k}\;\Varid{n}\;\mathbin{:}\;\Conid{ℕ}\mskip0.5mu\}\;{}\<[E]%
\\
\>[12]{}\Varid{→}\;{}\<[15]%
\>[15]{}(\Varid{x}\;\mathbin{:}\;\Conid{E}\;\Varid{α}\;\Varid{β})\;{}\<[49]%
\>[49]{}\mbox{\onelinecomment  \textbf{let} \ensuremath{\Varid{x}}}{}\<[E]%
\\
\>[12]{}\Varid{→}\;{}\<[15]%
\>[15]{}(\Varid{f}\;\mathbin{:}\;\Conid{Label}\;\Varid{k})\;(\Varid{v}\;\mathbin{:}\;\Conid{Args}\;\Varid{α}\;\Varid{m}\;\Varid{k})\;{}\<[49]%
\>[49]{}\mbox{\onelinecomment  $\mathbf{=}$ \ensuremath{\Varid{f}\;(\Varid{v})} }{}\<[E]%
\\
\>[12]{}\Varid{→}\;{}\<[15]%
\>[15]{}(\Varid{u}\;\mathbin{:}\;\Conid{TG′}\;\Conid{E}\;\Varid{β}\;\Varid{m}\;\Varid{n})\;{}\<[49]%
\>[49]{}\mbox{\onelinecomment  \textbf{in} \ensuremath{\Varid{u}}}{}\<[E]%
\\
\>[12]{}\Varid{→}\;{}\<[15]%
\>[15]{}\Conid{TG′}\;\Conid{E}\;\Varid{α}\;\Varid{m}\;\Varid{n}{}\<[E]%
\ColumnHook
\end{hscode}\resethooks

\noindent
We first show that primitive and wiring graphs are easily programmed:

\begin{hscode}\SaveRestoreHook
\column{B}{@{}>{\hspre}l<{\hspost}@{}}%
\column{3}{@{}>{\hspre}l<{\hspost}@{}}%
\column{5}{@{}>{\hspre}l<{\hspost}@{}}%
\column{E}{@{}>{\hspre}l<{\hspost}@{}}%
\>[3]{}\Varid{prim}\;\mathbin{:}\;\{\mskip0.5mu \Varid{k}\;\mathbin{:}\;\Conid{ℕ}\mskip0.5mu\}\;\Varid{→}\;\Conid{Label}\;\Varid{k}\;\Varid{→}\;\Conid{TG′}\;\Varid{\char95 ↼→\char95 }\;\Varid{ø}\;\Varid{k}\;\Varid{1}{}\<[E]%
\\
\>[3]{}\Varid{prim}\;\{\mskip0.5mu \Varid{k}\mskip0.5mu\}\;\Varid{f}\;\mathrel{=}\;\Conid{Let}\;\Varid{strongOf}\;\Varid{f}\;(\Conid{Vec.map}\;\Conid{Input}\;(\Conid{Vec.allFin}\;\Varid{k}))\;(\Conid{Output}\;[\mskip1.5mu \Conid{V}\;\Varid{nameOf}\mskip1.5mu]){}\<[E]%
\\
\>[3]{}\hsindent{2}{}\<[5]%
\>[5]{}\Keyword{where}\;\Keyword{open}\;\Conid{FreshPack}\;\Varid{freshø}{}\<[E]%
\\[\blanklineskip]%
\>[3]{}\Varid{wire}\;\mathbin{:}\;\{\mskip0.5mu \Varid{k}\;\Varid{n}\;\mathbin{:}\;\Conid{ℕ}\mskip0.5mu\}\;\{\mskip0.5mu \Conid{E}\;\mathbin{:}\;\anonymous \mskip0.5mu\}\;\{\mskip0.5mu \Varid{α}\;\mathbin{:}\;\Conid{World}\mskip0.5mu\}\;\Varid{→}\;\Conid{Vec}\;(\Conid{Fin}\;\Varid{k})\;\Varid{n}\;\Varid{→}\;\Conid{TG′}\;\Conid{E}\;\Varid{α}\;\Varid{k}\;\Varid{n}{}\<[E]%
\\
\>[3]{}\Varid{wire}\;\Varid{v}\;\mathrel{=}\;\Conid{Output}\;(\Conid{Vec.map}\;\Conid{Input}\;\Varid{v}){}\<[E]%
\\[\blanklineskip]%
\>[3]{}\Varid{idWire}\;\mathbin{:}\;\{\mskip0.5mu \Varid{k}\;\mathbin{:}\;\Conid{ℕ}\mskip0.5mu\}\;\{\mskip0.5mu \Conid{E}\;\mathbin{:}\;\anonymous \mskip0.5mu\}\;\{\mskip0.5mu \Varid{α}\;\mathbin{:}\;\Conid{World}\mskip0.5mu\}\;\Varid{→}\;\Conid{TG′}\;\Conid{E}\;\Varid{α}\;\Varid{k}\;\Varid{k}{}\<[E]%
\\
\>[3]{}\Varid{idWire}\;\{\mskip0.5mu \Varid{k}\mskip0.5mu\}\;\mathrel{=}\;\Varid{wire}\;(\Conid{Vec.allFin}\;\Varid{k}){}\<[E]%
\\[\blanklineskip]%
\>[3]{}\Varid{dup}\;\mathbin{:}\;\{\mskip0.5mu \Varid{k}\;\mathbin{:}\;\Conid{ℕ}\mskip0.5mu\}\;\{\mskip0.5mu \Conid{E}\;\mathbin{:}\;\anonymous \mskip0.5mu\}\;\{\mskip0.5mu \Varid{α}\;\mathbin{:}\;\Conid{World}\mskip0.5mu\}\;\Varid{→}\;\Conid{TG′}\;\Conid{E}\;\Varid{α}\;\Varid{k}\;(\Varid{k}\;\Varid{+}\;\Varid{k}){}\<[E]%
\\
\>[3]{}\Varid{dup}\;\{\mskip0.5mu \Varid{k}\mskip0.5mu\}\;\mathrel{=}\;\Varid{wire}\;(\Conid{Vec.allFin}\;\Varid{k}\;\plus \;\Conid{Vec.allFin}\;\Varid{k}){}\<[E]%
\\[\blanklineskip]%
\>[3]{}\Varid{term}\;\mathbin{:}\;\{\mskip0.5mu \Varid{k}\;\mathbin{:}\;\Conid{ℕ}\mskip0.5mu\}\;\{\mskip0.5mu \Conid{E}\;\mathbin{:}\;\anonymous \mskip0.5mu\}\;\{\mskip0.5mu \Varid{α}\;\mathbin{:}\;\Conid{World}\mskip0.5mu\}\;\Varid{→}\;\Conid{TG′}\;\Conid{E}\;\Varid{α}\;\Varid{k}\;\Varid{0}{}\<[E]%
\\
\>[3]{}\Varid{term}\;\mathrel{=}\;\Varid{wire}\;[\mskip1.5mu \mskip1.5mu]{}\<[E]%
\ColumnHook
\end{hscode}\resethooks

\noindent
With these definitions,
we can reconstruct the term graph \ensuremath{\Conid{TG0}} from above
via the gs-monoidal interface,
with sequential composition \ensuremath{\Varid{seqTG′}} and parallel composition \ensuremath{\Varid{parTG′}} defined below:

\begin{hscode}\SaveRestoreHook
\column{B}{@{}>{\hspre}l<{\hspost}@{}}%
\column{E}{@{}>{\hspre}l<{\hspost}@{}}%
\>[B]{}\Varid{tg0}\;\mathrel{=}\;\Varid{seqTG′}\;(\Varid{parTG′}\;(\Varid{seqTG′}\;(\Varid{prim}\;\Conid{F})\;\Varid{dup})\;(\Varid{prim}\;\Conid{G}))\;(\Varid{prim}\;\Conid{H}){}\<[E]%
\ColumnHook
\end{hscode}\resethooks

\noindent
For the analogous function to \ensuremath{\Varid{strengthenTG}},
which replaces each link \ensuremath{\Varid{x}} in a \ensuremath{\Conid{Let}} construct
with a fresh link,
we present an easy generalisation
to serve dual purposes:
\begin{itemize}
\item Starting from weak links, \ensuremath{\Varid{strengthenTG′}\;\{\mskip0.5mu \Varid{\char95 ↼\char95 }\mskip0.5mu\}\;\Varid{id}} is proper
  strengthening;
\item starting from strong links,
  \ensuremath{\Varid{strengthenTG′}\;\{\mskip0.5mu \Varid{\char95 ↼→\char95 }\mskip0.5mu\}\;\Conid{StrongPack.weakOf}} is renaming with fresh
  names with respect to the new world \ensuremath{\Varid{α′}}.
\end{itemize}

\begin{hscode}\SaveRestoreHook
\column{B}{@{}>{\hspre}l<{\hspost}@{}}%
\column{3}{@{}>{\hspre}l<{\hspost}@{}}%
\column{8}{@{}>{\hspre}l<{\hspost}@{}}%
\column{11}{@{}>{\hspre}l<{\hspost}@{}}%
\column{18}{@{}>{\hspre}l<{\hspost}@{}}%
\column{42}{@{}>{\hspre}c<{\hspost}@{}}%
\column{42E}{@{}l@{}}%
\column{45}{@{}>{\hspre}l<{\hspost}@{}}%
\column{E}{@{}>{\hspre}l<{\hspost}@{}}%
\>[3]{}\Varid{strengthenTG′}\;{}\<[18]%
\>[18]{}\mathbin{:}\;\{\mskip0.5mu \Conid{E}\;\mathbin{:}\;\anonymous \mskip0.5mu\}\;\Varid{→}\;(\Conid{E}\;\Varid{⇒}\;\Varid{\char95 ↼\char95 })\;{}\<[E]%
\\
\>[18]{}\Varid{→}\;\{\mskip0.5mu \Varid{α}\;\Varid{α′}\;\mathbin{:}\;\Conid{World}\mskip0.5mu\}\;\Varid{→}\;\Conid{Fresh}\;\Varid{α′}\;\Varid{→}\;\Conid{CEnv}\;(\Conid{Name}\;\Varid{α′})\;\Varid{α}\;{}\<[E]%
\\
\>[18]{}\Varid{→}\;\{\mskip0.5mu \Varid{m}\;\Varid{n}\;\mathbin{:}\;\Conid{ℕ}\mskip0.5mu\}\;\Varid{→}\;\Conid{TG′}\;\Conid{E}\;\Varid{α}\;\Varid{m}\;\Varid{n}\;\Varid{→}\;\Conid{TG′}\;\Varid{\char95 ↼→\char95 }\;\Varid{α′}\;\Varid{m}\;\Varid{n}{}\<[E]%
\\
\>[3]{}\Varid{strengthenTG′}\;\Varid{weak}\;\Varid{fr}\;\Conid{Γ}\;(\Conid{Output}\;\Varid{as})\;\mathrel{=}\;\Conid{Output}\;(\Varid{mapVarArgs}\;(\Varid{lookupCEnv}\;\Conid{Γ})\;\Varid{as}){}\<[E]%
\\
\>[3]{}\Varid{strengthenTG′}\;\Varid{weak}\;\Varid{fr}\;\Conid{Γ}\;(\Conid{Let}\;\Varid{x}\;\Varid{f}\;\Varid{as}\;\Varid{u})\;{}\<[E]%
\\
\>[3]{}\hsindent{5}{}\<[8]%
\>[8]{}\mathrel{=}\;{}\<[11]%
\>[11]{}\Keyword{let}\;\Conid{Γ′}\;\mathrel{=}\;\Varid{mapCEnv}\;\Varid{importWith}\;\Conid{Γ}{}\<[42]%
\>[42]{},{}\<[42E]%
\>[45]{}\Varid{weak}\;\Varid{x}\;\Varid{↦}\;\Varid{nameOf}{}\<[E]%
\\
\>[11]{}\Keyword{in}\;\Conid{Let}\;\Varid{strongOf}\;\Varid{f}\;(\Varid{mapVarArgs}\;(\Varid{lookupCEnv}\;\Conid{Γ})\;\Varid{as})\;(\Varid{strengthenTG′}\;\Varid{weak}\;\Varid{nextOf}\;\Conid{Γ′}\;\Varid{u}){}\<[E]%
\\
\>[3]{}\hsindent{5}{}\<[8]%
\>[8]{}\Keyword{where}\;\Keyword{open}\;\Conid{FreshPack}\;\Varid{fr}{}\<[E]%
\ColumnHook
\end{hscode}\resethooks

\noindent
Both sequential and parallel composition
are implemented by inserting the material of one graph
between the innermost \ensuremath{\Conid{Let}} and the \ensuremath{\Conid{Output}} of the other graph.
We define a general helper function for this purpose:

\begin{hscode}\SaveRestoreHook
\column{B}{@{}>{\hspre}l<{\hspost}@{}}%
\column{3}{@{}>{\hspre}l<{\hspost}@{}}%
\column{11}{@{}>{\hspre}l<{\hspost}@{}}%
\column{14}{@{}>{\hspre}l<{\hspost}@{}}%
\column{17}{@{}>{\hspre}l<{\hspost}@{}}%
\column{20}{@{}>{\hspre}l<{\hspost}@{}}%
\column{70}{@{}>{\hspre}l<{\hspost}@{}}%
\column{E}{@{}>{\hspre}l<{\hspost}@{}}%
\>[3]{}\Varid{inLet′}\;{}\<[11]%
\>[11]{}\mathbin{:}\;{}\<[14]%
\>[14]{}\{\mskip0.5mu \Varid{α}\;\Varid{β}\;\mathbin{:}\;\Conid{World}\mskip0.5mu\}\;\Varid{→}\;(\Varid{s}\;\mathbin{:}\;\Varid{α}\;\Varid{★↼→}\;\Varid{β})\;\Varid{→}\;\Conid{Fresh}\;\Varid{β}\;\Varid{→}\;\{\mskip0.5mu \Varid{m}\;\Varid{n}\;\Varid{n′}\;\mathbin{:}\;\Conid{ℕ}\mskip0.5mu\}\;{}\<[E]%
\\
\>[11]{}\Varid{→}\;{}\<[14]%
\>[14]{}({}\<[17]%
\>[17]{}\{\mskip0.5mu \Varid{γ}\;\mathbin{:}\;\Conid{World}\mskip0.5mu\}\;\Varid{→}\;(\Varid{s′}\;\mathbin{:}\;\Varid{α}\;\Varid{★↼→}\;\Varid{γ})\;\Varid{→}\;\Conid{Fresh}\;\Varid{γ}{}\<[E]%
\\
\>[17]{}\Varid{→}\;{}\<[20]%
\>[20]{}\Conid{Args}\;\Varid{γ}\;\Varid{m}\;\Varid{n}\;\Varid{→}\;\Conid{TG′}\;\Varid{\char95 ↼→\char95 }\;\Varid{γ}\;\Varid{m}\;\Varid{n′})\;{}\<[E]%
\\
\>[11]{}\Varid{→}\;{}\<[20]%
\>[20]{}\Conid{TG′}\;\Varid{\char95 ↼→\char95 }\;\Varid{β}\;\Varid{m}\;\Varid{n}\;\Varid{→}\;\Conid{TG′}\;\Varid{\char95 ↼→\char95 }\;\Varid{β}\;\Varid{m}\;\Varid{n′}{}\<[E]%
\\
\>[3]{}\Varid{inLet′}\;\Varid{s}\;\Varid{fr}\;\Conid{F}\;(\Conid{Let}\;\Varid{x}\;\Varid{f}\;\Varid{t}\;\Varid{u})\;\mathrel{=}\;\Conid{Let}\;\Varid{x}\;\Varid{f}\;\Varid{t}\;(\Varid{inLet′}\;(\Varid{s}\;\Varid{▻}\;\Varid{x})\;\Varid{fr′}\;\Conid{F}\;\Varid{u})\;{}\<[70]%
\>[70]{}\Keyword{where}\;\Varid{fr′}\;\mathrel{=}\;\Conid{StrongPack.nextOf}\;\Varid{x}{}\<[E]%
\\
\>[3]{}\Varid{inLet′}\;\Varid{s}\;\Varid{fr}\;\Conid{F}\;(\Conid{Output}\;\Varid{as})\;\mathrel{=}\;\Conid{F}\;\Varid{s}\;\Varid{fr}\;\Varid{as}{}\<[E]%
\ColumnHook
\end{hscode}\resethooks

\noindent
We first implement fork,
which walks the only primitively available fresh link
\ensuremath{\Varid{freshø}} past all the \ensuremath{\Conid{Let}}s of \ensuremath{\Varid{g₁}},
uses the resulting fresh link \ensuremath{\Varid{fr}} to rename \ensuremath{\Varid{g₂}},
and afterwards adapts the output list \ensuremath{\Varid{as₁}} of \ensuremath{\Varid{g₁}}
to the inner world of the renamed \ensuremath{\Varid{g₂}},
so that the two output lists can be concatenated:

\begin{hscode}\SaveRestoreHook
\column{B}{@{}>{\hspre}l<{\hspost}@{}}%
\column{3}{@{}>{\hspre}l<{\hspost}@{}}%
\column{6}{@{}>{\hspre}l<{\hspost}@{}}%
\column{12}{@{}>{\hspre}l<{\hspost}@{}}%
\column{26}{@{}>{\hspre}l<{\hspost}@{}}%
\column{E}{@{}>{\hspre}l<{\hspost}@{}}%
\>[3]{}\Varid{forkTG′}\;{}\<[12]%
\>[12]{}\mathbin{:}\;\{\mskip0.5mu \Varid{m}\;\Varid{n₁}\;\Varid{n₂}\;\mathbin{:}\;\Conid{ℕ}\mskip0.5mu\}\;{}\<[E]%
\\
\>[12]{}\Varid{→}\;\Conid{TG′}\;\Varid{\char95 ↼→\char95 }\;\Varid{ø}\;\Varid{m}\;\Varid{n₁}\;{}\<[E]%
\\
\>[12]{}\Varid{→}\;\Conid{TG′}\;\Varid{\char95 ↼→\char95 }\;\Varid{ø}\;\Varid{m}\;\Varid{n₂}\;{}\<[E]%
\\
\>[12]{}\Varid{→}\;\Conid{TG′}\;\Varid{\char95 ↼→\char95 }\;\Varid{ø}\;\Varid{m}\;(\Varid{n₁}\;\Varid{+}\;\Varid{n₂}){}\<[E]%
\\
\>[3]{}\Varid{forkTG′}\;\{\mskip0.5mu \Varid{m}\mskip0.5mu\}\;\{\mskip0.5mu \Varid{n₁}\mskip0.5mu\}\;\{\mskip0.5mu \Varid{n₂}\mskip0.5mu\}\;\Varid{g₁}\;\Varid{g₂}\;\mathrel{=}\;\Varid{inLet′}\;\Varid{ε}\;\Varid{freshø}\;{}\<[E]%
\\
\>[3]{}\hsindent{3}{}\<[6]%
\>[6]{}(\Varid{λ}\;\{\mskip0.5mu \Varid{γ}\mskip0.5mu\}\;\Varid{s′}\;\Varid{fr}\;\Varid{as₁}\;\Varid{→}\;\Varid{inLet′}\;\Varid{ε}\;\Varid{fr}\;{}\<[E]%
\\
\>[6]{}\hsindent{20}{}\<[26]%
\>[26]{}(\Varid{λ}\;\Varid{s′′}\;\anonymous \;\Varid{as₂}\;\Varid{→}\;\Conid{Output}\;(\Varid{mapVarArgs}\;(\Varid{import⊆}\;(\Varid{★↼→-⊆}\;\Varid{s′′}))\;\Varid{as₁}\;\plus \;\Varid{as₂}))\;{}\<[E]%
\\
\>[6]{}\hsindent{20}{}\<[26]%
\>[26]{}(\Varid{strengthenTG′}\;\{\mskip0.5mu \Varid{\char95 ↼→\char95 }\mskip0.5mu\}\;\Conid{StrongPack.weakOf}\;\Varid{fr}\;\Varid{emptyCEnv}\;\Varid{g₂}){}\<[E]%
\\
\>[3]{}\hsindent{3}{}\<[6]%
\>[6]{})\;\Varid{g₁}{}\<[E]%
\ColumnHook
\end{hscode}\resethooks

\noindent
The implementation of parallel composition
then relies on fork in the same way as that for TG:

\noindent
\begin{hscode}\SaveRestoreHook
\column{B}{@{}>{\hspre}l<{\hspost}@{}}%
\column{3}{@{}>{\hspre}l<{\hspost}@{}}%
\column{11}{@{}>{\hspre}l<{\hspost}@{}}%
\column{26}{@{}>{\hspre}l<{\hspost}@{}}%
\column{E}{@{}>{\hspre}l<{\hspost}@{}}%
\>[3]{}\Varid{parTG′}\;{}\<[11]%
\>[11]{}\mathbin{:}\;\{\mskip0.5mu \Varid{m₁}\;\Varid{n₁}\;\mathbin{:}\;\Conid{ℕ}\mskip0.5mu\}\;{}\<[26]%
\>[26]{}\Varid{→}\;\Conid{TG′}\;\Varid{\char95 ↼→\char95 }\;\Varid{ø}\;\Varid{m₁}\;\Varid{n₁}\;{}\<[E]%
\\
\>[11]{}\Varid{→}\;\{\mskip0.5mu \Varid{m₂}\;\Varid{n₂}\;\mathbin{:}\;\Conid{ℕ}\mskip0.5mu\}\;{}\<[26]%
\>[26]{}\Varid{→}\;\Conid{TG′}\;\Varid{\char95 ↼→\char95 }\;\Varid{ø}\;\Varid{m₂}\;\Varid{n₂}\;{}\<[E]%
\\
\>[26]{}\Varid{→}\;\Conid{TG′}\;\Varid{\char95 ↼→\char95 }\;\Varid{ø}\;(\Varid{m₁}\;\Varid{+}\;\Varid{m₂})\;(\Varid{n₁}\;\Varid{+}\;\Varid{n₂}){}\<[E]%
\\
\>[3]{}\Varid{parTG′}\;\{\mskip0.5mu \Varid{m₁}\mskip0.5mu\}\;\Varid{g₁}\;\{\mskip0.5mu \Varid{m₂}\mskip0.5mu\}\;\Varid{g₂}\;\mathrel{=}\;\Varid{forkTG′}\;(\Varid{extendTG′}\;\Varid{m₂}\;\Varid{g₁})\;(\Varid{shiftTG′}\;\Varid{m₁}\;\Varid{g₂}){}\<[E]%
\ColumnHook
\end{hscode}\resethooks

\noindent
Sequential composition follows the same pattern as \ensuremath{\Varid{forkTG′}},
and first traverses the declarations of \ensuremath{\Varid{g₁}}, which are preserved,
but uses the helper function \ensuremath{\Varid{mapArgsTG′}}
to properly replace any occurrence of inputs
in argument and output lists of the renamed \ensuremath{\Varid{g₂}}
with the corresponding output nodes of \ensuremath{\Varid{g₁}},
after adapting them to the respective nested world.

\begin{hscode}\SaveRestoreHook
\column{B}{@{}>{\hspre}l<{\hspost}@{}}%
\column{3}{@{}>{\hspre}l<{\hspost}@{}}%
\column{6}{@{}>{\hspre}l<{\hspost}@{}}%
\column{11}{@{}>{\hspre}l<{\hspost}@{}}%
\column{26}{@{}>{\hspre}l<{\hspost}@{}}%
\column{E}{@{}>{\hspre}l<{\hspost}@{}}%
\>[3]{}\Varid{seqTG′}\;{}\<[11]%
\>[11]{}\mathbin{:}\;\{\mskip0.5mu \Varid{k}\;\Varid{m}\;\Varid{n}\;\mathbin{:}\;\Conid{ℕ}\mskip0.5mu\}\;{}\<[E]%
\\
\>[11]{}\Varid{→}\;\Conid{TG′}\;\Varid{\char95 ↼→\char95 }\;\Varid{ø}\;\Varid{k}\;\Varid{m}\;{}\<[E]%
\\
\>[11]{}\Varid{→}\;\Conid{TG′}\;\Varid{\char95 ↼→\char95 }\;\Varid{ø}\;\Varid{m}\;\Varid{n}\;{}\<[E]%
\\
\>[11]{}\Varid{→}\;\Conid{TG′}\;\Varid{\char95 ↼→\char95 }\;\Varid{ø}\;\Varid{k}\;\Varid{n}{}\<[E]%
\\
\>[3]{}\Varid{seqTG′}\;\Varid{g₁}\;\Varid{g₂}\;\mathrel{=}\;\Varid{inLet′}\;\Varid{ε}\;\Varid{freshø}\;{}\<[E]%
\\
\>[3]{}\hsindent{3}{}\<[6]%
\>[6]{}(\Varid{λ}\;\{\mskip0.5mu \Varid{γ}\mskip0.5mu\}\;\Varid{s′}\;\Varid{fr}\;\Varid{as₁}\;\Varid{→}\;\Varid{mapArgsTG′}\;\Varid{ε}\;{}\<[E]%
\\
\>[6]{}\hsindent{20}{}\<[26]%
\>[26]{}(\Varid{λ}\;\Varid{s′′}\;\Varid{as}\;\Varid{→}\;\Varid{seqArgs}\;(\Varid{mapVarArgs}\;(\Varid{import⊆}\;(\Varid{★↼→-⊆}\;\Varid{s′′}))\;\Varid{as₁})\;\Varid{as})\;{}\<[E]%
\\
\>[6]{}\hsindent{20}{}\<[26]%
\>[26]{}(\Varid{strengthenTG′}\;\{\mskip0.5mu \Varid{\char95 ↼→\char95 }\mskip0.5mu\}\;\Conid{StrongPack.weakOf}\;\Varid{fr}\;\Varid{emptyCEnv}\;\Varid{g₂}){}\<[E]%
\\
\>[3]{}\hsindent{3}{}\<[6]%
\>[6]{})\;\Varid{g₁}{}\<[E]%
\ColumnHook
\end{hscode}\resethooks

\noindent
Finally, it is also reasonably easy to convert
a \ensuremath{\Conid{TG'}} term graph into a \ensuremath{\Conid{Jungle}}
with \ensuremath{\Conid{Fin}\;\Varid{k}} as \ensuremath{\Conid{Inner}} node set and as \ensuremath{\Conid{Edge}} set,
where \ensuremath{\Varid{k}} is the number of \ensuremath{\Conid{Let}} declarations.


\section{Conclusion and Outlook}

Formalising mathematical definitions of term graphs and their
operations in Agda
is a remarkably straight-forward exercise,
and, due to the dependent typing of Agda,
also carries over to typed term graphs much more easily than
in the more restricted type systems of Haskell or higher-order logic.

The remarkable abstract interface to variable binding provided by
Pouillard and Pottier's \ensuremath{\Conid{NotSoFresh}} Agda library
\cite{Pouillard-Pottier-2010}
also makes name-binding representations of term graphs
conveniently accessible to mechanised reasoning and programmed manipulation.
Typing is easily added to our \ensuremath{\Conid{TG}} and \ensuremath{\Conid{TG′}} datatypes ---
the original \ensuremath{\Conid{Tm}} datatype provided as \ensuremath{\Conid{NotSoFresh}} example
includes typing, but we omitted it here to improve readability.

Implementing additional term graph operations, manipulations, and conversion functions,
and proving the algebraic properties of the term graph operations
is ongoing work.

Future work will strive to base
code-graph based optimised-code generation algorithms
for the Coconut project \cite{Anand-Kahl-2009b}
on our Agda formalisations of code graphs,
with a fully verifying tool chain as ultimate goal.




\end{document}